
\input harvmac
\noblackbox
\def \smd {$\s$-model.\ }

\def \FM {$F$-model\  }
\def \KM {$K$-model\  }

\def \eq#1 {\eqno {(#1)}}

\def\bl{\left}
\def\br{\right}

\def \ra {\rightarrow}
\def\k{\kappa}
\def\r{\rho}
\def\a{\alpha}
\def\b{\beta}

\def\g{\gamma}

\def\d{\delta}

\def\e{\epsilon}

\def\p{\phi}

\def\th{\theta}

\def\m{\mu}
\def\n{\nu}

\def\l{\lambda}

\def\s{\sigma}

\def\cA{{\cal A}}
\def \sm {$\s$-model\ }

\def \bd {\bar \del}
\def \bh { {\bar h} }

\def \A { {\bar A} }

\def \ov {\over}

\def\const{{\rm const}}
\def \p {\phi}
\def \vp {\varphi}

\def\bl{\bigl}
\def\br{\bigr}

\def \sms {$\s$-models\ }

\def \bd {\bar \del}

\def \ra {\rightarrow}

\def \na {\nabla }

\def \a {\alpha}
\def \b {\beta}

\def \Tr {{\ \rm Tr \ }}

\def \ln {{\rm \ ln \  }}
\def \det {{\ \rm det \ }}
\def \ch {{\rm cosh \ }}
\def \th {{\rm tanh  }}
\def \l {\lambda}
\def \p {\phi}

\def \m {\mu }
\def \n {\nu}
\def \ep {\epsilon}
\def\g {\gamma}
\def \r {\rho}
\def \k {\kappa }
\def \d {\delta}
\def \o {\omega}
\def \s {\sigma}
\def \t {\theta}

\def \e#1 {{{\rm e}^{#1}}}
\def \const {{\rm const }}

\def \eq#1 {\eqno {(#1)}}
\def \sm {$\s$-model\ }

\def \bd  {{ \bar \del }}

\def \bd  { \bar \del }

\def \A { \bar A}


\def \o {\omega}

\def \p {\phi}
\def \ep {\epsilon}
\def \s {\sigma}
\def \ha {{\textstyle 1\over 2 }}

\def \r {\rho}
\def \d {\delta}
\def \l {\lambda}
\def \m {\mu}

\def \g {\gamma}
\def \n {\nu}

\def \fourth {{1\over 4}}

\def \e#1 {{{\rm e}^{#1}}}
\def \const {{\rm const }}\def \vp {\varphi}

\def \m {\mu}  

\def \ep {\epsilon}

\def \ra {\rightarrow}

\def \const {{\rm const} }

\def \eq#1 {\eqno{(#1)}}
\def \e {\rm e}
\def \ra {\rightarrow }
\def \e#1 {{\rm e}^{#1}}
\gdef \jnl#1, #2, #3, 1#4#5#6{ {\sl #1~}{\bf #2} (1#4#5#6) #3}

\def \ch {\ {\rm cosh} \ }
\def \th {\ {\rm tanh} \ }
\def \ln { {\rm ln } }
\def \sin {\ {\rm sin} \ }
\def \cos { {\rm cos}  }

\def \l {\lambda}
\def \p {\phi}
\def \vp {\varphi}
\def  \g {\gamma}
\def \o {\omega}
\def \r {\rho}

\def\({\left (}
\def\){\right )}
\def\[{\left [}
\def\]{\right ]}

\def \F {{\cal F}}

\def\bd {{\bar \del}}\def \ra {\rightarrow}
\def \vp {\varphi}

\def \eq#1 {\eqno{(#1)}}

\def \a {\alpha}

\def \b {\beta}
\def \k {\kappa}

\def \o {\omega}

\def \p {\phi}
\def \ep {\epsilon}
\def \s {\sigma}
\def \r {\rho}
\def \d {\delta}
\def \l {\lambda}
\def \m {\mu}
\def \g {\gamma}
\def \n {\nu}

\def \fourth {{1\over 4}}

\def \e#1 {{\rm e}^{#1}}
\def \const {{\rm const }}

\def \vp {\varphi}

\def \ov {\over}

\def \lr { \lref}

\def \smd {$\s$-model.\ }

\def \four{{\textstyle{1\over 4}}}

\def \fourth{{{1\over 4}}}

\def\np {  Nucl. Phys. }
\def \pl { Phys. Lett. }
\def \mpl { Mod. Phys. Lett. }
\def \prl { Phys. Rev. Lett. }
\def \pr  { Phys. Rev. }
\def \ap  { Ann. Phys. }
\def \cmp { Commun. Math. Phys. }
\def \ijmp { Int. J. Mod. Phys. }
\def \ijmp { Int. J. Mod. Phys. }
\def \jmp { J. Math. Phys. }

\def \cqg {Class. Quant. Grav.}
\def \lr {\lref}

\lref \tss { A.A. Tseytlin, \pl  B317(1993)559. }
\lref \jack {I. Jack, D.R.T. Jones and N.  Mohammedi, \np B332(1990)333.}
\lref \calg { C.G. Callan and Z. Gan, \np B272(1986)647. }
\lref \ttss {A.A. Tseytlin, \pl B178(1986)34. }
\lref \fri  {D.H. Friedan, \prl 45(1980)1057; Ann. Phys. (NY) 163(1985)318. }
\lref \tsss { A.A. Tseytlin, \ijmp A4(1989)1257. }
\lref \osb { H. Osborn, Ann. Phys. (NY) 200(1990)1.}
\lref \jackk {I. Jack, D.R.T. Jones and D.A. Ross, \np B307(1988)130.}
\lref \jackkk {I. Jack, D.R.T. Jones and N. Mohammedi, \np B332(1990)333.}
\lref \shts  {A.S. Schwarz and A.A. Tseytlin, \np B399(1993)691.}

\lr \sfeets { K. Sfetsos and A.A. Tseytlin, \jnl \np, B427, 245, 1994. }


\lref \berg {E.A. Bergshoeff, R. Kallosh and T. Ortin, \pr D47(1993)5444. }

 \lref \kumar {A. Kumar,
\pl B293(1992)49; D. Gershon, preprint TAUP-2005-92.}
\lref  \hussen {  S. Hussan and A. Sen,  \np B405(1993)143. }
\lref \kiri {E. Kiritsis, \np B405(1993)109. }
\lref \alv { E. Alvarez, L. Alvarez-Gaum\'e, J. Barb\'on and Y. Lozano,
preprint CERN-TH.6991/93.}

\lref  \rocver { M. Ro\v cek and E. Verlinde, \jnl \np, B373, 630, 1992.}

\lref \brink{ H.W. Brinkmann, {\sl Math. Ann.} {\bf 94} (1925) 119.}
\lref \guv {R. Guven, Phys. Lett. B191(1987)275.}

\lref \hor { G. Horowitz and A.R. Steif, Phys.Rev.Lett. 64(1990)260;
Phys.Rev. D42(1990)1950.}

\lref \horr {G. Horowitz, in: {\it Proceedings
of  Strings '90},
College Station, Texas, March 1990 (World Scientific,1991).}


\lr \mel { M.A. Melvin, \jnl \pl,  8, 65,  1964.
  }

\lref \duval { C. Duval, G.W. Gibbons and P.A. Horv\'athy, \jnl \pr,  D43,
3907, 1991;
C.Duval, G.W. Gibbons, P.A. Horv\'athy and M.J. Perry, unpublished (1991).}
\lref \duvall {
C. Duval, Z. Horvath and P.A. Horvathy, \jnl \pl,  B313, 10, 1993.}

\lref \tsnuu { A.A. Tseytlin,  \jnl \pl, B288, 279,  1992; \jnl \pr,  D47,
3421, 1993.}

\lref \dunu { G. Horowitz and A.A. Steif, \pl B250(1990)49.}
\lref \pol{ E. Smith and J. Polchinski, \pl
B263(1991)59.}

 \lref \busc { T.H. Buscher, \pl B194(1987)59 ; \pl B201(1988)466.}
\lref \pan  { J. Panvel, \pl B284(1992)50. }

\lref \mye {  R. Myers, \jnl \pl,  B199, 371, 1987.}

  \lr \antm{   I. Antoniadis, C. Bachas, J. Ellis, D. Nanopoulos,
\jnl \pl,  B211, 393, 1988;
\jnl  \np,  B328, 115, 1989. }
\lref \givpas { A. Giveon and A. Pasquinucci, \jnl \pl,  B294, 162, 1992.  }
\lref \GK {A. Giveon and E. Kiritsis,  preprint CERN-TH.6816/93,
RIP-149-93.}
\lref \givroc {A. Giveon and M. Ro\v{c}ek, Nucl. Phys. B380(1992)128.}
\lref \giv {  A. Giveon, \jnl \mpl, A6, 2843, 1991. }

\lref \bars {I. Bars, preprint USC-91-HEP-B3;
 E. Kiritsis, \mpl A6(1991)2871. }

\lref \GRV {A. Giveon, E. Rabinovici and G. Veneziano, Nucl. Phys.
B322(1989)167;
A. Shapere and F. Wilczek, \np B320(1989)669.}
\lref \GMR {A. Giveon, N. Malkin and E. Rabinovici, Phys. Lett. B238(1990)57.}
\lref \vene   { K. Meissner and G. Veneziano, \pl B267(1991)33;
M. Gasperini, J. Maharana and G. Veneziano, \pl B272(1991)277; \pl
B296(1992)51.}


\lref \VV    {  G. Veneziano, \jnl \pl,   B265,  287, 1991.}

\lref \tsmpl {A.A. Tseytlin, \mpl A6(1991)1721.}

\lref \nsw {  K.S. Narain, M.H. Sarmadi and E. Witten, \np B279(1987)369. }

\lref \cec { S. Cecotti, S. Ferrara and L. Girardello, \ jnl \np,  B308, 436,
1988. }


\lref \tsdu { A.A. Tseytlin, \pl B242(1990)163; \np B350(1991)395.  }


\lref \wi { E. Witten, unpublished (1991). }
\lref \kik {  K.~Kikkawa and M.~Yamanaka,  \jnl \pl, B149, 357, 1984;
N.~Sakai and I.~Senda, {{\sl  Progr. Theor. Phys.}}
{{\bf 75}} (1986) 692;
M.~B.~Green,
 J.~H.~Schwarz and L.~Brink, {{\sl Nucl.Phys}}.  {{\bf B198}} (1982) 474. }
\lref\nair{ V.~Nair, A.~Shapere, A.~Strominger and F.~Wilczek,
{{\sl Nucl. Phys}}. {{\bf B287}} (1987) 402.
}

\lref \vaf {R. Brandenberger and C. Vafa, \jnl \np, B316,  391, 1988; A.A.
Tseytlin and C. Vafa, \jnl \np,  B372, 443, 1992.}
\lref \por{ A. Giveon, M. Porrati and E. Rabinovici, \jnl {\sl Phys. Rept.}
{\bf 244} (1994) 77.}
\lref \hht{G. Horowitz and A. Steif, \jnl \pl,  B258, 91, 1991.}

\lr \sts{ A.S. Schwarz and A.A. Tseytlin, \np B399 (1993) 691. }

\lr \gtwo {E. Del Giudice, P. Di Vecchia and S. Fubini, Ann. Phys. 70
(1972) 378; K. A. Friedman and C. Rosenzweig, Nuovo Cimento 10A (1972) 53;
S. Matsuda and T. Saido, Phys. Lett. B43 (1973) 123; M. Ademollo {\sl et al},
Nuovo Cimento A21 (1974) 77;
S. Ferrara, M. Porrati and V.L. Teledgi, Phys. Rev. D46
(1992) 3529.}

\lr \plane { D. Amati and C. Klim\v c\'\i k,
\jnl \pl, B219, 443, 1989;
 G. Horowitz and A. Steif,  \jnl \prl, 64, 260, 1990; \jnl \pr,
D42, 1950, 1990;
 G. Horowitz, in: {\it
 Strings '90}, eds. R Arnowitt et. al.
 (World Scientific, Singapore, 1991).}

\lr \tsett {A.A. Tseytlin, \pl B317 (1993) 559.}
\lr \wit { E. Witten,  \prd D44 (1991) 314. }
\lref \dvv  { R. Dijkgraaf, H. Verlinde and E. Verlinde, \np B371 (1992) 269. }

\lr \gibma {
G.W.  Gibbons, in: {\it Fields and Geometry}, Proceedings of the 22-nd Karpacz
Winter School of Theoretical Physics, ed. A. Jadczyk (World Scientific,
Singapore,  1986).}

\lref \tsnul { A.A. Tseytlin, \jnl \np, B390, 153, 1993.}

\lr\ant{I. Antoniadis, C. Bachas and A. Sagnotti, \pl B235 (1990) 255.}
\lr\bak{C. Bachas and E. Kiritsis, \pl B325 (1994) 103. }

\lr \quev {C. Burgess and F. Quevedo,  \np B 421 (1994) 373. }
  \lr \nahm { W. Nahm, \np B124 (1977) 121. }
 \lr \kkkk {  E. Kiritsis and C. Kounnas,  ``Curved four-dimensional spacetime
as infrared regulator in superstring theories", hep-th/9410212. }
\lr \senn{A. Sen, \jnl \pr, D32, 2102,  1985.}
\lr \hulw { C. M. Hull and E.  Witten,  \jnl \pl, B160, 398, 1985. }

\lr\hult { C. Hull and P. Townsend, \jnl \pl, B178, 187, 1986. }
\lr \gps {S.  Giddings, J. Polchinski and A. Strominger, \jnl  \pr,  D48,
 5784, 1993. }

\lr\horrt{G.T. Horowitz and A.A. Tseytlin,\jnl  \prl,  73, 3351,  1994.}

\lref \klts {C. Klim\v c\'\i k  and A.A. Tseytlin, \jnl \np, B424, 71, 1994.}

\lr \attick {J.J. Attick  and E. Witten, \np B310 (1988) 291. }
\lr\fund{A. Dabholkar, G. Gibbons, J. Harvey and F. Ruiz Ruiz, \jnl \np, B340,
33, 1990;
D. Garfinkle, \jnl \pr, D46, 4286, 1992; A. Sen, \jnl \np, B388, 457, 1992;
D. Waldram, \jnl \pr, D47, 2528, 1993. }

\lr \dabha{A. Dabholkar and J. Harvey, \jnl \prl, 63, 719, 1989. }

\lr \duality { L. Brink, M. Green and J. Schwarz, \np B198 (1982) 474;
K. Kikkawa and M. Yamasaki, \pl B149 (1984) 357;
N. Sakai and I. Senda, { Progr. Theor. Phys. }  75 (1984) 692. }

\lr \canon {A. Giveon, E. Rabinovici and G. Veneziano, \np B332 (1989)167;
K. Meissner and G. Veneziano, \pl B267 (1991) 33;
E. \' Alvarez, L. \' Alvarez-Gaum\' e and Y. Lozano, \pl B336 (1994) 183. }

\lr\vene{G. Veneziano, \jnl \pl,  B265, 287,  1991;
K. Meissner and G. Veneziano, \jnl \pl,  B267, 33, 1991.}

\
\lr \kkl {E. Kiritsis, C. Kounnas and D. L\" ust, \pl B331 (1994)321.}
\lr \kk {E. Kiritsis and  C. Kounnas,  \pl B320(1994)361.}
\lref \kallosh { E. Bergshoeff, R. Kallosh and T. Ort\' \i n, \jnl \pr,  D47,
5444,
1993; E. Bergshoeff, I. Entrop and R. Kallosh,
\jnl \pr, D49, 6663, 1994.   }

\lref \duff{M. Duff, B. Nilsson and C. Pope, \jnl \pl, B163, 343,  1985;
R. Nepomechie, \jnl \pr, D33, 3670, 1986. }

\lr\frats { E.S. Fradkin and A.A. Tseytlin , \pl B163 (1985) 123. }
\lr\tsey{  A.A. Tseytlin, \pl B202 (1988) 81.  }
\lr\tset { A.A. Tseytlin, \np B350 (1991) 395.}
\lr\metsa {R.R. Metsaev and A.A. Tseytlin, \np B298 (1988) 109. }
\lr\abo { A. Abouelsaood, C. Callan, C. Nappi and S. Yost, \np B280 (1987) 599.
}
\lr\mono{T. Banks, M. Dine, H. Dijkstra and W. Fischler, \pl B212 (1988) 45.}
\lr\rb {I. Robinson, Bull. Acad. Polon. Sci. 7 (1959) 351;
B. Bertotti, \pr 116 (1959) 1331. }

\lr\gaun { H.F. Dowker, J.P. Gauntlett, D.A. Kastor and J. Traschen,
\prd D49 (1994) 2909;
 H.F. Dowker, J.P. Gauntlett, S.B. Giddings and G.T. Horowitz, \prd D50 (1994)
2662; S.W. Hawking, G.T. Horowitz and  S.F. Ross, ``Entropy, area and black
hole pairs", NI-94-012, DAMTP/R 94-26, UCSBTH-94-25, gr-qc/9409013. }

 \lr \napwi {C. Nappi and E. Witten, \jnl \prl,  71, 3751,  1993.}
\lr\tset{A.A. Tseytlin, \pl B317 (1993) 559.}
\lr \ruts { J.G.  Russo and A.A. Tseytlin, ``Constant magnetic field in closed
string theory: an exactly solvable model", CERN-TH.7494/94,
Imperial/TP/94-95/3, hep-th/9411099. }
\lr \givki{A. Giveon and E. Kiritsis,  \np B411 (1994) 487.}
\lr\sft{  K. Sfetsos, \pl B324 (1994) 335.}
\lr\sfts{K. Sfetsos and  A.A. Tseytlin, \np B427 (1994) 325.}
\lr\kum{A. Kumar, \pl B293 (1992) 49; S. Hassan and A. Sen,
\np B405 (1993) 143; E. Kiritsis, \np B405 (1993) 109. }

\lr \tsee{A.A. Tseytlin, ``Exact string solutions and duality", to appear in:
{\it Proceedings of the 2nd Journ\' ee Cosmologie}, ed. H. de Vega  and N. S\'
anchez (World
Scientific), hep-th/9407099. }
\lr \hot { G.T. Horowitz and A.A. Tseytlin, \pr D50 (1994) 5204. }

\lr\busch{T.H. Buscher, Phys. Lett. B194 (1987) 51; B201 (1988) 466.}

\lr \dab {A. Dabholkar, \jnl \np, B439, 650, 1995, hep-th/9408098;
D.A. Lowe and A.  Strominger, \jnl \pr, D51, 1793, 1995,  hep-th/9410215.}

\lr\hrt {G.T. Horowitz and A.A. Tseytlin,  \jnl \pr,  D50, 5204,  1994. }
\lr\bag{ J.A. Bagger, C.G.  Callan and J.A.  Harvey, \np B278 (1986) 550. }

\lr\ant{I. Antoniadis, C. Bachas and A. Sagnotti, \pl B235 (1990) 255.}
\lr\bak{C. Bachas and E. Kiritsis, \pl B325 (1994) 103. }
\lr\rutsn{J.G.  Russo and A.A. Tseytlin,  ``Exactly solvable string models
of curved space-time backgrounds", CERN-TH/95-20, Imperial/TP/94-95/17,
hep-th/9502038. }
\lr\rqsn{J.G.  Russo and A.A. Tseytlin,  ``Heterotic strings in uniform
magnetic field", CERN-TH/95-106, Imperial/TP/94-95/29,
hep-th/9506071. }

\lr \kltspl { C. Klim\v c\'\i k and A.A. Tseytlin, \pl B323 (1994) 305.}

\lr \gps {S.  Giddings, J. Polchinski and A. Strominger,   \pr D48 (1993)
 5784. }

\lr \hoho {J. Horne and G.T. Horowitz, \np B368 (1992) 444. }
\lr \sftse { K. Sfetsos and A.A.  Tseytlin,  \pr D49 (1994) 2933.}

\lr\khu{R. Khuri, \pl B259 (1991) 261; \np B387 (1992) 315.}

\lr \nels {W. Nelson, \pr D49 (1994) 5302.}
\lr \kalor { R. Kallosh and T. Ort\'\i n, \jnl \pr, D50, 7123, 1994.  }

\lr \los { D.A. Lowe and A. Strominger, \jnl \prl,  73, 1468,  1994.}

\lr\khu{R. Khuri, \jnl \pl, B259, 261,  1991.}

\lr\ant{I. Antoniadis, C. Bachas and A. Sagnotti, \jnl \pl,  B235, 255, 1990. }
\lr\bak{C. Bachas and E. Kiritsis, \jnl \pl,  B325, 103,  1994. }

\lr\anto{I. Antoniadis and N. Obers, \jnl \np, B423, 639, 1994}
\lr \oliv { D. Olive, E. Rabinovici and A. Schwimmer,  \jnl \pl, B321, 361,
1994.}
\lr \nons{A.A.  Kehagias and P.A.A.  Meesen, \pl B331 (1994) 77;
J.M. Figueroa-O'Farril and S. Stanciu, \pl B327 (1994) 40;
A. Kehagias, ``All WZW models in $D\leq 5$", hep-th/9406136. }

\lr \busc {T.H. Buscher, \pl  B194 (1987) 59; \pl B201 (1988) 466.}

\lr\givroc{A. Giveon and M. Ro\v{c}ek, Nucl. Phys.  B380 (1992) 128.}

\lref \hhtt{ J. Horne, G. Horowitz and A. Steif, \prl 68(1992)568.
}
\lref \loveo {C. Lovelace, \jnl \pl,  B135, 75, 1984;
P. Candelas, G. Horowitz, A. Strominger and E. Witten,
\jnl\np, B258, 46, 1985.}
\lref \lovet {C. Lovelace, \jnl \np,  B273, 413, 1986;
B. Fridling and A. Jevicki, \jnl\pl, B174, 75, 1986.}

\lr \scherk {J. Scherk  and J.H. Schwarz,
\jnl \np, B81, 118, 1974;
T. Yoneya, {\sl Progr. Theor. Phys.} {\bf 51} (1974) 1907.}
\lr \honer { J. Honerkamp, \jnl \np, B36, 130, 1972;
D.  Friedan, \jnl\prl, 51, 334, 1981; \jnl \ap, 163, 318, 1985;
T. Curthright  and C. Zachos,  \jnl \prl, 53,  1799, 1984.}

\lr \tseytlin { A.A. Tseytlin, \jnl \ijmp, A4, 1257, 1989. }
\lr \zamo { A.B. Zamolodchikov, {\sl JETP Lett.} {\bf 43} (1986) 730;
A.M. Polyakov, {\sl Phys. Scripta }  {\bf T15} (1986) 191.}

\lr \tseyy {A.A.  Tseytlin, \jnl \pl, B208, 221, 1988.}
\lr\osb {H. Osborn, \jnl \np, B308, 629, 1988. }

\def\ha{{\textstyle{1\over2}}}

\lref \vene    { K. Meissner and G. Veneziano, \jnl \pl,  B267, 33, 1991;
A. Sen,  \jnl \pl,   B271,  295, 1991.}


\baselineskip8pt
\Title{\vbox
{\baselineskip 6pt{\hbox{Imperial/TP/94-95/28  }}{\hbox
{hep-th/9505052 }}{\hbox{ revised }} } }
{\vbox{\centerline {  Exact  solutions  of   closed   string theory }
 }}
\vskip  15 true pt
\centerline{   A.A. Tseytlin\footnote{$^{\star}$}{\baselineskip8pt
e-mail address: tseytlin@ic.ac.uk}\footnote{$^{\dagger}$}{\baselineskip5pt
On leave  from Lebedev  Physics
Institute, Moscow, Russia.} }

\smallskip\smallskip
\centerline {\it  Theoretical Physics Group, Blackett Laboratory}

\centerline {\it  Imperial College,  London SW7 2BZ, U.K. }
\bigskip\bigskip
\centerline {\bf Abstract}
\bigskip
\baselineskip6pt
\noindent
We review explicitly known exact $D=4$ solutions with  Minkowski
signature in closed bosonic string theory. Classical string solutions
with space-time interpretation are represented by conformal sigma models.
Two large (intersecting) classes of solutions are described by
gauged WZW models and `chiral null models' (models with conserved
chiral null current). The latter class includes  plane-wave type
backgrounds (admitting a covariantly constant null Killing vector)
and backgrounds with two null Killing vectors (e.g., fundamental
string solution). $D>4$ chiral null models describe some exact $D=4$
solutions with electromagnetic fields, for example, extreme electric
black holes, charged fundamental strings and their generalisations.
In addition, there exists a class of conformal models representing
axially symmetric stationary magnetic flux tube backgrounds
(including, in particular, the dilatonic Melvin solution).
In contrast to spherically symmetric chiral null models for which
the corresponding conformal field theory is not known  explicitly,
the magnetic flux tube models (together with some non-semisimple
WZW models) are among the first examples of  solvable unitary
conformal string models with  non-trivial $D=4$  curved
space-time interpretation. For these models one is able to
express the quantum hamiltonian in terms of free fields and to
find  explicitly the physical spectrum and string partition function.

\bigskip

{\centerline {\it To appear as a review in   Classical and Quantum Gravity}}
\medskip
\Date {May 1995}

\noblackbox
\noblackbox

\vfill\eject

\def \lr { \lref}

\lr \mans { P. Mansfield and J. Miramontes, \jnl \pl, B199, 224, 1988;
A.A. Tseytlin, \jnl \pl, B208, 228, 1988; \jnl \pl, B223, 165, 1989.}

\lr \kalmor{R. Kallosh and A. Morozov,  \jnl \ijmp,  A3, 1943, 1988.}

\lr \ghrw{J. Gauntlett, J. Harvey, M. Robinson, and D. Waldram,
\jnl \np, B411, 461, 1994.}
\lr \garf{D. Garfinkle, \jnl \pr, D46, 4286, 1992.}

\lr \onofri { V. Fateev, E. Onofri and Al. Zamolodchikov, \jnl \np, B406,
 521, 1993.}

\lref \tspl {A.A. Tseytlin, \jnl \pl, B317, 559, 1993.}
\lref \tssfet { K. Sfetsos and A.A.  Tseytlin, \jnl  \pr, D49, 2933, 1994.}

\lr \sfexac {K. Sfetsos,  \jnl \np, B389, 424,  1993.}

\lr \tsmac{A.A. Tseytlin, \jnl \pl,  B251, 530, 1990.}

\lr \cakh{C. Callan and R. Khuri, \jnl \pl, B261, 363, 1991;
R. Khuri, \jnl \np, B403, 335, 1993.}
\lr \dgt{M. Duff, G. Gibbons and P. Townsend, ``Macroscopic superstrings
as interpolating solitons", DAMTP/R-93/5, hep-th/9405124.}

\lref \ger {A. Gerasimov, A. Morozov, M. Olshanetsky, A. Marshakov and S.
Shatashvili, \jnl \ijmp,
A5, 2495,  1990. }

\lr \hutow{C. Hull and P. Townsend, \jnl \np, B274, 349, 1986.}
\lr \mukh {S. Mukhi,    \jnl \pl,  B162, 345, 1985;
S. De Alwis, \jnl \pl, B164, 67, 1985. }


 \lr \lov  {C. Lovelace,  \jnl \np,  B273, 413,  1986.}
\lr \call{C. Callan, D. Friedan, E. Martinec and  M. Perry, \jnl \np, B262,
593, 1985.}
\lr \frts {E.S.  Fradkin  and A.A. Tseytlin, \jnl \pl, B158, 316, 1985;
\jnl \np, B261, 1, 1985.}
\lr \tsred{A.A. Tseytlin,
\jnl  \pl, B176, 92, 1986; \jnl  \np, B276, 391, 1986.}
\lr \grwi {D. Gross and E. Witten, \jnl \np, B277, 1, 1986.}

\lr \gps {S.  Giddings, J. Polchinski and A. Strominger, \jnl  \pr,  D48,
 5784, 1993. }

\lr \tsppl  {A.A. Tseytlin, \jnl   \pl,  B208, 221, 1988.}
\lr\rabi  {S. Elitzur, A. Forge and E. Rabinovici, \jnl \np, B359, 581, 1991;
 G. Mandal, A. Sengupta and S. Wadia, \jnl \mpl,  A6, 1685, 1991. }
 \lr \witt{ E. Witten, \jnl \pr, D44, 314, 1991. }
 \lr \dvv { R. Dijkgraaf, H. Verlinde and E. Verlinde, \jnl \np, B371,
269, 1992.}
\lr \hoho { J. Horne and G.  Horowitz, \jnl \np, B368, 444, 1992. }
\lr \horwel{G. Horowitz and D. Welch, \jnl \prl, 71, 328, 1993;
N. Kaloper,  \jnl \pr,  D48, 2598, 1993. }
\lr \host{ G. Horowitz and A. Steif,  \jnl \prl, 64, 260, 1990; \jnl \pr,
D42, 1950, 1990;  G. Horowitz, in: {\it
 Strings '90}, (eds. R Arnowitt et. al.)
 World Scientific (1991).}
\lr \busch {T.  Buscher, \jnl \pl, B194, 59, 1987; \jnl \pl,
 B201, 466, 1988.}
\lr \kallosh {E. Bergshoeff, I. Entrop, and R. Kallosh, ``Exact Duality in
String Effective Action", SU-ITP-93-37; hep-th/9401025.}

\lr \tsmpl {A.A. Tseytlin, \jnl  \mpl, A6, 1721, 1991.}

\lr \kltspl { C. Klim\v c\'\i k and A.A. Tseytlin, \jnl \pl, B323, 305, 1994.}
\lr \shwts { A.S.  Schwarz and A.A. Tseytlin, \jnl \np, B399, 691, 1993.}
\lr \callnts { C. Callan and Z. Gan, \jnl  \np, B272, 647, 1986;  A.A.
Tseytlin,
\jnl \pl,
B178, 34, 1986.}

\lr \guv { R. G\"uven, \jnl \pl, B191, 275, 1987.}

 \lr \desa{ H. de Vega and N. Sanchez, \jnl
\pr, D45, 2783, 1992; \jnl \cqg, 10, 2007, 1993.}
\lr \desas{ H. de Vega and N. Sanchez, \jnl
\pl, B244,  215, 1990.}
\lref \tsnul { A.A. Tseytlin, \jnl \np, B390, 153, 1993.}

\lr \gauged {I. Bars and K. Sfetsos, \jnl  \mpl, A7, 1091, 1992;
 P. Ginsparg and F. Quevedo, \jnl \np, B385, 527, 1992. }

\lr \bsfet {I. Bars and K. Sfetsos, \jnl \pr, D46, 4510, 1992; \jnl \pr,
 D48, 844, 1993. }
\lr \tsnp{ A.A. Tseytlin, \jnl \np, B399, 601, 1993;  \jnl \np, B411, 509,
1994.}
\lr \gibb{A. Dabholkar, G. Gibbons, J. Harvey, and F. Ruiz, \jnl \np, B340,
33, 1990.}
\lr \hhs{J. Horne, G. Horowitz, and A. Steif, \jnl \prl, 68, 568, 1992;
G. Horowitz, in:  {\it  String theory and quantum gravity '92}, ed. J. Harvey
{\it et al}. (World Scientific, Singapore, 1993); hep-th/9210119.}

\lr \jack {I. Jack, D.  Jones and J. Panvel, \jnl \np, B393, 95, 1993.}
\lr \mettstwo {R.R.  Metsaev and A.A. Tseytlin, \jnl \pl, B185, 52, 1987.}
\lr \banks {T. Banks, M. Dine, H. Dijkstra and W. Fischler, \jnl \pl, B212,
45, 1988.}

\lr \horstr{G. Horowitz and A. Strominger, \jnl \np, B360, 197, 1991.}

\lr \givkir {A.  Giveon and E. Kiritsis, \jnl \np, B411, 487, 1994.  }

\lr \jac{I. Jack and D. Jones, \jnl \pl, B200, 453, 1988.}
\lr \metts{R.R. Metsaev and A.A. Tseytlin, \jnl \np,  B293, 385, 1987;
 C. Hull and P. Townsend, \jnl \np,  B301, 197, 1988. }
\lr \horv{ P. Horava, \jnl \pl,
B278, 101, 1992.}
\lr \callkleb { C. Callan, I.  Klebanov and M. Perry,
 \jnl \np,  B278, 78,  1986.   }

\lref \FT {E.S. Fradkin and A.A. Tseytlin, Phys.Lett. B158(1985)316; Nucl.Phys.
B261(1985)1. }
\lref \love {C. Lovelace, \pl B135(1984)75; \np B273(1986)413.}
\lref \sch {J. Scherk and J.H. Schwarz, \np B81(1974)118.}

\lr \hort { G. Horowitz and A.A. Tseytlin,  unpublished (1994).}

\lref \gris {  M.T. Grisaru, A. van de Ven and D. Zanon, \jnl \np, B277,  409,
1986. }

\lr\frats { E.S. Fradkin and A.A. Tseytlin , \jnl \pl, B163, 123,  1985. }
\lr \burg{C.P. Burgess, \jnl \np, B294, 427,  1987; V.V. Nesterenko, \jnl
\ijmp,  A4, 2627,  1989.}
\lr\abo { A. Abouelsaood, C. Callan, C. Nappi and S. Yost, \jnl \np,  B280,
599, 1987.
}
\lr\ferrara{S. Ferrara and M. Porrati, \jnl \mpl,  A8, 2497,  1993.}
\lr \sfees{K. Sfetsos, \jnl \ijmp,  A9, 4759, 1994.}
\lr \kounn {C. Kounnas, in: {\it  Proceedings of the International
Europhysics Conference on High Energy Physics}, Marseille, 22-28 July,  1993;
hep-th/9402080.}

\lr \palla { P. Forg\' acs, P.A. Horv\' athy, Z. Horv\' ath and L. Palla,
``The Nappi-Witten string in the light-cone gauge", hep-th/9503222.}

\lref \wit {E. Witten, \jnl \cmp,  92, 455, 1984.}
\lref\bra{ E. Braaten, T.L. Curtright and C.K. Zachos, \jnl  \np,  B260,  630,
1985.}

\lr\gepner{  D. Gepner, \jnl \np, B296,  757, 1988. }

\lr \nemsen{D. Nemeschanski and A. Sen, \jnl \pl, B178, 365, 1986.  }
\lr\cres{ K. Bardakci, M. Crescimanno and E. Rabinovici,
{\sl Nucl. Phys.}    {\bf B344} (1990) 344.}

\lref \gwzw {      K. Bardakci, E. Rabinovici and
B. S\"aring, \jnl \np, B299, 157, 1988;
 K. Gawedzki and A. Kupiainen,
\jnl \np, B320, 625, 1989;  D. Karabali, Q-Han Park, H.J.
Schnitzer and
Z. Yang, \jnl \pl,  B216,  307,  1989;  D. Karabali and H.J. Schnitzer, \jnl
\np,  B329,  649, 1990.
}
\lref \bha { K. Bardakci and M.B. Halpern, \jnl \pr,  D3, 2493,  1971;
M.B. Halpern, \jnl \pr, D4, 2398, 1971.}

\lref\GKO{P. Goddard, A. Kent and D. Olive, \jnl \pl,  B152,   88,  1985.}

\def \bh{\tilde h}

\lref\kou{C. Kounnas and D. L\"ust, \jnl \pl,  B289, 56,  1992.}

\lref\barsf { I. Bars and K. Sfetsos, \jnl \pl,  B277, 269,  1992;
 \jnl \pr,  D46,  4495, 1992.}

\lref\brsne{ I. Bars and D. Nemeschansky, {\sl  Nucl. Phys.}  {\bf B348} (1991)
89.}
\lref \napwit { C. Nappi and E. Witten, \jnl \pl,  B293, 309, 1992.}

\lr\bergsh { E. Bergshoeff, R. Kallosh and T. Ort\'in, \jnl \pr,  D47, 5444,
1993.}
\lref \penr {R. Penrose,  ``Any space-time has a plane wave as a limit",
in: {\it Differential Geometry and Relativity}, A. Cahen and M. Flato eds.,
(Reidel, Dordrecht-Holland), 1976, p. 271. }
\lr\sen{A. Sen, \jnl \np, B388, 457, 1992. }
\lr \garf{D. Garfinkle, \jnl \pr, D46, 4286, 1992.}
\lr \wald {D. Waldram, \jnl \pr, D47, 2528, 1993.}

\lr \callmy { R.C. Myers, \jnl \np, B289, 701, 1987; C.G. Callan, R.C.  Myers
and M.J.  Perry,  \jnl \np, B311, 673, 1989.}

\lr \howe {P. Howe and G. Papadopoulos,  \jnl \pl, B263, 230, 1991; \jnl \cmp,
151, 467, 1993.}
\lr \tsem { A.A. Tseytlin, \jnl \pl, B346, 55, 1995. }

\def \tvp {\tilde \vp}
\def \td {\tilde }

\lr\kk{
E. Kiritsis and C. Kounnas,  \pl B320 (1994) 264;
E. Kiritsis, C. Kounnas and  D. L\"ust, \pl B331 (1994) 321.}
\lr \oliv{ D.  Olive,
 E. Rabinovici and A. Schwimmer,  \jnl \pl, B321, 361, 1994.}
\lr\sfetso{ K. Sfetsos,  \jnl \pl, B324, 335, 1994;
 \jnl \pr,  D50, 2784,  1994.}
\lr \hrts { G.T. Horowitz and A.A. Tseytlin, \jnl \pr, D51, 2896, 1995.}

\lr\figu{
N. Mohammedi, \jnl \pl,  B325, 379,  1994;  {\bf B331} (1994) 93; {\bf  B337}
(1994) 279;
 J.M. Figueroa-O'Farrill and S. Stanciu, \jnl \pl, B327, 40,  1994;
``Nonreductive WZW models and their CFTs", hep-th/9506157;
A.  Kehagias and P.  Meessen, \jnl  \pl,  B331, 77, 1984;
A.  Kehagias, NTUA-45/94, hep-th/9406136.}

\lref \amkl {D. Amati and C. Klim\v c\'\i k,
\jnl \pl, B219, 443, 1989.
}

\lref \hor { G. Horowitz and A.R. Steif, Phys. Rev. Lett. 64 (1990) 260;
Phys. Rev. D42 (1990) 1950.}

\lref \horr {G. Horowitz, in: {\it Proceedings
of  Strings '90},
College Station, Texas, March 1990 (World Scientific,1991).}

\lref \rudd {  R. Rudd, \jnl \np, B352, 489, 1991.}

\lr \deve{H. de Vega and N. Sachez, \pr D45 (1992) 2783; \cqg 10 (1993) 2007.}

\def \cB {{\cal B}}

\lref \hal { M.B. Halpern and E. Kiritsis, \jnl \mpl,  A4, 1373, 1989;
A.Yu. Morozov, A.M. Perelomov, A.A. Rosly, M.A. Shifman and A.V.
Turbiner, \jnl  \ijmp,
A5, 803, 1990.}

\lr \halp { M.B. Halpern, E. Kiritsis, N.A. Obers and K. Clubok, ``Irrational
conformal field theory", hep-th/9501144.}

\lr\tss{A.A. Tseytlin, \jnl\np, B418, 173, 1994.}
\lr\deb{J. de Boer, K. Clubok and M.B. Halpern, \jnl \ijmp,
A9, 2451, 1994.}
\lr\bard{K. Bardakci, \jnl\np, B431, 191, 1994;
K. Bardakci and L. Bernardo, ``Conformal points of the generalized Thirring
model", hep-th/9503143.}

\lr \polyakov { A. Polyakov, \jnl \pl, B103, 207, 1981.  }

\lr \jons { C. Johnson and R.  Myers, \jnl \pr, D50, 6512, 1994.
 }
\lr \galts {D. Galtsov and  O. Kechkin, \jnl \pr, D50, 7394, 1994.}
\lr \gurs{M. G\"urses, \jnl \pr, D46, 2522, 1992.}

\lr \iwp{W. Israel and G. Wilson, \jnl \jmp, 13, 865, 1972;
Z. Perj\'es, \jnl \prl, 27, 1668, 1971.}

\lr \maha{J. Maharana and J. Schwarz, \jnl \np, B390, 3, 1993.}

\lr \jooo{C. Johnson, \jnl \pr, D50, 4032, 1994;
C. Johnson and R. Myers,  ``Conformal field theory of a rotating dyon",
hep-th/9503027.}

\lr \khga {R. Khuri, \jnl \np, B387, 315, 1992;  \jnl \pl, B294,
325, 1992; J. Gauntlett, J. Harvey and J. Liu,
\jnl \np, B409, 363, 1993.}

\lr \dukh {M. Duff, R.  Khuri,  R. Minasyan and J. Rahmfeld, \jnl \np, B418,
195, 1994. }

\lr \nats{M. Natsuume, \jnl \pr, D50, 3949, 1994.}

\lr \nels{W. Nelson, \jnl \pr, D49, 5302, 1994.}

\lr \chs{C. Callan, J. Harvey and A. Strominger, \jnl \np, B359,
 611, 1991;
in {\it Proceedings of the 1991 Trieste Spring School on String Theory and
quantum Gravity}, ed. J. Harvey {\it et al.} (World Scientific, Singapore
1992).}

\lr \khga {R. Khuri, \jnl \np, B387, 315, 1992;  \jnl \pl, B294,
325, 1992; J. Gauntlett, J. Harvey and J. Liu, \jnl \np, B409, 363, 1993.}

\lr \dukh {M. Duff, R.  Khuri,  R. Minasyan and J. Rahmfeld,
\jnl \np, B418, 195, 1994. }
\lr \sus{ P. Howe and G. Papadopoulos, \jnl \np, B289, 264, 1987. }

\lr \sss {T. Banks and L. Dixon, \jnl \np, B307, 93, 1988.}
\lr \gaun{J. Gauntlett, talk presented at the conference
``Quantum Aspects of Black Holes",
Santa Barbara, June 1993.}
\lr   \fudd {M.J. Duff, R.R. Khuri and J.X. Lu, ``String Solitons",
NI-94-017, hep-th/9412184.}
\lr \sus{ P. Howe and G. Papadopoulos, \jnl \np, B289, 264, 1987;
\jnl \np, B381, 360, 1992. }
\lr \cct{C. Callan and L. Thorlacius, in: {\it  Proceedings of TASI 1988
School},
Providence, 1989, p. 795.}

\lr\behr{K. Behrndt, \jnl \pl, B348, 395, 1995.}
\lr\behrr{K. Behrndt,
``About a class of exact string backgrounds", hep-th/9506106.}
 \lr\park{J. Park, ``Black hole solutions of Kaluza-Klein supergravity theories
and string theory", hep-th/9503084.}

\lr \gib { G. Gibbons, \jnl \np, B207, 337, 1982. }
\lr \pet {A. Peterman,  Marseille preprint (1993); A. Peterman and  A.
Zichichi,  {\sl Nuovo Cim.} {\bf 106A} (1993) 719. }

\lr \gim { G. Gibbons and K. Maeda,  \jnl \np, B298, 741, 1988. }
\lr\kall { R. Kallosh, D. Kastor, T. Ort\'in   and T. Torma,
\jnl \pr, D50, 6374, 1994.  }
\lr\gar { D. Garfinkle, G. Horowitz and A. Strominger, \jnl \pr,  D43, 3140,
1991; {\bf D45} (1992) 3888(E).}

\lr\kall { R. Kallosh, D. Kastor, T. Ort\'in   and T. Torma,
\jnl \pr, D50, 6374, 1994.  hep-th/9406059. }


\lr \tseer{A.A. Tseytlin, in: {\it  ``From superstrings to supergravity"}, ed.
by M.J. Duff, S. Ferrara and R.R. Khuri (World Scientific, 1994),  p.3;
hep-th/9303054.}

\lref \ora {P. Forg\'acs, A. Wipf, J. Balog, L. Feh\'er and L.
O'Raifeartaigh,
\jnl \pl,  B227, 214, 1989; {\sl Ann. Phys.} {\bf 203} (1990) 76. }

\lr \barss { I. Bars, in: {\it  ``Perspectives in mathematical physics"}, vol.
3, eds. R. Penner and  S.-T. Yau (International Press, 1994), p.51;
hep-th/9309042. }

\lr \kirko { E. Kiritsis  and C.  Kounnas, in: {\it  Proceedings  of 2nd
Journee
Cosmologique}, Paris,  2-4 Jun 1994, ed. by H. de Vega and N. Sanchez;
 hep-th/9407005.}
\lr \hor{G. Horowitz and A. Steif,  \jnl \prl, 64, 260, 1990; \jnl \pr,
D42, 1950, 1990.}
\lr \kk {E. Kiritsis and C. Kounnas, \jnl \pl, B320, 264, 1994;
E. Kiritsis, C. Kounnas and  D. L\"ust, \jnl \pl,  B331, 321,  1994.}
\newsec{Introduction}
String theory is a remarkable extension of  General Relativity
which is consistent at the quantum level.
One of the  important   problems  is to  study
 the set of  exact classical solutions to string equations of motion.
This may clarify its  formal  aspects   but also may be relevant for
understanding the implications
of string theory  for cosmology and black holes (assuming  that  in certain
regions,  e.g., at small times/scales, the effective string coupling is small
so that perturbative and non-perturbative corrections to string equations of
motion can be ignored).
To be able to  address   issues of singularities and
strong field behavior
one  should determine  not just    solutions of the leading-order
low-energy  string effective equations derived under the assumption of small
field  gradients ($|\a'R|\ll  1$, etc.)  but    solutions   that  are exact
 to all orders in  $\a'$  (i.e.,  the corresponding  exact conformal
$\s$-models),  and,   ultimately,   how  first-quantized string modes propagate
in a given background (i.e., the  underlying conformal field theories).

\lr \weh{C.M. Hull and P.K. Townsend, \jnl \np, B274, 349, 1986;
A.A. Tseytlin, \jnl \pl, B178, 34, 1986; \jnl \np, B294, 383, 1987.}

\lref \por{ A. Giveon, M. Porrati and E. Rabinovici, {\sl Phys. Rept.}  {\bf
244} (1994) 77.}

\lr \horry {G.T. Horowitz and D. Marolf, ``Quantum probes of spacetime
singularities", gr-qc/9504028.}

\subsec{Sigma model approach}
In addition to its dependence on  extra   `massless' fields (antisymmetric
tensor and  dilaton in bosonic string case)
string theory `massless' effective action contains terms of all orders in
derivatives
multiplied by powers of $\a'$.   When  this effective
action is derived   by starting from the string S-matrix \scherk\
its full  structure  is hard  to  determine explicitly
and the problem of finding its extrema   exactly to all orders in $\a'$ looks
hopeless.  A remarkable possibility  to find
exact backgrounds  without knowing explicit  form of the effective action
is based   on the correspondence between  classical
string solutions and conformal $\s$-models.

  Let us briefly recall the  basic features of the \sm approach.
The  recognition of the role of \sms in string theory was
prompted by  the covariant path integral representation
for the string-theory scattering amplitudes  \polyakov\
and  by studies of \sms\ as geometrical
2-dimensional quantum field theories \refs{\honer,\wit}.
It was observed \frts\ that the generating functional
for the string scattering amplitudes can be interpreted as
a partition function  $Z$ of a \sm  with couplings being the string space-time
fields. That suggested a relation between  $Z$ and  the string effective action
 $S$ leading to the expression for  $S$ which, in contrast to the one
in the S-matrix approach,  was manifestly covariant and non-perturbative in
number of external field `legs'.
An independent development was  the realisation that a necessary condition
on  a string solution  is that the \sm describing   string propagation
in  this background should be conformally invariant \loveo.
It was checked  \refs{\call,\senn}
that the  leading order conditions of conformal invariance of a \sm do indeed
correspond to extrema of the leading  order term in the effective action.
The proportionality  between the Weyl anomaly coefficients
(`$\bar \b $-functions') \weh\  and  derivatives of the effective action
was suggested to hold also off shell \refs{\zamo,\tsred}
and the field redefinition (scheme choice) ambiguity in the effective action
(`$\bar \b $-functions')  was understood  \refs{\tsred,\grwi}.
That made possible to carry out the non-trivial test   of  equivalence
between the \sm  `$\bar \b $-functions'
and the effective action derived from the string  S-matrix
at  the next to leading order \refs{\callkleb, \metts}.
The observation  that a  renormalisation of the \sm corresponds to  a
subtraction
of massless poles in the string scattering amplitudes \lovet\
led to the expression  for  the   (S-matrix derived)  closed string
effective action $S$
in terms of the  renormalised derivative of the \sm partition function $Z$
over the 2d cutoff \tseyy\
    (correcting the original ansatz $S \sim Z$ of \frts\ which is valid only in
the open string theory).
This representation provided  a basis for a proof
that the extrema of the effective action indeed correspond
to the Weyl invariant \sms\  (\refs{\tseyy,\osb,\tseytlin} and refs. there).
 For reviews of \sm approach see, e.g.,  \refs{\tseytlin,\cct}.

To summarize,
in the  `first-quantized'   (or `$\s$-model')  approach
a string background  is represented   by a 2-dimensional
$\s$-model.  The
fundamental feature of the closed string theory is that
a  large-scale  classical approximation to the geometry is thus  described
not only by the metric but also by the antisymmetric tensor and  the  scalar
dilaton field,
 \eqn\stri{ I= {1\over { 4 \pi \a' }} \int d^2 z \sqrt {\g} \ [  \del_m x^\m
\del^m x^\n G_{\m
\n}(x)\  + \ i \ep^{mn} \del_m x^\m \del_n x^\n  B_{\m\n}(x) \ } $$  +  \ \a'
R^{(2)}  \p (x)
\ ] \  .$$
$\g$ is a  world sheet metric
which decouples once the background fields satisfy the  classical string  field
equations.\foot{The classical string  follows the `geodesic' equations
corresponding to the action  \stri.
One of the consequences  of the  extended nature of  strings is that
while a point-like  string mode
follows the geodesics of the metric,  other string modes
 feel also the antisymmetric tensor background.
First-quantised test string  also feels the dilaton background \refs{\frts}.}

In the  classical   (leading order in string coupling)    approximation
the string field equations are the conditions of conformal (Weyl) invariance of
\sm \stri\ which  are equivalent to the conditions of stationary
of the  tree-level effective action  $S$
\eqn\rele{  \bar \b ^i = \k^{ij}(\vp)  {\delta  S \ov \delta \vp^i} \ ,
\ \ \ \   \vp^i = {(G_{\m\n},B_{\m\n},\p)} , }
where $\k^{ij}$ is (in general) non-degenerate  and
\eqn\actt{  S = \int d^{D}x \sqrt {\mathstrut G} \ {\rm e}^{- 2
\phi} \big[ (D-26)  -
  {3\over 2} \a' \big(   R  + 4 \na^2 \p - 4 (\na \p )^2 - {1\ov 12}
(H_{\m\n\l})^2   \big)   +    O(\a')\big] ,   }
reproduces the `massless' sector of the string S-matrix.
Under a special choice of definitions of the fields (i.e. in a special
`scheme')
$S$ admits also the following  \sm representations
\eqn\reps{ S= \int d^{D}x \sqrt {\mathstrut G} \ {\rm e}^{- 2
\phi} \ \tilde \b^\p \ , \  \ \ \  \  \
S = \big({\del Z\ov \del t}\big)_{t=1} \ , }
where $\tilde \b^\p $ is the  \sm Weyl anomaly coefficient
$\tilde \b^\p =\bar  \b^\p  -\four G^{\m\n} \bar  \b^G_{\m\n}, $  $\ Z$ is the
\sm partition function on a 2-sphere
and $t$ is the logarithm of the renormalisation parameter.
The tree-level string dynamics is thus encoded in the
quantum 2d \sm with the classical string solutions corresponding  to conformal
2d $\s$-models.
It should be noted that this correspondence
may break down  for certain singular solutions:
since $\k^{ij} $ in \rele\ is field-dependent
it may vanish at the points where the metric (or dilaton) is singular.
As a result, a  background which corresponds to a conformal \sm may
solve the string effective equations only in the presence of additional sources
(see \hrts\ and  Section 5 below).

\subsec{Conformal  models  as string solutions}

The remarkable correspondence between
   the stationary points of the effective action and conformally invariant \sms
makes  it
possible to determine  the exact solutions
of  the effective  field equations which are not  even known in an explicit
form.  All one needs to do is to find a conformal \sm (e.g.,  proving that all
$\a'$ corrections are absent in a special scheme). It is   natural from string
theory point of view
 to describe a string background
not  by  a collection of  space-time  fields but  directly
by the \sm itself.  This  has an important consequence:
since 2d quantum \sms\ may  be related by certain equivalence transformations
(e.g.,   world-sheet / target space duality), differently looking collections
of background fields may actually represent
 the same string  solution.
More generally, one may  define a string solution as the
2d conformal field theory (CFT)  corresponding to a given conformal \smd
CFT  contains, of course, much more information
 than just one set of local background  fields parametrising
a conformal \smd   CFT   gives an adequate
interpretation of a given  string solution: it describes, in particular,
 how all modes of
a first-quantised string  propagate in  this background.
It should be emphasized,  at the same time,  that
if one is interested in string solutions  which should have  a space-time
interpretation,  the knowledge  of CFT at an algebraic  level  only is not
sufficient:
one  should be able to  relate it to  a lagrangian  field  theory,
 i.e. there should exist  weak field regions
where a conformal \sm description is adequate.

Thus, there are several problems
 one is   to  solve in order to give a string-theoretic  description of  a
particular  string background:

(1) First,  one needs to  specify some basic long-distance properties
of a solution one is looking for. This may involve finding a solution of  the
leading-order string effective equations.

(2) Second, one is to promote this leading-order solution to an exact (all
order in $\a'$) string solution
by identifying   a conformal \sm to which it corresponds in the  weak field (or
 $\a' \to 0 $) approximation.
 It may happen that this conformal
\sm is the same as the  leading-order  one (i.e. its couplings are not modified
by $\a'$ corrections) but in general  the  background fields   are expected  be
 given by power series in $\a'$.

(3) Finally, one is to  define
  a  $2d$ conformal field theory  associated with  the  conformal \sm
and to solve it, determining the spectrum of string excitations on a given
background and their scattering amplitudes. In particular,   CFT encodes
information about equations for all string modes
in a given background.

The last step is, of course, the most difficult  one.
Unfortunately, only very few
 exact string solutions have known  CFT which corresponds to  them
and only in  few cases this CFT is solvable.
Given a conformal \sm  (and thus exact  expressions for  the massless
background fields) but   having no  explicit
 information about  the underlying   CFT,   it is, in general, not possible
to determine  the exact form of the linearised equations for the  tachyon and
other string excitations
since they may receive $\a'$-corrections.
As a result, one is unable, for example,  to give  a definite answer about
(non)singularity of a  string  background as probed by test first-quantised
strings.

\subsec{Classes  of exact solutions}

This paper   is devoted to a discussion
of exact solutions of (bosonic)  string theory, mainly  at the \sm level.
While the solutions of the leading-order effective  equations are numerous
and straightforward to find,  little is known about their exact counterparts.
For example, the  exact form of the string analog of the Schwarzschild solution
 remains  unknown.
It turns out that  there exist  a subset  of solutions  (and these are, in
fact, the only exact solutions known at present) which are not modified by
$\a'$-corrections in a special scheme, i.e. retain their leading-order
form.\foot{This does not mean, however,
 that the corresponding CFT Hamiltonian
and thus equations for other  string modes (e.g. tachyon) do not contain
non-trivial $\a'$-corrections.}
It  seems  remarkable that this can happen at all, given that string effective
action is an infinite series of terms of all orders in $\a'$.
The reason for that, of course, is
that string  effective action is special, being related to \sm conformal
anomaly coefficients
by \rele. Thus,   given a  conformal
 \sm (e.g., $(4,n), \   n\geq 1$  supersymmetric one
or a special `chiral null'  bosonic one)  one gets a string solution
which is not modified by $\a'$-corrections.

As was mentioned above,  to obtain exact solutions in  string theory  it is
rather hopeless to  start
with the
field equations expressed as a power series in $\a'$, and  try  to solve
them explicitly.
As was demonstrated in  \refs{\hrt,\hrts, \rutsn},
one can find exact solutions by starting  directly
with some specific \sms
and  explicitly proving that they are conformal.
This  \sm approach  leads to new solutions   which
 look quite complicated  and  hard to find by explicitly   solving the string
effective equations even
at the leading order.
Thus by  choosing an ansatz  for the string
world sheet action
   which yields simple equations for the \sm  $\b$-functions,
one can easily find new solutions of even the leading order equations.

There are  several   approaches that  were    used
to construct exact string solutions:

(i) Start with a particular leading-order solution and show that
$\a'$-corrections
are absent (or take some explicitly known form)
 due to some special properties of this background.  Such are some
`plane wave'-type  backgrounds, or, more generally, certain backgrounds  with
a covariantly constant Killing vector, see, e.g.,
\refs{\guv,\amkl,\hor,\horr,\rudd,\desa,\tsnul,\hrt}.

(ii) Start with a known lagrangian CFT and represent it as a conformal
$\s$-model.
Essentially the only known example of such construction is based
on gauged WZW models, see, e.g.,  \refs{\witt,\dvv,\gauged,\horv}.

(iii) Start  directly with  a  leading-order \sm path integral
and prove that there exists such a definition of the \sm couplings (i.e. a
`renormalisation scheme') for which this \sm is  conformal  to all  loop
orders.
This strategy works  for `$F$-models'  \refs{\klts, \hrt},  their  `chiral null
model' generalisations \hrts\ and the magnetic flux tube models \rutsn.
A particular example of \FM is the `fundamental string'  solution \gibb.
 Chiral null models describe  also  travelling waves along the
fundamental string, charged fundamental strings
and some electromagnetic backgrounds. Special  $D=5$  chiral null models
can be interpreted (in a Kaluza-Klein way) as extreme $D=4$  charged black
holes \horrt.

The above  three classes of exact solutions  have common subsets.
For example, a subclass of plane waves  can be identified with gauged WZW
models based on
non-semisimple groups \refs{\napwi,\oliv,\kk,\sfetso,\figu,\anto, \sfeets}.
Also, a different subclass of backgrounds with a covariantly constant null
Killing vector  (called `$K$-models' in \hrt)
are  dual to $F$-models. Finally,  a subclass of $F$-models  can be interpreted
as  special  $G/H$ gauged WZW models  with $G$ being a maximally non-compact
real group
and  $H$ being a nilpotent subgroup  \klts;  similar interpretation is true for
a generic $D=3$ \FM \hrt.

 Conformal field theories  corresponding to those of chiral null models which
cannot be identified with gauged WZW models  seem hard to solve explicitly.
One particular approach one may follow in order to  construct exact solutions
described by   conformal \sms
which are solvable as string theories (i.e. as CFT's)
was suggested in  \rutsn.   It was  demonstrated there
 (on the example of the
magnetic flux tube models) that
the  world-sheet
duality transformations  relating  complicated \sms to simpler ones
are  a useful  tool for finding conformal \sms with {\it solvable}  CFT's
behind them.
 For example, starting with  a dimension $D\geq d+n$ flat model
with  a number $d$ of periodic coordinates and making  formal
$O(d+n,d+n;R)$  world-sheet duality transformations (see e.g.
\refs{\cec,\vene,\por})
 with {\it  continuous } parameters
one obtains new {\it inequivalent}  conformal theories (with  $O(d,d;Z)$
dualities as symmetries),  corresponding to
 complicated  space-time backgrounds which solve
the  string effective equations.


The backgrounds we shall consider below are
 exact solutions  of both bosonic and superstring  (and heterotic string)
theories.\foot{For  reviews of some  superstring solutions
see, e.g.,  \refs{\chs,\kounn,\fudd} and references there.}
 In addition, there are special exact solutions of superstring theory only
(corresponding to \sms with extended $(4,n)$  world-sheet
 and partial space-time supersymmetry).  Even  having in mind superstring
applications,  it is  still important to understand mechanisms
leading to exactness of  particular solutions which  may  not be directly based
on extended supersymmetry (some physically interesting solutions may   have
broken space-time supersymmetry).\foot{In particular, in claims about
finiteness of supersymmetric  \sms it is  a special holonomy of  the bosonic
background that often plays the crucial role.  For example,   it was
 pointed  out in  \howe\   that   since the   \sm on a  Calabi-Yau space has a
special
holonomy it  thus has an extra infinite-dimensional non-linear classical
symmetry which (if it were not anomalous  at the quantum level)
would  rule out all higher-loop corrections to the  $\b$-function.
In  the case of the  bosonic chiral null models  the analogous symmetry is
linear and  is the affine symmetry
generated by  the chiral  null  current.}

\subsec{Outline }

In Section 2  we shall   briefly  discuss  the
exact solutions corresponding to gauged WZW models,
pointing out the existence of  a  `leading-order' scheme in which the
`massless' background fields are $\a'$-independent.
 We shall also  mention  some  known solutions in  $D=4$ dimensions.

Gauged WZW models seem not to be  sufficient to describe  physically
interesting $D\geq 4$
string backgrounds.
In Section 3 we shall make some general remarks about another class of
solutions with Minkowski signature which are described by
\sms  with conserved chiral null currents. Particular subclasses
of such backgrounds will be  the topics of  Sections 4-7.

Section 4 will be devoted to solutions with a covariantly constant null Killing
vector. We shall start with  simplest plane wave backgrounds
(some of which  will be   related to gauged WZW models based on non-semisimple
groups)
and discuss several generalisations,
e.g., a   `hybrid'  $D=4$ model with the transverse part  represented by the
euclidean $D=2$ background.

Exact solutions with two null Killing vectors ($F$-models) will be presented
in   Section 5. In particular, we  shall  consider   $F$-model with  a flat
transverse part   which describes the fundamental string background and its
generalisations.
We shall also
mention   a  subclass of $F$-models  (with  $D=4$ and Minkowski signature)
which
correspond to specific nilpotently gauged WZW models and thus admit
direct CFT interpretation.

Chiral null models  generalising
  both the  plane-wave models and $F$-models will be introduced in Section 6.
In Section 7 we shall  consider   charged $D=4$ backgrounds which are described
by $D > 4$ chiral  null models. They include, in particular,
extreme electric black holes and charged fundamental strings.
We shall  explain the
procedure of  Kaluza-Klein-type reduction
and  review  the known  extreme magnetic and electric black holes
 which are    exact (super)string solutions  to all orders in $\a'$.
We shall also consider  more general  electromagnetic   backgrounds (including
IWP and Taub-NUT type ones), as well as a special    `constant  magnetic field'
solution which has no analog in Einstein-Maxwell theory.

In Section 8 we shall study   a special $D\geq 5$  conformal \sm which does not
belong to the gauged WZW or chiral null model classes.
Its $D=5$ version (with one of the coordinates being  a compact Kaluza-Klein
one)  describes  $D=4$ stationary axisymmetric  magnetic flux tube
backgrounds  generalising both the  dilatonic   Melvin and the
constant  magnetic field solutions.

This model provides a remarkable example
of a non-trivial   curved space-time solution  with an explicitly solvable
conformal string  theory corresponding to it.  Its solution  \ruts\ will be
reviewed  in Section 9 where
 we  shall first explain  the relation
of the magnetic flux tube model to a flat space model via angular
duality. As a result,  it will be   possible  to solve the classical  string
equations    explicitly,   expressing the string coordinates in terms of free
fields  satisfying `twisted' boundary conditions.
After  straightforward operator quantisation   one then determines the quantum
Hamiltonian, spectrum of states and partition function,  in direct  analogy
with how this is done in simpler models like
closed  string on a torus or an  orbifold,
or open string in a constant magnetic field.

In  Section 10  we shall make  some concluding remarks.

\newsec{Exact solutions corresponding to gauged WZW models}
In this section we shall briefly discuss exact string solutions
described by gauged WZW models (for more detailed reviews of different aspects
of these solutions see, e.g., \refs{\barss,\tseer,\kirko}).  An important
property of these solutions is
that the corresponding  CFT is, in principle,  known explicitly.
At the same time, it should be noted that  most of the  \sm backgrounds
obtained from  gauged WZW models do not have non-abelian
isometries and thus do not seem to be relevant for description
of  physically interesting  spherically symmetric  $D=4$
cosmological and black hole backgrounds.

The general bosonic \sm describing string propagation in a `massless'
background is given by \stri, i.e., in conformal gauge,
\eqn\mod{ \ L=  (G_{\m\n} + B_{\m\n})(x)\ \del x^\m \bd x^\n
+ {\cal R}\p (x)\ ,  }
$$I= {1\over {  \pi \a' }} \int d^2 z \  L  \ , \ \ \
\ \ \ \   {\cal R} \equiv \four \a'\sqrt \g R^{(2)} \ ,   $$
where $G_{\m\n}$ is the metric,  $B_{\m\n}$ is the antisymmetric tensor and
$\p$ is the dilaton.
The action of the ungauged WZW model for a group $G$ \wit\
 \eqn\unga{  I=kI_0\ , \ \ \   I_0 \equiv  {1\over 2\pi }
\int d^2 z  \Tr (\del g^{-1}
\bd g )  +  {i\over  12 \pi   } \int d^3 z \Tr ( g^{-1} dg)^3   \ ,
}
can be put into the general \sm form \mod\
by introducing the coordinates on the group manifold. Then
$G_{\m\n}$ is the group space metric,  $H_{\m\n\l} = 3 \del_{[\m}B_{\n\l]}$
is the parallelising torsion  and dilaton $\p=\p_0=\const$ (and $k=1/\a'$). In
a special scheme (where the $\beta$-function of the general model \mod\ is
proportional to the generalised curvature)
the  resulting \sm is finite  at each order of $\a'$-perturbation theory
\refs{\bra,\mukh} and corresponds to the well-known current algebra CFT \wit.

The action of the gauged WZW model  \gwzw\
\eqn\gau{
I (g,A) = kI_0(g)  +{k\over \pi }
 \int d^2 z \Tr \bigl( - A\,\bd g g\inv +
 \bar A \,g\inv\del g   + g\inv A g \bar A  - A \A \bigr)
   \ ,  }
is invariant under  the    vector  gauge transformations
with parameters taking values in a subgroup $H$ of $G$. Parametrising  $A$ and
$\A$ in terms of $\  h$ and $\bh$  from $H$,
$ A = h \del h\inv \  , \  \ \A = \bh
\bd \bh\inv $ one can  represent \gau\   as the
difference of the two manifestly gauge-invariant terms: the ungauged  WZW
actions
corresponding to the group
$G$
and  the subgroup $H$,
\eqn\gact{ I (g,A) =k I_0 (h\inv g \bh ) -  kI_0 (h\inv \bh)  \  . }
This representation  implies that the  gauged WZW  model corresponds to  a
conformal
theory (coset CFT \refs{\GKO,\bha}).
Fixing a  gauge on $g$ and changing the variables to $g' = h\inv g{\tilde h} ,
\
 h' = h\inv {\tilde h}$  we get a \sm on the group space $G\times H$ which is
conformal to all orders in a particular  `leading-order'  scheme.
That means that the 1-loop  group space solution remains  exact solution in
that scheme.
Replacing \gact\ with  the `quantum' action with renormalised levels
$k\ra  k + \ha c_G$ and $k\ra  k + \ha c_H$  does not change this conclusion.
This replacement   corresponds to  starting with the theory formulated in
the `CFT' scheme  in which, e.g.,  the  exact central  charge of the WZW model
is reproduced by the first non-trivial correction \refs{\tspl, \tssfet}
 and the metric $( k + \ha c_G ) G_{\m\n}$ is the one that appears in the CFT
Hamiltonian $L_0$  considered as a Klein-Gordon operator.

\subsec{`CFT' and `leading-order' schemes}
To obtain the corresponding  \sm in the `reduced'  $G/H$
configuration space (with coordinates being parameters of gauge-fixed $g$) one
needs to integrate out $A,\A$ (or, more precisely, the WZW fields $h$ and
${\tilde h}$).  This is a non-trivial step and the form of the result depends
on a choice of a scheme in which the original
 `extended' $(g,h,{\tilde h})$  WZW theory is formulated.

Suppose first   the latter is taken in the leading-order  scheme
with the action \gact.
Then the   result of integrating out $A,\A$
and fixing a gauge takes the form of the \sm \mod\ where the \sm metric and
dilaton are then given by (see \hrt\ and refs. there)
\eqn\resm{ G'_{\m\n} = G_{\m\n}  - 2 \a' \del_\m \p \del_\n \p \ , \ \ \
\p=\p_0  - \ha \ln \det F \ . }
$G_{\m\n}$ is the metric obtained by solving for $A,\A$ at the classical
level
and $\p_0$ is the original constant dilaton.
Since the  $\a'$-term in the metric can be  eliminated by a field redefinition
we conclude that there exists a {\it leading-order scheme}  in which the
leading-order  gauged WZW
\sm
background $(G,B,\p)$ remains an  exact solution. The leading-order scheme
for the ungauged  WZW \sm   is thus   related to the leading-order scheme for
the gauged  WZW $\s$-model
 by an extra $2 \a' \del_\m \p \del_\n \p$ redefinition of the metric.
This   provides a general explanation for the observations in \refs{ \tspl,
\tssfet} about the existence of a leading-order scheme
 for  particular $D=2,3$ gauged WZW models.

If instead we start with the  $(g,h,{\tilde h})$  WZW theory in the CFT scheme,
i.e
 with the action
\eqn\gactt{I(g,A) = (k+ \ha c_G)\big[ I_{0} (h\inv g{\tilde h})- {k+ \ha
c_H\ov k+ \ha c_G}
I_{0} (h\inv {\tilde h})\big] \ , }
then the resulting  \sm couplings will explicitly depend on $1/k\sim \a' $
(and will agree with the  coset CFT operator approach results
\refs{\dvv,\barsf,\bsfet,\tsnp}).
While in  the WZW model the transformation from the  CFT to the leading order
scheme is just a simple rescaling of couplings,  this  transformation  becomes
non-trivial    at the level of gauged WZW $\s$-model.   It is the  `reduction'
of the configuration space
resulting  from   integration over the gauge fields
$A,\A$  that is
 responsible for  a  complicated form of the transformation law between the
`CFT' and `leading-order' schemes in the gauged WZW $\s$-models  (in
particular,  this transformation  involves  dilaton terms of  all orders in
$1/k$, see  below).\foot{An exception is provided by some  $\s$-models
obtained by nilpotent
gauging:
here the second term in \gactt\ is absent by construction \klts. The background
fields do not receive non-trivial $1/k$ corrections even in the CFT scheme,
i.e.  the relation
between the leading-order and CFT schemes is equivalent to the one  for the
ungauged WZW model.
The  same is true for the $D=3$ \FM or the extremal limit of the $SL(2,R)\times
R/R$ coset.}

The basic example is  the $SL(2,R)/R$ gauged WZW model (or $D=2$ black hole)
\refs{\witt,\brsne,\cres}.
 The  (euclidean) background in the leading-order scheme
\eqn\exann{ ds^2 =    dx^2 + \ { {\tanh}^2 bx }   \ d{\t}^2   \ ,  \ \ \
 \p = \p_0  -  {\ln \ch }  bx  \ ,  }
is related to the  background in the CFT scheme \dvv
\eqn\exann{ ds^2 = dx^2 + \ { {\tanh}^2 bx \ov 1 \ - {\ p \ }
{\tanh}^2 bx }   \ d{\t}^2   \ ,  \ \ \  }  \eqn\eeee{
 \p = \p_0  -  {\ln \ch }  bx   -
\four  \ln \bl( 1 \ - \ p \ {\tanh}^2 bx )\ ,  }
\eqn\exxxx{  \  p\equiv {2\ov k} \ , \ \
\a'b^2 = {1 \ov k-2}\ , \ \ \ D - 26  + 6\a'b^2  =0  \ , \ \ D=2\  , }
by the following local covariant redefinition \tspl
\eqn\cftf{   G^{(lead)}_{\m\n}  = G_{\m\n}  - { 2\a'  \ov 1  + \ha \a' R  }  [
\del_\m \p \del_\n
\p  -    (\del \p)^2  G_{\m\n}]
\ , }
$$  \p^{(lead)}=\p  - \four  \ln ( 1  +
{\textstyle {1\ov 2}} \a' R  ) \ .   $$
It should be emphasized that the two backgrounds related by \cftf\
describe the same  string geometry since the probe of the geometry is the
tachyon field
and the tachyon equation  remains the same  differential equation
implied by coset CFT (even though it looks
different being expressed in terms of
different $G$ and $\p$).

Similar remarks apply to
 a  particular  $D=3$ solution  -- the `charged black string'
corresponding to $[SL(2,R)\times R]/R$ gauged WZW model \hoho.
Its form in the leading-order scheme can be obtained by
making a duality rotation of the neutral black string, i.e. the direct product
of the $D=2$ black hole and a free scalar theory (see \refs{\tssfet,\givkir}).

\subsec{Some $D=4$ solutions}

In general, the backgrounds obtained from  the gauged WZW models
have very few (at most,  abelian) symmetries and thus cannot directly describe
 non-trivial $SO(3)$-symmetric  backgrounds which are of interest in connection
with
 $D=4$ cosmology and black holes.
Let us mention the explicitly known $D=4$  solutions with $(-,+,+,+)$
signature.
A class of axisymmetric black-hole-type and anisotropic cosmological
backgrounds is found
by starting with $[G_1\times G_2]/[H_1\times H_2]$ gauged WZW model with
various combinations of $G_i=SL(2,R)$ or $ SU(2)$ and $H_i= R$ or $U(1)$
\refs{\horv,\kou,\napwit,\givpas}.
The leading-order form of these backgrounds can be obtained by applying the
$O(2,2)$ duality transformation
to the direct product of the Euclidean and  Minkowski $D=2$ black holes or
their
analytic continuations.
The stationary black-hole-type background is \horv
\eqn\dualih{ds^2=  -{g_1 (y) \ov g_1 (y)g_2 (z)  - q^2 }\
dt^2 +
 {g_2 (z)\ov g_1 (y)g_2(z)  - q^2} \ dx^2   + dy^2 + dz^2 \ , }
$$B_{tx }= {q \ov  g_1 (y) g_2 (z) - q^2}   \ , \ \ \
\phi= \p_0 -\ha \ln \bl(\sinh^2y \sinh^2z  [ g_1(y) g_2(z) - q^2]\br)  \ ,  $$
where $g_1= \coth^2 y, \ g_2= \coth^2 z$.
The cosmological background  \napwit\ can be obtained by the analytic
continuation
and renaming the coordinates
\eqn\dualiw{ds^2= -dt^2 + dx^2 +  {g_1 (t) \ov g_1 (t)g_2 (x)  + q^2 }\
dy^2 +
 {g_2 (x)\ov g_1 (t)g_2(x)  + q^2} \ dz^2    \ , }
$$B_{yz }= {q \ov  g_1 (x) g_2 (t) + q^2}   \ , \ \ \
\phi= \p_0 -\ha \ln \bl(\cos^2t \cosh^2 x  [ g_1(t) g_2(x) + q^2]\br)  \ ,  $$
where $g_1= \tan^2 t, \ g_2= \tanh^2 x$.
A  non-trivial (`non-direct-product') $D=4$ background without isometries is
obtained  from  $SO(2,3)/SO(1,3)$ gauged WZW model \barsf.

Two other classes of $D=4$ Minkowski signature   `coset' solutions will be
discussed in more detail
in the following sections.  The first includes  plane-wave-type backgrounds
which are obtained from non-semisimple versions of  $R\times SU(2)$ and the
above
$[SL(2,R)\times SL(2,R)]/[R\times R]$, etc., `product' cosets
\refs{\napwi,\oliv,\kk,\sfetso,\figu,\anto, \sfeets}.
The second contains $F$-models (backgrounds with 2 null Killing vectors)
 which follow from nilpotent gauging of rank 2 maximally non-compact groups
\klts.

Some other exact $D=4$ solutions  which  are not described by gauged WZW models
as a whole but  incorporate  $D=2$  euclidean
black hole background  as a two-dimensional  part
will be mentioned below in Sections 4.3, 5.2, 8.

\newsec{$\s$-models  with  conserved
chiral null  currents: general remarks}
It appears that gauged WZW models are not sufficient
in order to describe  physically interesting
$D\geq 4$ string solutions,
e.g.,  black holes  with asymptotically flat regions  (cf. \gps)
and non-trivial isotropic cosmological solutions (cf. \refs{\mye,\antm}).
We are  thus to  study  other  types  of  conformal \sms
which  have physical signature and dimension.
One possible  direction of investigation  (which remains largely unexplored)
is to look
for  conformal \sms \refs{\tss,\deb,\bard,\halp}
 related   to generalisations
of coset models which  solve
of the  affine-Virasoro master equation \refs{\hal,\halp}.
Here we shall  discuss  other  classes of \sms which are conformal because
of their   special {\it Minkowski} signature  structure.

Every Killing vector  generating a symmetry of  spacetime fields
gives rise to a conserved current
on  the string world sheet. If the antisymmetric tensor field is related to the
spacetime metric in  a certain way, these currents are chiral. The
existence of such chiral currents  turns out to simplify the search for
exact conformal models. One example
 is the WZW model which describes string propagation on a group
manifold where  all the associated
currents
are chiral (since the gauged WZW models can be represented in terms of the
difference between
two WZW models for a group and a subgroup, a similar statement applies there).
Another example is provided by  the $F$-models discussed in
\refs{\klts,\hrt,\hrts}
which have two null Killing vectors and two associated chiral
currents. Also,  the known plane wave solutions  and their
generalizations \refs{\brink,\amkl,\hor,\horr} are characterized by the
existence of
a  covariantly constant null Killing vector.

 $F$-models and generalized plane waves
are both special cases of a larger class of exact solutions --
 `chiral null
models'  \hrts\
 which have a
null Killing vector and an associated conserved chiral current.
   The presence of a null chiral  current  is a consequence of an
infinite-dimensional affine symmetry of  the \sm action. This symmetry is
present if spacetime fields
have certain special properties.
The generalized connection with torsion equal to the antisymmetric field
strength plays an important role since it is the one that appears
in the  classical
string equations of motion. This connection turns out to have
reduced holonomy.
A   `balance' between the metric and the antisymmetric tensor resulting in
 chirality  of the action
is the crucial property of the chiral null  models which is in the core of
their exact
conformal invariance.

In addition to the plane-wave type solutions  (Section 4)  as well
as  the $F$-models \hrt\  which contain the fundamental string solution
\gibb\ as a special case (Section 5), the class of chiral null models include
 generalizations of these
solutions, e.g.,  the travelling
waves along the fundamental string \garf\  (Section 6).
Although the bosonic string does not have fundamental gauge fields,
effective gauge fields can arise from dimensional reduction.
In  particular,
 charged fundamental string solutions \refs{\sen,\wald}
 and  four dimensional extreme electrically
charged black holes \refs{\gib,\gim,\gar} can  also be obtained from the
dimensional reduction of
a chiral null model and hence are exact \horrt. Similarly,  the generalizations
of
the extremal black holes which include  NUT charge and rotation
\refs{\kall,\jons,\galts} are also exact \hrts\  (Section 7). Finally, the
chiral null models
  and their generalisations  describe also (electro)magnetic flux tube
backgrounds \refs{\hrts,\ruts,\rutsn} (Section 8).

If one considers only the leading order string equations, many of these
solutions arise as an extremal limit of a family of solutions with a regular
event horizon.
It is clear, of course, that
not all of the solutions of the leading order equations can be obtained
from chiral null models. The chiral coupling  leads to a no-force condition on
the solutions, i.e. to the
possibility of linear superposition. This  happens  only for a  certain
charge to mass ratio which  typically characterizes  extreme black holes
or black strings. The  balance between the spacetime
metric and the antisymmetric tensor field necessary for the existence of a
chiral current results upon dimensional reduction
 in a relation between the charge and the mass.
 Furthermore, one obtains only  four dimensional
black-hole type
solutions with {\it electric}  charge; extreme magnetically charged black holes
do not appear  to be
described by chiral null models.

 In general, the non-extremal solutions are not of the chiral null form
and  receive  $\a'$-corrections in all renormalization schemes.
Finding the exact analogs of these solutions (which include as a special case
 the
Schwarzschild metric) remains an outstanding open problem.

Below we shall first  discuss   the plane wave and $F$-model
 subclasses of the  class of chiral null models and then
discuss   some   special cases  and  generalisations.


\newsec{Solutions with covariantly constant null Killing vector}

The string action in a  generic background with a covariantly constant null
Killing vector
can be represented in the following   form, cf. \mod\  ($i,j=1,...,N$)
\eqn\str{   L =   \del u \bd v +
  K(u,x)\del u \bd u  +  2A_i(u,x) \del u  \bd  x^i  + 2\A_i(u,x) \bd u \del
x^i }
$$+ \  (G_{ij} + B_{ij})(u,x)\ \del x^i \bd x^j  \  +  \   {\cal R}  \p (u,x)\
. $$
$K, A_i, \A_i$ can be eliminated (locally) by a coordinate and  $B_{\m\n}$-
gauge transformation \refs{\brink,\tsnul}
so that the general form of the Lagrangian is
\eqn\sstr{ L =   \del u \bd v
 +   (G_{ij} + B_{ij})(u,x)\ \del x^i \bd x^j  \  +  \   {\cal R}  \p (u,x)
\  ,  }
with $K, A_i, \A_i$ been now `hidden' in a  possible  coordinate and
gauge transformation. If the `transverse' space is trivial (for fixed $u$)
one may use \sstr\  with `flat' $G_{ij}, B_{ij}$ taken in the most general
frame, or may
choose a special frame where $G_{ij}=\delta_{ij}, B_{ij}=0$ and  then   go back
to  \str\
(which may be preferable for a global coordinate choice).

There are two possibilities to   satisfy the conditions of conformal
invariance.
 The first (which will be discussed in the following subsections)  is realised
in the case when  the `transverse' theory
is conformally  invariant.
The second is realised when
the `transverse'  $N$-dimensional  part of the model is
not conformal  but one may  still find a   conformal model  by  adding  the
term linear in $v$  to the dilaton field \tsnuu.
Then the  conformal invariance  conditions are satisfied provided
the `transverse' couplings $G,B,\p$ depend on $u$ according to the standard RG
equations
 with the $\b$-functions of the `transverse' model, $\del G_{ij}/\del u =
\b_{ij}^G. $ etc.
  Since  in general these $\b$-functions are not known in a closed form
one is unable to determine the explicit all-order form of the solution.
An exception is provided by  the  theories  with transverse part represented by
  $(2,2)$ supersymmetric  Einstein-K\"ahler \sms \tsnuu\  (see  also \pet) when
the  exact  $\b$-function is given by the  one-loop
term.
The  simplest non-trivial example
corresponds to the case when the transverse theory is   the
$O(3)$ supersymmetric $\s$-model. The resulting metric
is that of $D=N+2=4$ dimensional space  with the transverse part
being  proportional to the metric  on $S^2$,
 \eqn\metty{ds^2  =  dudv
+  u (d\t^2 + {\rm sin}^2 \t d\varphi^2 )    \  .   }
To get a solution of the conformal invariance conditions one should
add the following dilaton
\eqn\ttty{\p (v,u)  =  \p_0 +  v +  \four
\ln\ u  \ .  }
The  spacetime \metty\   is conformal to the
the  direct product of   two-dimensional Minkowski space and two-sphere.
Though the exact metric  has apparent curvature singularity at $u=0$ the
corresponding  CFT may be non-singular. Note  also  that the string coupling
$ e^{\phi} = A u^{1/4} e^{v}\  $  goes to zero
 in the  strong coupling region $u \ra 0$
of the transverse sigma model, i.e. is {small} near the singularity $u=0$.
Similar remarks apply  to  some other exact solutions discussed below, e.g., to
the fundamental string one.

\subsec{Plane waves }
The simplest  special case of \sstr\ is that of the `plane-wave' backgrounds
for which  $G,B,\p$ in \sstr\ do not depend on $x$
(so that, in particular, the transverse metric is flat).
The conformal invariance condition then reduces to one equation
\eqn\eeeq{ \   -{\ha} G^{ij}  {\ddot G}_{ij} + \four  G^{ij} G^{mn}(\dot
G_{im}\dot G_{jn}
 -{\dot B}_{im}{\dot B}_{jn} )
 + 2 \ddot \p =0\  \  .  }
 $G_{ij } (u), B_{ij} (u) $  and $\p (u)$ satisfying \eeeq\
  represent  { exact}   solutions of string theory \kltspl\ (which transform
into each other under the full duality symmetry  group  $O(N,N;R)$).

A subclass of  such  models  admits a coset CFT interpretation in terms
of gauged WZW models for  non-semisimple groups \refs{\napwi, \oliv,\kk,
\sfetso,\sfees,\figu, \anto, \sfeets}.
The latter can be obtained from standard  semisimple gauged WZW models by
taking special singular limits.  The singular procedure  involves a coordinate
transformation and a rescaling of $\a'$ and is carried out directly at the
level of the string action \refs{\sfetso,\sfeets}.\foot{It is similar  to  the
transformation  in \penr\ which maps
any  space-time  into some plane wave.}
Using the singular limit one  is able  to obtain the plane-wave background
fields and  formal CFT operator algebra relations
 but   to construct the full `non-semisimple' coset CFT  from a `semisimple'
one  may be   a more subtle problem\foot{This singular limit changes the
boundary conditions  or topology (mapping, e.g.,  a compact coordinate into a
non-compact one),  so it is not a priori clear that the correspondence between
the two conformal models  holds also
at the full CFT level, i.e.
at the level of states and correlation functions.} so that  one may need  to
start directly from the `non-semisimple' coset CFT
 in order  to give
a CFT description to this  subclass of plane-wave solutions (see \kk).

In particular,  a  set of  $D=4$ plane wave  solutions that can be obtained in
this way from the
gauged  $[SU(2)\times SL(2, R)]/[U(1)\times R]$ WZW models is given by \sfeets\
(cf.\dualiw)
\eqn\duali{ds^2= dudv +  {g_1 (u') \ov g_1 (u')g_2 (u)  + q^2 }\
dx_1^2 +
 {g_2 (u)\ov g_1 (u')g_2(u)  + q^2} \ dx_2^2  \ , }
$$B_{12 }= {q \ov  g_1 (u') g_2 (u) + q^2}   \ , \ \ \
\phi=\p_0  -\ha \ln \bl(f_1^2 (u')f_2^2 (u) [ g_1(u') g_2(u) + q^2]\br) \ ,  $$
where $u'= au + d \ $ ($a,d=\const$) and the functions $g_i, f_i$ can take any
pair of the following values
 \eqn\hhh{\eqalign{&g(u) = 1\ , \ \ u^2\  ,  \ \  \tanh^2 u \ , \ \  \tan^2 u \
, \ \
 u^{-2}\ , \ \ \coth^2 u\ , \ \ \cot^2 u \ , \cr
 &f(u) = 1\ , \ \ 1\ , \ \ \ \  \cosh u\ , \ \ \ \ \cos\ u \ , \ \
\ u\ , \ \ \ \  \sinh u \ , \ \  \  \ \sin u \ . \cr } }
A particular case is  $g_1=1, \ g_2 =u^2$
(this    background   is dual to  flat space).
 Another special case is the
 $E^c_2$  WZW model of \napwi\foot
{$E^c_2$ is  a  $D=4$ non-semisimple algebra  which is a central extension of
the Euclidean  algebra  in two dimensions admitting  a non-degenerate invariant
bilinear form.}
\eqn\acd{L = \del u\bd v + \del x_1 \bd x_1
+\del x_2 \bd x_2 + 2\ \cos\ u\ \del x_1 \bd x_2 \ , }
which   can be obtained by a singular limit from
 the WZW action for
$SU(2)\times R$.

The special case of \sstr\  with $G_{ij}=G_{ij}(u), \ B_{ij}=0, \ \p=\p(u)$ or,
equivalently,   \str\ with $G_{ij}=\delta_{ij}, \ A_i=\A_i=B_{ij}=0, \ \p=\p
(u) $
is
\eqn\sstry{ L =   \del u \bd v +  K(u,x) \del u \bd u
 +   \del x^i \bd x_i  \  +  \   {\cal R}  \p (u)
\  .   }
The fact that this model  has a covariantly constant null vector
can be used to give a simple geometrical argument \refs{ \hor,\horr} that
  such  leading order solutions  are exact.
The curvature contains two powers of the
constant null vector $l$, and derivatives of $\p$ are also
proportional to $l$. One can thus show that all
higher order terms in the equations of motion vanish identically.
The one-loop conformal invariance condition
\eqn\tacc{
 - \ha  \del^i\del_i K    +
2 \del_u^2  \p =0 \  , }
is then the exact one and is solved, e.g.,  by  the standard plane-wave ansatz
\refs{\guv,\hor,\horr}   $K= w_{ij} (u) x^i x^j $.
A different rotationally symmetric solution  exists for $\p=\const$
($ r^2 \equiv  x_ix^i > 0  $)
\eqn\ffss{ K = 1 + {M\ov r^{D-4}}  \ , \   \ \ D> 4  \ ; \ \  \ \ \ \  \ K= 1 -
\m\ \ln {r\ov r_0}  \ , \ \   D=4\
   .  }
This background \gib\  is dual \hhs\ to the  fundamental string  solution  and
describes a
string boosted to the speed of light.\foot{Note that this solution needs a
source to support it at the origin.}

\def \F {{\cal F}}
\subsec{Generalised plane waves}
If one sets $G_{ij}=\d_{ij}, \ B_{ij}=0, \ \p=\p(u)$ in \str\  but keeps $A_i,
\A_i$
non-vanishing the conditions of conformal invariance
take the form  \refs{\tsnul, \hrts} ($\F_{ij} \equiv
\del_i A_j - \del_j A_i , \  \bar \F\equiv  \del_i \bar A_j - \del_j\bar  A_i$)
\eqn\con{\del^j \F_{ij} =0 \ , \ \ \ \  \ \ \del^j \bar \F_{ij} =0 \ ,  }
\eqn\coni{- \ha \del^i\del_i   K  +   \F^{ij} \bar \F_{ij}
+   \del^i{\del_u( A_i + \A_i)}
+ 2 {\del^2_u\p}  +  \sum_{s=1}^\infty c_s{\a'}{}^{2s} O(\del^s \F \del^{s}\bar
\F)
=0  .  }
These equations admit straightforward generalisation to the case of $\p=\p(u) +
b_i x^i$.
The  higher order  terms  do not vanish in general
so it is not true that any model with a covariantly constant
Killing vector does not receive $\a'$ corrections.
However, $\a'^n$-terms in \coni\  have  particular structure and thus
vanish, e.g., for $\A_i=0$. The special
property of the `chiral'  model  with $\bar A_i=0$ or $A_i=0$ (i.e. with
$G_{ui}=\pm B_{ui} $) resulting in cancellation of the   vector-dependent
contributions to
the
$\beta_{uu}
$-function  corresponding to the renormalisation of  $K$  was  noted  at the
one-loop
level in \tsnul\ and extended to the two-loop level in \duvall.\foot{As
explained  in \tsnul,  the  exact solutions of the string equations with
$G_{ij}=\d_{ij} , \ A_i= - \ha \F_{ij} (u) x^j ,\  K=K_0 + k_{ij} x^i x^j, $
considered in
\refs{\amkl,\hor,\horr}  are  the same as  the solutions  with
$g_{ij} = g_{ij}(u)$ in \refs{\rudd,\duval,\duvall}.
Equivalent representations for $B_{\m\n}$  are  $B_{iu}= - \ha {\dot b}_{ij}x^j
, \ B_{ij} =0 $
and  $B_{iu}= 0 , \ B_{ij} = b_{ij} (u) $.
}
  It was further  shown  \bergsh\ that such backgrounds
are (`half') supersymmetric  when embedded in $D=10$  supergravity  theory and
it was suggested    that these  `supersymmetric string waves'
  remain  exact
heterotic  string solutions
to all orders in $\a'$ when supplemented  with some   gauge field background.
  It was  shown in   \hrts\ that  `chiral'  plane wave backgrounds
with $\A_i=0$ (or $A_i=0$)
\eqn\stre{   L =   \del u \bd v +
  K(u,x)\del u \bd u  +  2A_i(u,x) \del u  \bd  x^i + \del x^i \bd x^i
   +    {\cal R}  \p (u)\  , }
 are exact solutions of bosonic string  theory  provided
$K$,$A_i, \p$ satisfy the one-loop conformal invariance conditions,
\eqn\sses{- \ha  \del^2 K
+   \del^i\del_u A_i
+ 2 {\del^2_u\p} =0 \ , \ \ \ \del^j \F_{ij} =0 \ . }
Furthermore,   due to the  chiral structure of the model \stre,  they  can be
promoted to  the
exact superstring and heterotic string solutions even without an extra  gauge
field background \hrts.

The models \stre\ with trivial  $K$  and $A_i$ with  a constant field strength,
\eqn\trii{ K=0\ , \ \ \  \ \  \ \    A_i=-\ha \F_{ij} x^j\  , \  \ \ \
\F_{ij} =\const \ , }
 can be interpreted \hrts\ as boosted products of group spaces,
or, equivalently, as spaces corresponding to
WZW models for non-semisimple groups\foot{In even dimensional cases these are
Heisenberg groups with the algebra containing generators $ (e_0, e'_0, e_i,
e'_i)$ with  non-trivial  commutation relations ($e_0$ belongs to the center):
$ [e_0', e_i]= -e_i, \ [e_0', e'_i]= e'_i, \ [e_i, e'_i]= \delta_{ij} e_0$
 (see Kehagias and Meessen, ref. \figu).},
with the $D=4$ model of \napwi\ being the
 simplest ($D=4$, $N=2$)  example ($A_i=-\ha \b \epsilon_{ij} x^j$;  $\b$ can
be set equal to 1 by  rescaling $u$ and $v$).
In particular, the corresponding connection with torsion is flat.
These \sms  (or corresponding non-semisimple  WZW models) can be obtained by
singular boosts and rescalings of levels
from semisimple WZW models
based on direct products of $SL(2,R)_{-k}, \ SU(2)_k$ and $R$ factors.
It is interesting to note also  that all these  models  \stre,\trii,
can be  formally related  by $O(d,d;R)$  duality to the
flat space model  \hrts\ in the same way  as
 this was shown   \refs{\kk,\klts}  for the $D=4$ model
of \napwi\ (see also Section 9.1).

 The higher-loop corrections in \coni\  vanish  also
 in another  special case  when both
$A_i$ and $ \bar A_i$  have   field strengths\foot{More generally, the higher
loop corrections to \coni\ vanish when at least one of the two field strengths
is constant. To get a solution the non-constant field strength should still
satisfy the Maxwell equation \con; for example,  in $D=3$ it may  correspond to
  a monopole background (cf. (7.18),(7.19)).}
  constant in $x^i$ direction (in
general,
the field strengths and $\p$  may still depend on $u$ but for simplicity we
shall assume that they are constant)
\eqn\cobf{ A_i = -\ha  {\cal F}_{ij}  x^j \ , \ \ \ \ \
\bar A_i = -\ha  \bar {\cal F}_{ij}  x^j \ . }
Then  \coni\ takes the form
\eqn\conii{- \ha  \del^i\del_i  K  +   \F^{ij} \bar \F_{ij} =0 \ ,  }
\eqn\ook{  K  = K_0 (x) + m x^ix_i \ , \ \ \  m\equiv \ha  \F^{ij} \bar
\F_{ij}\ ,
\ \ \  \del^i\del_i  K_0=0 \ . }
One can choose, e.g., $K_0=k_0=\const$ or $K_0=k_0+  M/r^{D-4}$ as in \ffss.
This model
\eqn\stry{   L =   \del u \bd v +
  K(x)\del u \bd u  +  2A_i(x) \del u  \bd  x^i  + 2\A_i(x) \bd u \del  x^i
+ {\cal R}\p_0 \ , }
represents a simple and  interesting  conformal theory
which (for $K_0=\const$) can be solved explicitly \ruts.\foot{One particular
case of it corresponds to the $D=4$ $E^c_2$   WZW
model  \napwi, namely,
 $K= - x^i x_i , \  A_i  = - \bar A_i = -\ha
\ep_{ij} x^j , \  \p=\const$, which is obviously  a  solution of
\conii. This representation  is related by a $u$-dependent
coordinate transformation of $x^i$
to another one mentioned above:
 $K=0, \ A_i = - \ha \ep_{ij} x^j, \ \bar A_i =0$.}
This model and a similar one   obtained by  applying the duality transformation
in $u$ belong to a generalisation of chiral null models discussed  below in
Section 8.

\subsec{$K$-models with curved transverse space }
Another special case of \str\ is found when $K,G_{ij}, B_{ij},\p$ do not depend
on $u$  but  $G,B$ are, in general, nontrivial functions of $x^i$.
This is  a generalisation of the plane wave solutions called the `$K$-model' in
\hrt\foot{$K$-term cannot be eliminated as in \sstr\ since $G,B$ are assumed to
be $u$-independent.}
\eqn\mofk{ L_{K}= \del u \bd v +  K(x) \ \del u \bd u  + (G_{ij} +
B_{ij})(x) \ \del x^i \bd x^j     + {\cal R}\p (x)\ .  }
Suppose that the transverse space is  known to be an exact  string
solution  in some scheme. The exact form of the conformal invariance  equation
for $K$ turns out to be the following \hrt
\eqn\taac{  \
- \o K    +  \del^i \p \del_i K
=0 \ , \ \ \ \  \o = \ha  \na^2   +  O(\a') \ ,  }
where $\o$ is the scalar anomalous dimension operator
which in general contains $(G_{ij},B_{ij})$-dependent corrections to all
orders in $\a'$ (only  several  leading $\a'^n$-terms in it are known
explicitly, see,  e.g.,  \tspl\ and refs. there).
 The simplest possibility
 is that of the flat transverse space with linear  dilaton
 $( \p = \p_0  +
b_i x^i )$ which is an obvious generalisation of the plane wave case.
It is possible to  obtain more interesting  exact solutions when the
CFT behind the `transverse' space solution $(G_{ij},B_{ij} ,\p)$ is
nontrivial but  is  known explicitly \hrt:
in that case the structure of the  `tachyonic' operator
$\o$ is determined by the  zero mode part of the
CFT Hamiltonian $H=L_0 + \td L_0$.
Fixing a particular scheme (e.g.,  the `CFT' one
where $H $ has the standard
Klein-Gordon form with the dilaton term)
 one is able, in principle,  to  establish  the  exact form
of  the background fields $(G_{ij},B_{ij} ,\p)$
$and$ $K$. This produces  `hybrids' of  gauged WZW and plane wave solutions.

To obtain a {\it four} dimensional hybrid
 $K$-model solution \hrt\ one must start with a two
dimensional conformal $\s$-model.
Essentially the only (up to analytic continuation)
non-trivial  possibility
is
the   $SL(2,R)/U(1)$
gauged WZW model which describes the two dimensional euclidean black hole.
The exact background fields
of the  $SL(2,R)/U(1)$ model in the CFT  scheme were  given in \exann,\eeee.
In the CFT scheme the tachyonic equation has the
  standard uncorrected form,  so that
 the  function $K(x)$ must satisfy
\eqn\tach{ - \ha  \na^2  K  +   \del^i \p  \del_i K=
- {1\ov 2\sqrt G \e{-2\p} } \del_i (\sqrt G \e{-2\p} G^{ij} \del_j) K =  0 \ .
\  }
A particular solution of \tach\  with $K$ depending  only on $x$ and not on
$\t$
 is \eqn\taa{  K= a  + \  m  \ \ln \tanh bx \ . }
The constants $a, \ m $ can
 be absorbed into  a redefinition of $u$ and $v$,
 so that the full exact $D=4$ metric is \hrt
\eqn\mix{ds^2 = dudv + \  \ln \tanh bx \ du^2  + dx^2 +    { {\tanh}^2 bx \ov 1
\ - {\ p \ } {\tanh}^2 bx }\ d{\theta}^2 \ ,  }
while the dilaton is unchanged.
This metric is asymptotically flat,  being (at $x \to \infty$) a product of
$D=2$ Minkowski
space with
a cylinder.

The solution for $K$
 is the $same$ in the `leading-order' scheme
where the metric and dilaton do not receive $\a'$ corrections.
The reason  is that the differential  operator in the  tachyon   equation
remains the same,
it is only its expression in terms of the new $G, \p$ that changes.
Thus, in the `leading-order' scheme we get the following exact $D=4$ solution
\eqn\mixl{ds^2 = dudv +  \ln \tanh bx \  du^2  + dx^2 +
   { {\tanh}^2 bx }\ d{\theta}^2 \ , \ \ \ \p = \p_0  -  {\ln \ch }  bx \ . }
In addition to the    covariantly constant null vector $\del/ \del v$,
this solution  has
two
isometries  corresponding to  the shifts of $u$ and $\t$, i.e. one can consider
two different types of duals. The  $u$-dual background
 will be  given  in Section 5.

\newsec{Solutions described by $F$-models}
\subsec{Simple $F$-model: fundamental string solution}
The simplest \FM  \refs{\klts,\hrt}
\eqn\mof{  L_F=F(x) \del u \bd v +   \del x^i \bd x_i
+ {\cal R}\p (x)\ ,  }
describes
a family of backgrounds   with
metric and antisymmetric tensor
characterized by a single function $F(x)$ and dilaton $\p (x)$
(depending  only
on the transverse coordinates
$x^i$, $\ i=1,...,N$)
\eqn\ffff{ ds^2 = F(x) du dv + dx_i dx^i\ ,   \qquad B_{uv} = \ha F(x) \ . }
  The leading order string  equations
then reduce to  \klts
\eqn\fmod{ F^2(-\ha \del^2 F\inv  + b^i \del_i F\inv) =0 \ , \ \ \
  \p = \p_0 + b_i x^i  + \ha  \ln F (x) \ , \ }
where $b_i$ is a  constant vector.
At the points where $F$ is nonvanishing we get
\eqn\fmodd{  \del^2 F\inv =2 b^i \del_i F\inv \ . }
Some of the solutions to \fmod\  were  shown   to
correspond to gauged WZW models with the  gauged subgroup  being  nilpotent
\klts\  (see Section  5.3 below). It was
argued that like ungauged WZW models they should not receive non-trivial
higher order corrections even in
the CFT scheme (cf. Section 2.1). Though not
 all of  the  solutions  to  \fmod\ can be obtained
from  gauged WZW models
one can   show  \hrt\ that indeed  there
exists  a scheme in which all of them are  exact
and receive no $\a'$ corrections.\foot{The central observation  was that
according to
\shwts\ the determinant resulting from integration over $u$ and $v$
is local and (after a local change of a scheme) its only effect is to cancel
the nontrivial  dilaton term in \mof.}
 Since the equation \fmodd\
 for $F\inv$ is linear,
linear combinations of these solutions yield new exact solutions.

One of the most interesting solutions in this class  is the
`fundamental string'  (FS)  one \gibb\
 which
has $b_i=0$ and ($ r^2 = x_i x^i $, \  $D=2 + N$)
\eqn\fss{ F\inv ={ 1 + {M \ov r^{D-4}} }   \ , \ \ D>4 \ ; \ \ \
 \ \ \ \ F\inv ={ 1 - {\m  \  \ln { r \ov r_0} } } \ , \ \ \  D=4\ .  }
This solution can be interpreted  \refs{\dabha, \gibb}
as the field of a straight fundamental string located at
$r=0$.
Note that in contrast to \gibb\ where the  leading-order
equation
which follows from the effective action (and thus has  upper indices, $R^{\m\n}
+ ...=$source)  was solved,
  one  does not need  a source at the  origin to satisfy
the conformal invariance condition  \fmod\
($R_{\m\n} + ... =0$ everywhere, including $r=0$).
To prove  exactness of the fundamental string solution
\hrt\  it was important  to  write down the corresponding string  \sm
  which took  the very simple  $F$-model form \mof,  having
 two `null'
chiral currents.

\subsec{General  $F$-model }
The key property of the \KM \mofk\ is that it has a covariantly constant null
vector $\del/\del v $. The main features of the \FM
are that there are two
null Killing vectors corresponding to translations of $u$ and $v$, and that
the
coupling to $u,v$ is chiral (i.e. $G_{uv} = B_{uv}$). This means that
in addition to the three global Poincare transformations in the $u, v$ plane
the
\FM is invariant under the infinite dimensional symmetry $u' =u + f(\tau -
\s)$ and $v' = v+ h(\tau +\s)$. Associated with this symmetry are  two
conserved  world sheet chiral currents:
 ${\bar J}_u = F\bd v, \ J_v = F \del u$.
These properties are preserved if the transverse $x^i$-space  is curved,
i.e. the
general \FM   is defined  as  \hrt
\eqn\mofw{  L_F=F(x) \del u \bd v +  (G_{ij} + B_{ij})(x) \del x^i \bd x^j
+ {\cal R}\p (x)\ .  }
The all-order conformal invariance conditions for \FM
are satisfied \hrt\ provided one is given a
conformal `transverse' theory  $(G, B, \p')$  and
(assuming a special  `leading-order' scheme is used)
\eqn\summm{   \p= \p'   + \ha \ln F \ ,   }
\eqn\taccc{ -   \o F\inv +  \del^i \p' \del_i F\inv
=- \ha [\na^2  +  O(\a')]F\inv     +  \del^i \p' \del_i F\inv =0
\ . }
As in \taac\  $\o$  is the scalar anomalous dimension operator of the
transverse theory,   depending, in general,  on $G_{ij}$ and $B_{ij}$.
When   $(G,B,\p')$ correspond to  a  known (e.g.,  coset)  CFT
 this equation
can be put in an  explicit form  to all orders in $\a'$.

The  model \mofw\ is related by leading order duality to \KM \mofk: the dual
of  \mofk\   with respect
to $u$ is \mofw\ with $F=K\inv$.
In the leading-order
  scheme the leading-order duality is exact. In particular,   \mofw\  with
$G_{ij}=\d_{ij}, B_{ij}=0$  is  dual to the  $K$-model (cf. \mofk)
\eqn\feff{  L_{K}=\del u \bd v +   F\inv (x)\del u \bd u +  \del x^i \bd x_i
+ {\cal R}
 (\p_0 +b_ix^i) \  ,   }
which represents an exact  solution if $F$ solves \fmod.

Another exact $D=4$ \FM solution  is obtained  by taking the transverse
two dimensional theory to be non-trivial, i.e. represented by the $D=2$
euclidean black hole  background  \hrt. This solution  can be obtained from
\mofk\
with $K$ given by  \taa\  by the  $u$-duality transformation   (cf. \mixl)
\eqn\fmo{ ds^2  = (a +  m\ \ln \th bx)^{-1} dudv  + dx^2 +
 { {\rm tanh}^2 bx }\ d{\theta}^2 \ ,   } $$  \ \ B_{uv} = \ha (a+ m\ \ln \th
bx)^{-1} \ , \ \ \   \p = \p_0  -  {\ln \ch }  bx + \ha \ \ln\ F \ . $$
The  background  \fmo\  can be viewed as a generalisation of  the  fundamental
string  \fss\  in four dimensions (with $F\inv = 1 - \m  \ \ln{( r/ r_0)}$).
While the latter has,    in addition to
the usual singularity at $r=0$,   another singularity outside
the string at  $r=r_0$,  the solution  \fmo\ has the same singularity at  the
origin  but is regular elsewhere
and is  asymptotically flat.
The original  fundamental string solution  can be recovered by
taking the limit $b \ra 0$ which is consistent
since the central charge condition is now imposed
 only at the level of the full $D=4$ solution.

\subsec{$F$-models  obtained  by nilpotent  gauging of   WZW  models}
The conclusion   about the existence of  a scheme where the $F$-model
\mof,\fmod\
 represents an exact string solution  is consistent with the
result of \klts\  that
the  particular model \mof\  with
\eqn\klt{  F\inv =  {  \sum_{i=1}^N \ep_i {\rm e}^{  \a_i\cdot x  } }\ , \  \ \
\ \  \ \
\p =   \p_0 +   \r \cdot x   +
\ha\  \ln \ F
\   ,   }
can be obtained  from  a $G/H$ gauged WZW
model.
Here  the constants $\ep_i$ take values  $ 0$ or $ \pm 1$, $\ \a_i$
are simple roots of the   algebra of a maximally non-compact
real Lie group $G$ of rank $N=D-2 $
 and $\r= \ha
\sum_{s=1}^m  \a_s$ is
 half  of the sum of all positive roots.
 $H$ is  a  nilpotent subgroup of $G$
generated by $N-1$ simple roots (this condition on $H$
 is needed to get  a model with only one time-like  direction).
The flat transverse coordinates $x^i$ correspond to   the Cartan subalgebra
generators.

 This `null'  gauging \klts\ is based on the  Gauss decomposition   and thus
directly applies only to the
groups  with the algebras that are  the `maximally non-compact' real forms of
the Lie algebras
(real linear spans of the Cartan-Weyl   basis), i.e.   $sl(N+1,R), \
so(N,N+1),$ etc.  These WZW models can be considered as
natural generalisations of the $SL(2,R)$ WZW model.  For such  groups there
exists a real group-valued Gauss
decomposition
$$  g=  \exp ( \sum_{\Phi_+ } u^\a E_{\a }) \  \exp ( \sum_{i=1}^N  x^i H_i)  \
\exp ( \sum_{\Phi_+} v^\a E_{-\a }) \ .   $$
 $\Phi_+$ is  the set of positive  roots of a
complex algebra with the Cartan-Weyl  basis consisting of the  step operators
$E_\a, \ E_{-\a}  , \ \a \in \Phi_+ $  and  $N(=$rank$ G)$  Cartan  subalgebra
generators $ H_i$.

The {\it four}  dimensional  ($D= 2+N=4$)
models
are obtained  for each  of the rank 2 maximally non-compact groups:
  $SL(3,R), \ SO(2,3)=Sp(4,R)$, $SO(2,2)=SL(2,R)\oplus SL(2,R) $   and $G_2$.
In the rank 2 case the corresponding background  is parametrised by  a $2\times
2$ Cartan
matrix or by  two simple roots with components $\a_{1i}$ and $\a_{2i}$
 and  one  parameter $\ep =\ep_2/\ep_1$ with values $ \pm 1$,
\eqn\ffffw{ L = (  {\rm e}^{  \a_1\cdot x } + \ep  {\rm e}^{  \a_2\cdot x  }
)\inv
 \del u \bd v + \del x_1 \bd x_1 +
\del x_2 \bd x_2} $$ + \
\a' {\cal R} [\p_0  + \ha (\a_1\cdot x  +  \a_2\cdot x)
- \ha  \ln   \bigl(  {\rm e}^{  \a_1\cdot x  }   + \ep  {\rm e}^{  \a_2\cdot x
} \bigr)]  \ . $$
For example, in the case of $SL(3,R)$
$\a_1=(\sqrt 2 , 0), \  \a_2=(- {1/ \sqrt 2} , {3/ \sqrt 2}) $.
In addition to the Poincare symmetry in the $u,v$ plane and affine
transformations of $u$ and $v$   this model
is also invariant under a correlated constant shift of $x^i$
and $u$ (or $v$).
For $\ep=1$ the curvature is non-singular in the coordinate patch used in \mof,
but there is also a horizon so one should first consider the geodesic
completion.
It is likely that there is still no curvature singularity, as in the case of
the $SL(2,R)$ group space (in any case, the corresponding CFT seems to be
non-singular).

The presence of the two chiral currents implies \klts\ that the
classical equations following from \mof,\klt\  which describe   the string
propagation in these backgrounds reduce to the Toda equation for $x^i$
(cf. \ora)
\eqn\tod{  \del\bd x_i  + \ha \chi \del_i F\inv =0 \ , \ \ \  F \bd v = \n
(\bar z)  \
, \  \ F \del u = \m (z)
\ , \  \  \chi \equiv
\n (\bar z) \m (z) \  ,   }  where
$\chi$ can be made constant by the
 (conformal) coordinate transformation. Then
the  solutions  (including the solutions of the constraints)
can be  expressed in terms of the Toda model solutions.

Let us mention also that there exists another example of  \FM which
corresponds to a gauged WZW model:
the general  \FM in 3 dimensions. Solving \fmodd\ in $D=3$
one finds that $F\inv = a + m e^{bx}$.
As  explained  in  \refs{\hrt,\hrts}, this model can be obtained from a special
$SL(2,R)\times R /R$ gauged WZW model and, at the same time, is the extremal
limit
of the charged black string  solution \hoho.
The case of $a=0$, i.e. \mof\ with $F= e^{-bx}$,  corresponds to the ungauged
$SL(2,R)$ WZW model
(with the action written in the Gauss decomposition parametrisation).

\newsec{Chiral null models }
Chiral null models \hrts\ is a  class of \sms which generalize both plane wave
models and $F$-models  and have one conserved chiral null current (cf.
\stre,\mof)
\eqn\strem{   L =  F(x)  \del u \bd v +
  \td K(u,x)\del u \bd u  +  2\td A_i(u,x) \del u  \bd  x^i + \del x^i \bd x^i
  +   {\cal R}  \p (u,x)\  , }
or, with $\tilde K\equiv FK, \ \tilde A_i\equiv FA_i$,
 \eqn\strem{   L =  F(x)  \del u [\bd v +
   K(u,x) \bd u  +   2A_i(u,x)  \bd  x^i ] + \del x^i \bd x^i    +    {\cal R}
\p (u,x)\  . }
For simplicity we assumed that the transverse space is flat
but a generalisation to the case of  any conformal transverse theory is
straightforward. Note that $F$ does  not depend on $u$ since otherwise the
model is not conformal.  Like the $F$-term, the  vector coupling has a special
chiral structure: the
$G_{ui}$  and $B_{ui}$ components of the
metric and the antisymmetric tensor are equal to each other.
The affine symmetry
$v'=v + h(\tau + \s)$ of the action implies the existence of a conserved
chiral null current. In particular, the metric has  the null Killing vector
and the corresponding string background
has special holonomy properties.
The action is  also invariant under the  subgroup of coordinate transformations
combined with  a `gauge transformation' of the coupling functions
\eqn\ssa{  v'=v-
2\eta(x,u)\ , \ \ \ \ \  K' = K + 2\del_u \eta  \ , \ \ \ \ A'_i= A_i  + \del_i
\eta  \ .}
Using this freedom  one  may  choose a gauge in which
$K=0$. This is not possible in the special case  (which we shall mostly
consider below) when  $K,\ A_i$ and $\p$  do not depend on
$u$, i.e. when  $\del / \del u$ is a Killing vector and  $K$
cannot be set to zero without loss of generality.
In the latter case the chiral null model is `self-dual':
  performing  a
leading-order duality  transformation
along any non-null direction in the $(u,v)$-plane (e.g. setting  $v = \hat v +
a u, $ and dualizing with respect to $u$)  yields a $\s$-model
of exactly the same form with $F,\ K,\ A_i$ and $\p$ replaced by
\eqn\dudu{
 F'=  (K + a)^{-1}\ , \ \ \  K'=  F\inv
 \ , \ \ \  A_i'= {A_i} \ , \
\ \ \p'= \p - \ha \ln [F(K+ a) ] \  \ . }
 As in the case of the
 \FM,   there exists a scheme in which the conditions on the functions $F,\ K,\
A_i$ and $\p$  under which
\strem\ is conformal to all orders in $\a'$ turn out
to be equivalent
to the  leading-order equations \hrts\
 \eqn\qwqw{  F^2( -\ha \del^2
F\inv +  b^i \del_i F\inv ) =0 \ , \ \ \ \  F( -\ha  \del_i {\cal F}^{ij} + b_i
{\cal F}^{ij})  =0 \ , }
\eqn\uuu{
F( - \ha  \del^2 K  +  b^i \del_i K +    \del^i\del_u{  A_i} - 2b^i \del_u A_i)
+ 2 {\del^2_u \p}  =0\ , }
\eqn\qwqww{ \p(u,x)=  \p(u) + b_i x^i + \ha  \ln F(x)\ , }
$$  {\cal F}_{ij}\equiv \del_i A_j - \del_j A_i\ , \ \ \ \ \del^2\equiv
\del^i\del_i \ . $$
Eq. \qwqww\ implies that the central charge of the model
is  $c= D  + 6 b^ib_i$.
When $F,K,A_i,\p$ are  independent of $u$ and
 $b_i=0$, these equations take  simpler form (at the points where $F$ is
non-vanishing)
\eqn\eqqq{   \del^2 F\inv=0 \ , \ \ \  \del^2 K =0 \ , \ \ \   \del_i {\cal
F}^{ij} = 0 \ , \ \ \ \  \p=  \p_0 + \ha  \ln F(x) \ .  }
Since these equations are linear their
solutions
satisfy a solitonic no-force condition and can be superposed.

Special cases of the model \strem\ include  chiral plane waves ($F=1$)
and   generalisations of the fundamental string  solution, e.g.,
describing  travelling  waves along fundamental string,
$A_i=0, \ F\inv= 1 + M/r^{D-4}, \ K=K(u,x)$
(some of these solutions   were originally found   as the leading-order
solutions of the string effective equations  in \refs{\garf,\sen,\wald,\ghrw}).
In particular, in the $u$-independent spherically symmetric case
$K= a +  b/r^{D-4}=  a' + b'F\inv$  so that (redefining $u$ and $v$) the action
\strem\
can be put into the  simple form
\eqn\fsssw{ L= F(x) \del u \bd v +  n \del u \del u + \del x_i \bd x^i
+ {\cal R} \p (x) \ , \ \ \ \  n= 0, \pm 1\ ,  }
where the three values of $n$ correspond to inequivalent models.\foot{The
models
with $n=1$ and $n=-1$ are related by the analytic continuation
$u\to i u , \ v \to -i v$, suggesting that formal solutions
of the corresponding CFT's may be related. This
 does not imply, of course, the equivalence of the two models once the
periodicity  condition
on the spatial
part of $u$ and $v$  is imposed.}

Other related solutions will be discussed in detail in the following section.

\newsec{$D=4$ solutions  with electromagnetic fields
 corresponding to  $D>4$ chiral null models}
There exist several classes of exact $D=5$ solutions
which  `reduce' to  interesting $D=4$ backgrounds if the size
of the compact 5-th dimension is  very small.
Every solution with
an asymptotically flat
four dimensional space-time  must have a number of internal dimensions (to
have the total central charge 26 or 10).
Following  Kaluza-Klein idea,   we may
assume that  off-diagonal components of the higher
dimensional fields give rise to  extra four dimensional  fields, in particular,
gauge fields.
We shall consider  the closed
bosonic
string theory which has no fundamental gauge fields in the higher
dimensional space but the gauge fields do appear upon dimensional reduction
when the theory is  `viewed' from four dimensions.
The `reduction' does not  mean that we drop the dependence on the internal
direction: the conformal \sms   which  describe
these exact  $D=4$ solutions
are still  higher dimensional ones.

Below  we shall first review the
procedure of `dimensional reduction'  in string theory
and then summarize  the information  about string  black hole solutions
 which are exact to all orders in $\a'$.
All known exact  black hole solutions  turn out to be   the extreme ones,
with electric solutions
 being described by  $D=5$ chiral null models  \refs{\horrt,\hrts}.

\subsec{Kaluza-Klein  `reduction' }

The procedure of `dimensional reduction'
(e.g. from $D=5$ to $D=4$)
in string theory  is  reinterpreting  the $D=5$
 bosonic string \sm action
(with background fields  not depending on $x^5$)
 as an  action
of a  $D=4$ string  with an internal degree of freedom (represented  by the
Kaluza-Klein coordinate $x^5$)
which describes the   coupling  to  additional
vector (and scalar)  background fields.
Explicitly,
this corresponds to the following  `Kaluza-Klein'
rearrangement of terms in the  $D=5$  \sm action
    (see, e.g.,    \refs{\duff,\hrts})
$$ I_5={1\over \pi\alpha '}\int d^2 \s\big[ ( G_{MN} +B_{MN})(X)
\del X^M \bd  X^N  + {\cal R} \p
(X) \big]$$  $$
= {1\over \pi\alpha '}\int d^2 \s\big[  (\hat G_{\m\n} +B_{\m\n})(x)
\del x^\m \bd x^\n + \  e^{2\s(x) } [\del y+ {\cal A}_\m (x) \del x^\m][\bd y+
{\cal A}_\n (x)
\bd x^\n]  $$
\eqn\lagr{
 + \  {\cal B}_{\m   } (x) (\del x^\m \bd y- \bd x^\m \del y )  + {\cal R} \p
(x) \big]\ , }
where $X^M= (x^\m, x^5), \ x^\m=(t, x^i, x^3), \ x^5\equiv y$
and
\eqn\fgfg{\hat G_{\m\n} \equiv  G_{\m\n} - G_{55}{\cal A}_\m {\cal A}_\n
\ , \ \ \ G_{55}\equiv  e^{2\s}\ , \ \ \ \
 {\cal A} _\m\equiv   G^{55}  G_{\m 5}\ ,  \ \ {\cal B} _\m \equiv
B_{\m 5}\ .  }
These definitions of fields  guarantee
 manifest gauge invariance under $y'=y + \ep, \
 {\cal A}_\m' = {\cal A}_\m - \del_\m \ep, \
B_{\m\n}'= B_{\m\n} + \ep (\del _\m {\cal B} _\n - \del _\n{\cal B} _\m)$.
Similar decomposition of the string action can be done in the case of several
internal coordinates $y^a$.   From the point of view of the
low-energy effective  field theory,
  this decomposition corresponds to starting
 with the $D=5$ bosonic string effective action and
 assuming  that one spatial dimension $x^5$ is compactified on a small
circle. Dropping the massive Kaluza-Klein modes  one then
finds the following dimensionally reduced  $D=4$  action (see,  e.g., \maha)
\eqn\acttp{  S_4 = \int d^4 x \sqrt {\hat G }\  e^{-2\Phi}    \ \big[
  \   \hat R \ + 4 (\del_\m \Phi )^2  - (\del_\m \s )^2  } $$- {\textstyle 1\ov
12} (\hat H_{\m\n\l})^2\  -  \four e^{2\s} ({  F}_{\m\n}
({\cal A}))^2
-\four e^{- 2\s} (F _{\m\n} ({\cal B}))^2
  + O(\a')   \big]  \  , $$
where,   in addition to \fgfg,  we have  defined
 \eqn\jjj{  F_{\m\n} ({\cal A}) = 2\del_{[\m}
{{\cal A} _{\n]}  \ ,   \ \ F_{\m\n}({\cal B}) = 2 \del_{[\m} {\cal B}_{\n]} }
\  ,  } $$
\hat H_{\l\m\n} = 3\del_{[\l} B_{\m\n]} - 3 {\cal A}_{[\l} F_{\m\n]}
({\cal B})
\   , \ \ \ \ \Phi= \phi - \ha \s \ .    $$
The  \sm \lagr\ duality transformation in  the  $y$-isometry direction
induces the following  target space transformation
\eqn\daa{\cA \to \pm \cB\ , \  \ \
 \cB \to \pm  \cA \ , \ \   \
\s \to -\s\   , \  \  \ \p \to \p - \s\ , \  \ \  }
$$ \hat G_{\m\n}  \to \hat G_{\m\n}  \ , \  \ \
B_{\m\n}  \to B_{\m\n} + \cA_\m \cB_\n - \cB_\m \cA_\n\ , \ \ \
\hat H_{\m\n\l} \to \hat H_{\m\n\l}  \ , $$
which is obviously  an invariance of  the action \acttp.

Given a conformal $D=5$ \sm and representing it in the form \lagr\ one can
read off
the expressions for the corresponding $D=4$  background
fields which are then   bound to represent a solution
of the effective action  \acttp.
The string solution is,  of course,
 described by the full $D=5$ action \lagr.  Since $y$  was  assumed to be  a
compact isometry direction of the $D=5$ model,
dual $D=4$ backgrounds related by  \daa\    represent  the same string solution
(CFT).

In certain special
cases the nontrivial part of the action \acttp\ can be expressed in terms of
only one scalar and one vector and  takes   the familiar form (here we use  the
Einstein frame)
\eqn\actt{S_4 = \int d^4 x \sqrt {{\hat G_E}}   \ \big[
  \   \hat R_E \ - \ha (\del_\m \psi )^2 - \four e^{- a\psi} ({
F}_{\m\n})^2  + O(\a') \big]  \  . }
For example, if one sets $\p=0$  and $H_{MNK} = 0$ in the $D=5$ action  one
obtains \actt\ with $\psi= - a  \s$ and $ a=\sqrt 3$. This is
the standard Kaluza-Klein reduction of the Einstein action. Another possibility
is to set $\s =0\
(G_{55}=1),\ \ \hat H_{\m\n\l} =0$ and either the two vector fields
proportional
to each other, or let one of them vanish. This case corresponds to \actt\
 with $\psi= \Phi$  and $a =1$.

\subsec{ Exact  $D=4$ extreme  black hole solutions   }
Classical string theory, when considered  from the effective action point of
view,  is different from the Einstein theory at least in two respects:
(i) its action contains  terms of all orders  is $\a'$ expansion,
i.e. it is a one-parameter
`deformation' of the Einstein theory,  and
(ii) the action depends on other `massless' fields
which  should be treated  on an equal footing with the metric.
These differences are reflected in the structure of black hole solutions. For
example, the Schwarzschild solution is necessarily modified by $\a'$
corrections with dilaton becoming non-constant
at the next to the leading order \callmy.
Also, the presence of the dilaton coupling implies the
existence of new leading-order solutions describing charged black holes
\refs{\gib,\gim,\gar}.

Let us summarize what is known about {\it exact}  black hole
solutions. Exact black hole
solutions have been constructed in two \witt\    and three \horwel\
dimensions
using (gauged) WZW models based on the group $SL(2,R)$.
As for higher dimensions,
 there are  coset CFT or gauged WZW
 constructions of just the `throat' regions  of some
 four dimensional extreme magnetically charged dilatonic black holes
\refs{\gps,\los,\jooo}\foot{The fact that the form of the throat solution is
unchanged under string $\a'$
corrections was noticed earlier in  \gurs.}
but no connection  between complete asymptotically flat solutions and  coset
models was found.
 It appears that while  non-extreme
 (and some  non-supersymmetric extreme \nats)  leading-order charged black hole
solutions  are necessarily modified
by $\a'$-corrections,
   certain  extreme magnetic and electric black hole leading-order
 solutions are,  in fact,   exact. The magnetic and electric cases turn out to
be very different  and so  should be discussed separately.

(I) \  Extreme magnetic solutions:

(1a)\  The extremal limit of a black five-brane \horstr\
is an exact superstring solution \chs,  which upon dimensional
reduction gives  an exact
five dimensional extreme black hole.
This black hole has a magnetic type of charge associated
with the antisymmetric tensor field.

(1b) \ One can also
dimensionally reduce the exact solution
of \chs\ down to four dimensions \khga\ to obtain an
extreme magnetically charged  Kaluza-Klein ($a=\sqrt 3$) black hole
\dukh\ (see also \fudd).
The  $D=5$ string \sm  corresponding to  this `$H$-monopole' reduction is
\eqn\hmono{
L= - \del t\bd t + F\inv  (x) (\del x^i\bd x_i + \del y \bd y ) + A_i(x) (\del
y
\bd x^i  - \bd y \del x^i) + {\cal R} \p(x)
\ , }
where $y=x^5, \ i=1,2,3, \ r^2 = x_i x^i$, and
\eqn\hmonoh{ e^{2(\p-\p_0) } = e^{2\s}=  F\inv  \ , \ \ \   \ \ \  F\inv = 1 +
{M\ov r}
\ ,  } $$
  \cA_i=0\ , \ \ \ \  \cB_i =-A_i ={\rm monopole} \ ,  \ \ \
F_{ij} = H_{ij5}= M\ep_{ijk} {x^k\ov r^3} \  ,  $$
(and $H_{\a\b\g} = \ep_{\a\b\g\s}  \del^\s  e^{-2\p}  \ , \ \  \a,\b, ...,=
1,2,3,5$).
More precisely, this background is $T$-dual (in the sense of \daa)
 to the standard
extreme  Kaluza-Klein  black hole  solution \gib\ which has constant dilaton.
Its (1,1)  supersymmetric extension is  an exact conformal model
 having (4,4) worldsheet supersymmetry.
In the heterotic case, we achieve the same by  embedding
into the  superstring, i.e. by adding
$SU(2)$  internal gauge field  $V$  background
equal to  the  generalised connection  with torsion $\omega_+$.
Without embedding (for zero  $V$) one still has
(4,0) supersymmetry,  but it seems that finiteness is spoiled by anomaly (which
absent in the  similar   electric case model \hrts, see below).

(1c) \ The string \sm  corresponding to the embedding of the dilatonic ($a=1$)
extreme magnetic
  $D=4$ black hole  into $D=5$  \nels\   can be represented as
\eqn\hono{
L= - \del t\bd t + F^2(x)  \del x^i\bd x_i } $$
+ [ \del y  +  A_i(x)  \del x^i][\bd y  +A_i (x) \bd x^i ]
 + A_i (\del y \bd x^i  - \bd y \del x^i) + {\cal R} \p (x)
\ . $$
Here  the  modulus  $\s$  is constant (so that it is self-dual in the sense of
\daa)  and thus
 $\cA_i = - \cB_i= A_i$, while $F$, $A_i$  and $\p$ are the same as in
\hmonoh.
The $(1,0)$ supersymmetric version of this  model is  actually $(4,0)$
supersymmetric \nels\ (the corresponding background  breaks
 `one-half'
of space-time supersymmetry when embedded into  $D=10$ supergravity).
The same model
 was considered  by starting directly from $D=10$ (and  was generalised  to
multi-center case)  in \kalor\  where it was pointed out that it   becomes
(4,1) supersymmetric
 once one embeds the heterotic string solution into the  superstring theory,
i.e.
introduces also the  gauge field  background   $V= \omega_+$. Then
using the result of \sus\  one may  argue
that this model is an exact heterotic string solution.

(II) \ Extreme electric  solutions:

(2a) \ The extreme  electric `Kaluza-Klein' ($a=\sqrt 3$)
$D=4$ black hole   \gib\  can be obtained by
dimensional reduction  either of $D=5$  plane wave background \gib\  or of  the
dual $D=5$ fundamental string model \gaun.
Since these  two $D=5$ backgrounds are exact (bosonic) string solutions
\hrt,  this black hole solution is exact.

(2b)\  The extreme  electric  dilatonic  ($a=1$)
$D=4$ black hole \refs{\gim,\gar}  can be obtained {\horrt} by
dimensional reduction from  a generalised  $D=5$ fundamental string model
(found as a leading order solution in
\refs{\wald,\ghrw}) which is also an exact string solution
\refs{\horrt,\hrts}.

(3c)\   The analogs  and generalisations
  of the Israel-Wilson-Perj\'es (IWP) \iwp\ solution of the pure
Einstein-Maxwell theory, including as  special cases
Majumdar-Papapetrou-type  solution (a collection of extremal electric dilatonic
black holes)  and an extremal electric Taub-NUT-type
solution  are  dimensional reductions of the generalised
$D=5$ chiral null model \hrts\  and as such are exact string solutions
(some of these  backgrounds were  originally   found as
leading-order  string solutions in  \refs{\kall,\jons,\galts}).

Below we shall discuss in more detail  the  extreme electric black hole
solutions and their  generalisations. Let us  first
note   several important differences between
the  above magnetic and electric solutions.
The $D=5,4$  extreme magnetically charged
solutions  are  products of a time-like line
and
a euclidean solution, while the electric solutions are  non-trivial in the
time-like direction.
 Second,  these magnetic extreme black holes were shown to be exact solutions
only in the
superstring (or heterotic string) theory. Extended $(4,n)$  supersymmetry
(on the world-sheet,  related to  that on space-time) played a key role in the
arguments  that there are no $\a'$-corrections to the leading-order
backgrounds.
The  solutions we  discuss below, like the $D=2,3$  examples,  are  exact
already  in the bosonic string theory (they also have, of course,  exact
analogs  in the superstring and heterotic string
theories \hrts).

Let us start with  a simple  $D=5$
plane wave   model  (see \sstry,\taac,\ffss)
\eqn\ppw{ L= \del u \bd v + K(x) \del u \bd u + \del x_i \bd x^i
+ {\cal R} \p_0 \ , \ \ \ K = 1 + {M\ov r} \ .  }
With given $K$ this is a conformal model  (assuming  $r>0$;  one needs a
$\d$-function source at the origin to satisfy the Laplace equation at all
points).
Identifying $u$ with the internal coordinate $y$ and $v$  with $2t$
we find
that \ppw\ corresponds to the following $D=4$ background  (cf.\lagr)
\eqn\pla{ ds^2_4 = - F(r) dt^2  + dx_idx^i\  , \
\ \  \ e^{2\s } =F\inv (r) \ ,  \  \ \ \cA_t = F(r)  \ , }
$$ F \equiv K\inv =  (1 + {M\ov r})\inv \ .  $$
This is just the extreme electrically charged Kaluza-Klein black hole
\refs{\gib,\dukh}, which can thus be viewed as
an exact string solution \horrt.\foot{It should be noted that while the choice
of  a particular
direction $y= bu + c v$  of dimensional reduction
does not matter for the purpose of deriving the
dimensionally reduced $D=4$ background \pla\  (different reductions will give
gauge-equivalent backgrounds)
the corresponding string models \ppw\ with different choices
of the definition of the compact coordinate $y$ are, in general, inequivalent.
For example, the choices of $u=y-t, \ v=y +t$
and $u=y',  \ v= 2t'$  are not equivalent since  the relations
$t=t' - \ha y', \ y = \ha y' + t'$  are not consistent with having  $y$ and
$y'$ both compact while   $t$ and $t'$ both  noncompact.}
This  exact electric solution is related to  the exact  magnetic solution
\hmonoh\ by $S$-duality.

Similar background can be obtained by  using  another
conformal \sm which describes the $D=5$ fundamental string background
\eqn\fss{ L= F(x) \del u \bd v + \del x_i \bd x^i
+ {\cal R} \p (x) \ , \ }
\eqn\dife{   e^{2(\p-\p_0)}  = F(r) = (1 + {M\ov r})\inv \ .  }
With $u=y-t, \ v=y+t$ this leads again to the same  4-metric  as in \pla\ and
with $\cA_t\to - \cB_t, \ \s\to -\s, \ \p \to \p -\s$.  This is  consistent
with the duality transformation rule \daa.
Indeed,  \fss\ is $y$-dual to \ppw\ (i.e. represents the same CFT)
with $u$ and $v$  in \ppw\ being  indeed
$ u= \tilde y, \  v = 2t$.
Note that \fss\ is conformal (cf.\fmod)
without need for a source at the origin.
Thus the $a =\sqrt 3$ extreme electric  black hole background
can  be viewed as coming from an exact solution without sources.

This  is an example of  what  may  be a general rule:
for spherically symmetric charged solutions  the
duality  relates   `fundamental' (supported by a source)
 and  `solitonic' (sourceless)  solutions.
It appears that the usual duality transformation  may map solutions
of low-energy effective equations  into
solutions only when the metric is invertible. Solutions  with a singularity at
$r=0$
can be  mapped into configurations requiring sources.

Starting with  the  following conformal model  which is a  generalisation
\fsssw\  ($n=1$) of
the fundamental
string model \fss \
\eqn\fsss{ L= F(x) \del u \bd v +  \del u \del u + \del x_i \bd x^i
+ {\cal R} \p (x) \ , \ }
and setting $u=y$, $v=2t$ one finds upon dimensional reduction
to $D=4$
\eqn\did{ ds^2_4 = - F^2 (r) dt^2 + dx_idx^i \ ,  \ \ \ \ \p(x) = \p_0 + \ha
\ln F (r)\  , } $$
\cA_t=-\cB_t = F(r)\  , \ \ \
  \ \ \cA_i = \cB_i =  \s =  \hat H_{\l\m\n} =  0 \ .   $$
This  is the extremal limit of the dilatonic $a=1$  electric black hole
\refs{\gim,\gar}.
Again, $S$-duality relates this solution to the exact magnetic solution \hono.
Thus the (Kaluza-Klein embedded) $U(1)$  magnetic solutions of the heterotic
string are $S$-dual to   the extreme
electric solutions described by chiral null models (see also \behrr).

The $r=0$ singularity of this background  (and also of the fundamental string
 one \fss)
  is probably
absent at the level of the corresponding CFT (as is suggested by the  form
of the tachyon equation, see also \horry).
The model  \fsss\ is `self-dual' so the problem  with
sources in the dual version  does not  appear here.

Exactness of  the source-free
 extreme black hole solutions (supersymmetric in the  magnetic case and
Kaluza-Klein embedded in the bosonic,  superstring or  heterotic string in the
electric case)
supports their solitonic interpretation.

\subsec{More general  $D=4$ solutions obtained from  $D=5$ chiral null models}

Further  generalisation is provided by  the $D=5$ chiral null model
\eqn\yyy{ L=F(x) \del u \big[\bd v   +  K(x)   \bd u +  2A_i (x)  \bd x^i \big]
+  \del x_i \bd x^i     + {\cal R}\p (x)\   , }
which is conformally invariant if the functions  $F, \  K , \ \p $ and $A_i $
of the three coordinates $x^i$ satisfy \uuu,\qwqw,\qwqww, i.e.
\eqn\trt{ \del^2 F\inv=0 \ ,
\ \ \  \del^2 K =0 \ , \ \ \  \del_i F^{ij}=0\ , \  \ \ \p=\p_0 + \ha \ln F \ ,
}
\eqn\aaw{   \ep^{ijk}\del_j A_k= \del^i  T \ , \ \ \ \ \   \del^2 T =0 \ .  }
 The model is thus parametrised by 3 harmonic functions.
In  particular, one can  choose
\eqn\maj{ F\inv = 1+ \sum_{k=1}^{N} {M_k\ov |x - x_k|}\ , \ \ \ \  \ \  K = 1+
\sum_{k=1}^{N} {m_k\ov |x - x_k|}\ , \ \ \
T =   \sum_{k=1}^{N} {q_k\ov |x - x_k|}\ . }
With $u=y, \ v=2t$ the corresponding
 class of exact $D=4$ backgrounds  is \hrts
\eqn\didii{ ds^2_4 = - {F(x)  K\inv (x)  }  \big[ dt +  A_i(x)  dx^i\big]^2 +
dx_idx^i
\ , } $$
\cA_t=  K\inv (x) \ ,   \ \ \ \cA_i =  K\inv (x)  A_i(x) \ , \ \ \
\ \ \   \cB_t= -
F(x)\ , \ \ \ \cB_i =-   F(x)  A_i(x) \ , $$ $$   \  \s= \ha \ln [F(x)K(x)]\
, \ \ \  \ \ \p = \p_0 + \ha \ln F(x) \  , \ \ \ \ \
\hat H_{\l\m\n} =  - 6 \cA_{[\l} \del_\mu \cB_{\n]}\ . $$
This class is `self-dual' since  under the  duality  $F\to K\inv$
(cf.\daa). The above extreme electric
solutions \pla\ and \did\  are  obviously included as special cases.

The case  of $K=F\inv$
is of a particular interest.
 Since $\s=0$ and the two gauge fields differ only by a sign, these backgrounds
 are solutions  of  the equations following from  \acttp,\actt\ with $a=1$.
These are precisely the $D=4$  dilatonic IWP backgrounds  found  as
the leading-order solutions in \refs{\kall,\jons,\galts}. Adding a nonzero
$A_i$ to the
 solution   \did\ by setting $T = q/r$ has the effect
of adding a  NUT charge. The result is the extremal electrically
charged dilatonic Taub-NUT solution.  One can  also  get linear superposition
of an arbitrary
number of solutions of this type  by taking  $F$ and $ T$ to be `multicenter'
harmonic functions as in \maj.
The solutions of  the Laplace equations  in \trt\  which
are singular on circles, rather than points  correspond to adding angular
momentum.
The solutions with  generic  $K$ and thus   different
gauge fields ($\cA_\m\not= \pm \cB_\m$) were first found in \hrts.
Some  (IWP) solutions obtained from  higher  ($D=10$) dimensional  chiral null
models
were considered in  \refs{\behr,\behrr}.
Leading-order Kaluza-Klein black hole solutions  related
to the above solutions by  the electro-magnetic  duality were discussed in
\park.

Let us note again  that  even solving the leading order string analog of the
Einstein's equations
 can be rather complicated when
the dilaton and antisymmetric
tensor are nontrivial. By choosing an ansatz at the level of the string
world sheet action  which yields simple equations for the \sm  $\b$-functions,
one can easily find new solutions of  the leading
order equations. The chiral
null models  provide  an example of this. Another example
is magnetic flux tube solutions found in \rutsn\ (see Section 8).

The simplest example of  the solution with $K\not= F\inv$
is  found in the spherically symmetric  case
\eqn\mas{ F\inv = 1+ {M\ov r}\ , \ \ \ \  K= 1 - {M\ov r}\ , \ \ \
A_i=0\ , }
corresponding to the $D=5$ chiral null model \fsssw\ with $n=-1$ (and redefined
$v$) which, like \fsss\ describes $D=5$  fundamental string with linear
momentum flowing along the string.
The resulting $D=4$ background \didii\
\eqn\diiiq{ ds^2_4 = -  (1- {M^2\ov r^2})^{-1}   dt^2 +
dx_idx^i
\ , } $$
\cA_t=  (1 - {M\ov r})\inv  \ ,   \ \  \ \  \cB_t= -
(1 + {M\ov r})\inv\  , \ \ \ \cB_i =\cA_i=0  \ , \ \ \ \  \hat H_{\l\m\n} = 0 \
, $$
$$
   \s= \ha \ln [(1 - {M\ov r})/(1 + {M\ov r})]\
, \  \ \ \  \ \ \p = \p_0 -  \ha \ln( 1 + {M\ov r})    \ ,  $$ $$ \vp = 2\p
-\s= 2\p_0  - \ha \ln( 1 - {M^2\ov r^2}) \  ,
 $$
 was interpreted in \behrr\ as a `massless' extreme electric black hole (note
that there is no $1/r$-term in the metric coefficient; $M$ here plays the role
of an electric charge and not of the ADM mass which vanishes).
Note that under the duality \daa\  this background is transformed into itself
with $M\to -M$.
 The close similarity of the two $D=5$ chiral null models
(\fsssw\  with $n=1$ and $n=-1$) corresponding to the $a=1$ extreme electric
black hole \did\ and the `massless' black hole \diiiq\
suggests that the two $D=4$ solutions may  be in certain sense `complementary'.

Another interesting special case is   $F=K=1$.
The  corresponding model is a special case of \stre\   (cf. \yyy)
\eqn\yyye{ L=\del u \bd v   +  2A_i (x)  \bd x^i
+  \del x_i \bd x^i     + {\cal R}\p (x)\   , }
where $A_i$ satisfies \aaw. Particular  solutions for $A_i$
 are:

(1) $F_{ij} = p \ep_{ijk} x^k/r^3 $,  corresponding to extreme magnetically
charged
dilatonic Taub-NUT solution with  zero mass;

(2) $ A_i= -\ha F_{ij} x^j , \ F_{ij}= \const$ or  $F_{i3}=0, \ F_{ij }=\b
\ep_{ij }, \ i,j=1,2, $
 describing    a `rotating magnetic universe'  or `constant magnetic field
solution'  \refs{\hrts,\ruts} \  (see \didii)
\eqn\dyyq{ ds^2_4 = -  ( dt +  \ha \b \r^2 d\vp )^2 + d\r^2 + \r^2 d\vp^2
 + dx_3^2 \ , } $$
\cA =-\cB = A= \ha \b \r^2  d\vp \  , \ \ \ \s=\p-\p_0= 0\ , \   \ \ \ B=\ha
\b\r^2 d\vp\wedge dt\ , \ \
\hat H_{tij } = F_{ij} \ . $$
The  stationary metric  \dyyq\ is the metric  of a   homogeneous space --
the direct product of the group space of the Heisenberg group and
$x_3$-direction.
This solution  has  a non-trivial antisymmetric tensor (and constant dilaton)
and thus, in contrast to the dilatonic  Melvin solution \refs{\mel,\gim},  has
no analog in the Einstein-Maxwell  theory.

We shall  discuss
 this and other  uniform magnetic field solutions in Sections 8,9.

\subsec{Charged fundamental string  solutions }
Another example of an  exact solution obtained by dimensional reduction is
the charged  fundamental string   background found at the leading order
level in \refs{\sen,\wald}. Starting with the general chiral null
model in $D+N$ dimensions, requiring that all the fields are independent
of $u$ and $N$ of the transverse dimensions  $y^a$ and assuming
 that the vector coupling has only $y^a$-components we obtain
\def \K {\tilde K}
\eqn\mofkac{ L=F(x) \del u \bd v +  \tilde  K(x)  \del u \bd u  +
\del x_i \bd x^i     +  2 \tilde A_{a } (x) \del u \bd y^a    + \del y_a \bd
y^a +   {\cal R}\p (x)\ .   }
This model is  conformal to all orders provided  $F,\ K\equiv F\inv \K,\
 A_{a}\equiv F\inv \tilde A_{a}$ and $\p$ satisfy
\eqqq. Looking for solutions which are rotationally
symmetric
in $D-2$ coordinates $x^i$ and  solving  the Laplace equations  we can put
the functions $F,K, A_a$ in the form\foot{For $Q_a=0$
we get not just the fundamental string  solution of \gibb\ but its modification
\fsss\
which corresponds to   momentum
running  along the string.}
 \eqn\lapl{F\inv    = 1 + { M \ov r^{D-4}}\ , \ \ \
\p  = \p_0 + \ha \ln F(r) \ ,  \  \  K= c + { P \ov r^{D-4}}\ , \ \
 A_{a}  = {Q_a \ov r^{D-4}} \ .   }
Shifting $v$  we can  in general replace $\K$ in \mofkac\ by a constant.
To reinterpret \mofkac\ as a $D$-dimensional model coupled to $N$ internal
coordinates we  should rewrite it in the form \lagr\
\eqn\mofkacc{ L=F(r) \del u \bd v +   [\K  - (\tilde A_{a})^2] (r)  \del u \bd
u  +
\del x_i \bd x^i +   {\cal R}\p (r) }
 $$  + \   \tilde A_{a} (r)( \del u \bd y^a  - \del y^a \bd u)    + \  [\del
y^a  + \tilde A^{a}(r) \del u ][\bd y_a + \tilde A_{a}(r) \bd u]   \ . $$
The first four terms give the $D$-dimensional space-time metric, antisymmetric
tensor and dilaton  while the last two indicate  the presence of the
two equal vector field backgrounds (see \lagr).  Note that since $G_{ab} =
\delta_{ab}$, the modulus field is constant (so that the lower
dimensional dilaton is the same as the higher dimensional one).
In the case of just one internal dimension we get
 one  abelian  vector field  with non-vanishing $u$-component
and the
 resulting background becomes
that of the  charged  fundamental string  of \refs{\sen,\wald}.
Again, this model is conformal without need to introduce sources
suggesting a   solitonic interpretation of the corresponding background.

\newsec{Other  examples of exact solutions:
magnetic flux tube  backgrounds}
One may  look for other  exact conformal \sms
by trying to generalise the chiral null model Lagrangian \strem.
One  possibility  is to replace its   flat
transverse part  by a conformal model just as this was done in the particular
cases of  $K$- and $F$- models \mofk, \mof . If the total dimension is
 $D=4$ that means  again to
replace  the flat 2-space by  the   euclidean black hole $SL(2,R)/U(1)$ model
(or $SU(2)/U(1)$  background).
Examples of such   exact $D=4$
`hybrid' solutions  with  $A_i=0$  in \strem\
were already given in \mix,\mixl,\fmo.
Similar solutions exist when the vector coupling function $A_i$ is
non-vanishing  and satisfies  the  curved transverse space analogue of \qwqw,
$\del_i( e^{-2\p} \sqrt G {\cal F}^{ij}) =0$.
 In particular, the generalisation  of the constant magnetic field solution
\dyyq\  is \rutsn\
\eqn\htse{ ds^2_4 = - (  dt + \ha \b b^{-2} { \rm tanh}^2 b\r \  d\vp)^2  +
d\r^2 + b^{-2} { \rm tanh}^2 b\r \ d\vp^2 + dx_3^2 \ , }
$$ {\cal A} = -{\cal B}= A=  \ha \b   b^{-2} { \rm tanh}^2 b\r \ d\vp\   ,  $$
$$
 B=  \ha  \b  b^{-2} { \rm tanh}^2 b\r\ d\vp\wedge dt\ , \ \
\ e^{\p-\p_0} = (\cosh b\r)\inv \ , \ \ \s=0 \ . $$
The corresponding $D=5$  \sm  (cf. \yyye)
\eqn\yyyem{ L=\del u \bd v   + \b   b^{-2}
 { \rm tanh}^2 b\r\ \del u \bd \vp
+  \del \r \bd \r +     b^{-2} { \rm tanh}^2 b\r\  \del \vp \bd \vp
  + {\cal R}\p (\r)\   , }
is conformal to all orders in the leading-order  scheme.
The constant $b$  ($\a' b^2 = 1/k$)  is fixed by the condition  of the
vanishing of the total central charge, $ 2 + 3k/(k-2)  + N -26=0$,
where $N$ is a number of extra free bosonic dimensions.
The solution  \htse\  reduces to \dyyq\ in the limit $b\to 0$ (i.e. $N\to
22$).
 The introduction of the parameter $b$ changes the large $\r$  behaviour of the
background fields (while for small $\r$ the form of the fields
remains the same as for $b=0$).

One may also try to  extend  the chiral null model by adding extra
couplings, e.g., $ 2\tilde{ A}'_i (x) \del x^i\bd v$.
In general, such models will no longer be conformal invariant to all orders if
only the leading-order $\b$-function conditions are satisfied \hrts.
There exists, however, a special case when
the above  term  is supplemented by
 a particular  extra term in the `transverse' part of the  action,
\eqn\twvec{ L = F(x) \big[ \del u + 2 A'_i (x)  \del x^i ] [   \bd v
+ 2 A_i(x) \bd x^a\big] +
       \del x_i\bd x^i +  {\cal R} \phi(x) \ .    }
This  `generalised $F$-model'  can be obtained
 by $u$-duality  from the
`non-chiral'   generalization  of the plane-wave $K$-model  \stry\ with two
non-vanishing vector couplings (the relation  between  the functions  in
\twvec\ and in \stry\ is: $
F= K\inv , \ A'_i= -\bar A_i  , \ A_i = A_i , \ \p=\p_0 + \ha \ln
F$).
When $A_i$ and $A'_i$ have constant field strengths,
the theory \twvec\ can be shown to be conformally invariant
to all loop orders,
provided  \hrts\ (cf. \cobf,\conii)\foot{While the model obtained by adding
$\td K(x) \del u \bd u$ to \twvec\  is not  exactly conformal
for any $\td K$ there exists a generalisation
of this  exact
solution with one of the field strengths being constant and another
being an arbitrary
(e.g. monopole in $D=3$) solution of the  Maxwell equation.}
\eqn\sonm{ \ha  \del^2 F\inv
+    {\cal F}^{ij} {\cal F}'_{ij} =0 \  , \ \ \ \ \p= \p_0 + \ha \ln F \ \ .
}
A particular  $D=5$ solution  is  described by \rutsn\ ($\a,\b =\const$)
\eqn\lagge{ L=
 F(\r) (\del u  -\a  \r^2 \del \vp) (\bd v   +  \b \r^2 \bd \vp)
+ \del \r \bd \r +
   \r^2 \del \vp  \bd \vp   + \del x_3 \bd x_3  \ ,  }
$$
+\     {\cal R} (\p_0 +  \ha \ln F ) \ ,    \ \ \ \
 F\inv = F_0^{-1}  + \a\b \r^2\ ,  $$
where $F_0^{-1} $ is a solution of the homogeneous   $D=3$
$(\r,\vp,x_3)$ Laplace equation,  e.g. (cf. \ook)
 \eqn\geneg{   F\inv (\r)  = 1 - {\m \ \ln { \r\ov\r_0}}  + \a\b \r^2 \ ,   }
or
\eqn\gene{   { F}\inv (\r, x_3)  = 1 + {M\ov r}  + \a\b \r^2 \ , \ \ \ \ \ \ \
r^2\equiv  \r^2 + x_3^2 \ ,  }
or
\eqn\genev{   { F}\inv (r)  = 1 + {M\ov r}  + {2\ov 3}
\a\b r^2 \  .  }
Dimensional reduction   along the $y=\ha (u+v) $-direction (see \lagr)
leads to  (electro)magnetic flux tube backgrounds   \rutsn
\eqn\qqq{    ds^2_4=  -F(\r)[ dt +  \ha (\a + \b) \r^2 d\vp]^2   +     d\r^2
+  \r^2   d \vp^2 +  dx_3^2 \ ,  }
\eqn\salr{B
 =    \ha (\b -\a)F(\r)  \r^2 d\vp\wedge dt \ , \ \ \
 e^{2\s } =  e^{2(\p-\p_0) } =  F(\r)  \  ,  }
\eqn\veme{{\cal A}
 = \ha (\b-\a) \r^2  d \vp  \ , \ \  \
{\cal B}  = - \ha
{F(\r) \r^2 } [ (\a+\b)   d \vp  -2\a\b dt]  \ ,   }
with the `constant magnetic field' solution \dyyq\ as a special $F_0=1, \ \a=0$
case.
The vector field ${\cal A}$ has constant magnetic field strength
while ${\cal B}$ has both magnetic and electric components.
The curvature in non-singular for $\a\b \geq 0$.

The background  \qqq,\salr,\veme\ with $F$ given by \gene\
represent a black-hole type configuration in an external electromagnetic field,
namely, a  generalisation to the case of $\a\not=0$
of the solution  \yyy,\didii\   (with $K=1, \ F= 1/(1 + M/r)$)
  which  was  an extension ($\b\not=0$) of  the
extremal  Kaluza-Klein  ($a=\sqrt 3$) black hole.\foot{At the same time, the
solutions \lagge\   with $F$ given by \gene\
do not
include  a generalization of the extreme  electric
 dilatonic ($a=1$)  black hole since the model
\lagge\ does not contain  the  term $K\del u \bd u$ (or $K\del v \bd v $) which
is
necessary in order to obtain this black hole  background by dimensional
reduction {\horrt}.
In fact,  for
$\a\b\not=0$ such a term (or, e.g., its `gauge-invariant' generalization
$K(\del u -2\a A)(\bd u -2\a \bar A)$)
cannot  be added to \lagge\   without
spoiling the  conformal invariance of the model.}

It is possible also obtain a more general  globally inequivalent conformal
model  by
replacing $\vp\in (0,2\pi)$ in \lagge\ by $\vp'= \vp +  q  y, \ y\equiv  \ha
(u+v)$ \rutsn. Reducing in $y$ direction in the case of $F_0=1$ one finds a
3-parameter ($\a,\b,q$)
 class of   exact $D=4$ axially symmetric flux tube solutions \rutsn\
which generalise the constant magnetic field solution  \dyyq\
and  the dilatonic ($a=1$)
and  Kaluza-Klein ($a=\sqrt 3$)
Melvin solutions \refs{\gim,\gibma} \
which in the string \sm\ frame have the form  (cf. \dyyq)
\eqn\melv{a=1\ : \ \ \ \  ds^2_4=  - dt^2  + d\r^2 +  F^2(\r)  \r^2 d\vp^2 +
dx_3^2 \ , }
$$ {\cal A}= -{\cal B} =  \b  F(\r)   \r^2 d\vp\ , \ \ \ B=0\ , $$ $$
 \s=0\ , \ \ \ \ \ \  e^{2(\p-\p_0)}=  F  = (1 + \b^2 \r^2)\inv  \  ,    $$
and
\eqn\melvi{ a=\sqrt 3 \ : \ \ \ \ ds^2_4=  - dt^2  + d\r^2 +  \td F(\r)  \r^2
d\vp^2 + dx_3^2 \ , }
$$ {\cal A}=  q  \td F(\r)   \r^2 d\vp\ , \ \ \  \ {\cal B}=0\ , \ \ B=0\ , $$
 $$
 \p=\p_0   \ , \ \ \ \ e^{2\s} = {\td F}\inv = 1 +  q^2 \r^2 \  .    $$
The  resulting expressions for the $D=4$ background fields (which solve, in
particular,  the leading-order equations following from \acttp)  are given by
\rutsn\
\eqn\qqqs{  ds^2_4
=
 -dt^2  +      F(\r) \r^2 (d\vp -\a dt)(d\vp -\b dt)   }
 $$ -  \ \four F(\r )\tilde F(\r) \r^4  \big[ (\a-\b -2q ) d \vp + q(\a + \b)
dt\big]^2 +  d\r^2 +  dx_3^2 \  , $$
\eqn\ttts{{\cal A}  =- \ha  {{\tilde  F}}(\r)  \r^2[ (\a-\b  - 2q ) d \vp + q
(\a+\b)dt] \ ,  }
$${\cal B} = -\ha  F(\r) {\r^2 } [ (\a + \b ) d \vp - (2\a\b  + q\a - q\b) dt]
\ ,  $$
\eqn\sssu{ e^{2(\p-\p_0) } =  F (\r) \ , \   \ \ \  e^{2\s } =   {F(\r) \ov
{{\tilde  F}}(\r)}\ , \ \ \ \   B=
  - \ha   (\a-\b) F(\r)  \r^2 d\vp\wedge dt \ ,
  }
\eqn\dees{
F (\r) \equiv {1\ov 1 +  \a\b {\r^2} } \ , \ \ \
{{\tilde  F}} (\r) \equiv {1\ov 1 + q(q +\b-\a)
{\r^2}}\ .  }
The metric is  stationary  and, in general,  describes a rotating
`universe'.
For generic values of the parameters the  abelian  gauge fields contain  both
magnetic and electric components  with the former being more `fundamental'
(there are no  solutions when both of the  gauge fields are pure electric).
For simplicity we shall call these solutions `magnetic flux tube backgrounds'.
The uniform  pure magnetic field  solutions  \dyyq, \melv\  and \melvi\
are the  following special cases:  $q=\a=0, \ \b\not=0$ ;\
$\ \a=\b=q\not=0$,  and $\a=\b=0, \ q\not=0$.
In addition to the $q=0$ subclass of
magnetic backgrounds \qqq,\salr,\veme\
there are two other special subclasses:
$\a=q$ (stationary metric, non-zero $B_{\m\n}$, zero $\s$)
and $\a=\b$ (static metric, zero $B_{\m\n}$, non-zero $\s$).

It is interesting to note that performing the electromagnetic
$S$-duality on the   magnetic Melvin backgrounds (for any value of $a$)
  one finds   leading-order solutions
with {\it constant}  electric  field.
These electric  backgrounds do not,  however,   represent the exact
classical string solutions (i.e.  they are corrected at each order in $\a'$),
and thus, in contrast to the  magnetic Melvin  case,  the conformal \sm which
corresponds to them is not explicitly known.\foot{Let us mention also some
other  known exact magnetic solutions in string theory.
  String (electro)magnetic  backgrounds   obtained by dimensional
reduction  of  $SU(1,1)$
WZW model  were considered in  \ant.
An exact  $D=3$ monopole \khu\
magnetic background based on $SU(2)$ WZW model tensored with linear dilaton
 was  discussed     in \bak.
Heterotic string models with magnetic monopole type  or dyon type backgrounds
were constructed in \refs{\gps,\jooo}.
A  conformal embedding of
a monopole-type magnetic field into the gauge sector of the  heterotic string
theory
which uses  $SU(2)$ WZW model  was studied  in
\kkkk.  The  Robinson-Bertotti solution  has an exact string
 counterpart \los\
which is a product of the two conformal theories:
`(AdS)$_2$'   ($SL(2,R)/Z$ WZW)   and
`monopole'  ($SU(2)/Z_m$ WZW)  \gps\  ones.}

Though the   magnetic flux tube backgrounds \qqqs--\dees\ look quite
complicated  the conformal string model corresponding to them can be solved
explicitly.
This will be discussed in the next section.

\newsec{Exactly solvable string models corresponding
 to  magnetic flux tube backgrounds}

In contrast to  generic  chiral null models
(e.g., to fundamental string and
extreme electric  black hole solutions) for which the corresponding
CFT is  not known  (in particular, one does not know
the exact form of the equations for tachyon  and other string modes propagating
in these backgrounds)
the conformal string models which represent the
magnetic flux tube backgrounds (\lagge\ and its generalisations)
can be  defined
and solved explicitly. Namely,  one can find  the quantum Hamiltonian expressed
in terms of free oscillators, determine
the  string spectrum, compute the partition function, etc. \refs{\ruts,\rutsn}.
These models appear to be simpler than   coset CFT's corresponding to
semisimple gauged WZW solutions.\foot{For a review of solvable (super)string
models based on  semisimple coset  CFT's  see, e.g.,
\refs{\barss,\kounn}.}
For example, their
 unitary  is easy to demonstrate because of the existence of a light-cone
gauge.
 They provide (along with  non-semisimple coset models
\refs{\napwi,\kk,\oliv,\sfetso,\sfees}) one of the first  non-trivial
solvable examples
 of  a  consistent  string theory in  curved
space-times. Some features of these models (in particular, the presence of
tachyonic instabilities in the spectrum and existence of critical values of
magnetic field) are similar to those of  the conformal model describing open
string in a constant magnetic field \refs{\frats,\abo,\burg,\ferrara}.

Generalisations to the case of  superstring and heterotic string theories are
discussed in \refs{\ruts, \rqsn}.

\subsec{String actions and  angular duality}

As discussed in Section 7, to describe the coupling  of  a  closed bosonic
string to a  magnetic
gauge field  background one  introduces a  compact internal coordinate $y$.
Then the  string models which correspond to the $D=4$ magnetic backgrounds
\dyyq\ (`constant magnetic field'), \melv\ (`dilatonic $a=1$ Melvin flux tube')
and \melvi\   (`Kaluza-Klein  $a=\sqrt 3 $ Melvin flux tube') are represented
by the following $D=5$ \sms\ \refs{\ruts,\tsem,\rutsn}
  \eqn\lgg{ L_{\a=q=0}  = \del u \bd v   + \b \ep_{ij} x^i
\bd  x^j \del u   +  \del x_i \bd x^i +   \del x_3 \bd x_3 + {\cal R} \p_0\
 }
$$=  \del u \bd v   + \b \r^2 \bd \vp  \del u   +  \del \r \bd \r
+ \r^2 \del \vp \bd \vp +   \del x_3 \bd x_3 +   {\cal R} \p_0 \ ,
$$
\eqn\lggm{L_{\a=\b=q}    =  - \del t \bd t +  \del \r \bd \r +  {F(\r) \r^2}
(\del \vp  + 2 \b  \del y)  \bd \vp  +
\del y \bd y +
 \del x_3 \bd x_3  +   {\cal R} \p(\r)  \ ,  }
$$  e^{2(\p -\p_0)} = F(\r) =
(1+ \b^2 \r^2)\inv   \ ,  $$
\eqn\lggme{L_{\a=\b=0}    =  - \del t \bd t +  \del \r \bd \r +  {\r^2}
(\del \vp  +  q\del y) ( \bd \vp  + q \bd y) +
\del y \bd y +
 \del x_3 \bd x_3  +   {\cal R} \p_0  \ ,  }
where $x_1 + i x_2 = \r e^{i\vp}, \ \vp \in (0,2\pi)$ and
\eqn\defe{ u\equiv  y-t\ , \ \ \  \  v\equiv y + t \   , \ \ \  \
y\in (0,2\pi R) \ . }
All three models have free-theory central charge.
In the case of non-compact $y$, i.e. in the limit $R\to \infty$,  they
become equivalent to  other known models.
The constant field model \lgg\ becomes the $E^c_2$ WZW
model of \napwi\ (or `plane wave' \acd) with
the corresponding CFT discussed in  \refs{\kk,\ruts,\palla}.
The $a=1$ Melvin model \lggm\ with  coordinates  formally taken to be
non-compact  can be identified with a particular   limit
of  $[SL(2,R)\times R]/R$ gauged WZW  (`black string') model \tsem\foot{In this
limit $k\to \infty$ and the mass and charge
of `black string' \hoho\ vanish but simultaneous rescalings of coordinates
give rise to a nontrivial model.}
 or,  equivalently,  with the    $E^c_2/U(1)$ coset  theory
\sfetso. The $R= \infty$ case of the $a=\sqrt 3 $ Melvin model \lggme\
is identical  to the flat space model after the redefinition of $\vp$.

 \lr \kirr {E. Kiritsis, \jnl \np, B405, 109, 1993. }

The solvability of these and  more general 3-parameter $(\a,\b,q)$ class of
models can be  explained using
 their relation via duality and formal coordinate shifts to  flat models.
Consider, for example,  the \sm
which a direct product of $D=2$ Minkowski space and $D=2$ `dual 2-plane'
 \eqn\lagrq{ \tilde I={1\over \pi\alpha '}\int d^2 \s\big[ \del u \bd v
+ \del \r \bd \r + \r^{-2} \del \tvp\bd  \tvp  + {\cal R} (\p_0 - \ln \r)\big]
\  . }
 $\tvp$  should have
period $2\pi\a'$ to preserve equivalence of the `dual 2-plane' model to the
flat 2-plane CFT {\rocver}, i.e. to the  flat model\foot{The two models are
equivalent in the sense of a relation of classical solutions and equality of
the correlators  of certain operators  (e.g., $\del \tvp$ and $\r^2 \del \vp$)
 but the spectra of states  are formally different (cf. also
\refs{\dvv,\kirr}): the spectrum is continuous on 2-plane and discrete on dual
2-plane (with  duality  relating
 states with given orbital momentum on 2-plane   and  states with given winding
number on dual 2-plane).}
 \eqn\lfff{ I={1\over \pi\alpha '}\int d^2 \s\big(  \del u \bd v
+ \del \r \bd \r + \r^{2} \del \vp\bd  \vp  + {\cal R} \p_0  \big)\  . }
If we now  make coordinate shifts  and add a  constant antisymmetric tensor
term we obtain ($\a,\b, q$  are  free parameters of dimension
$cm^{-1}$)
 \eqn\lag{ I={1\over \pi\alpha '}\int d^2 \s \big[
( \del u + \a \del \tvp)  (\bd v + \beta  \bd \tvp)
+ \del \r \bd \r + \r^{-2} \del \tilde \vp \bd \tilde \vp
} $$ + \  \ha q [\del ( u  +v) \bd \tvp - \bd (u +v) \del \tvp]
+ {\cal R} (\p_0 - \ln \r ) \big] \ .   $$
The  two models \lagrq\ and \lag\  are of
course `locally-equivalent'; in particular,
\lag\  also  solves the  conformal invariance equations. However,
if $u$ and $v$ are given by \defe\
the
`shifted' coordinates $u + \a \td \vp $ and $v + \b \td \vp$
are not   globally defined  for generic $\a$ and $\b$  (since the periods of
$y=\ha (u + v)$ and $\vp$ are different) and the torsion term is non-trivial
for $q\not=0$. As a result,
the conformal field  theories corresponding to
\lagrq\ and \lag\  will  not be equivalent.
The $O(3,3;R)$ duality transformation  with  $continuous $ coefficients
which relates the model \lag\ to a flat
one \lfff\ is not a symmetry of the flat CFT, i.e. leads to a new conformal
model
which, however, is simple enough to be  explicitly solvable \ruts.

Starting with \lag\  and making the
duality transformation  in $\tvp$
one obtains a more complicated \sm\
 \eqn\lagger{ L=
 F(\r) (\del u  -\a  \r^2 \del \vp') (\bd v   +  \b \r^2 \bd \vp'  )
+ \del \r \bd \r +
   \r^2 \del \vp'  \bd \vp'  } $$  +  \  \del x_3 \bd x_3 +   {\cal R} (\p_0 +
\ha \ln F ) \   ,   $$
$$
 F\inv = 1 + \a\b \r^2\ , \ \ \ \   \vp'\equiv \vp + \ha q(u+v) \ .  $$
Here $\vp$  $ \in (0,2\pi) $ is  the periodic
coordinate dual to $\tvp$.
 The models \lgg--\lggme\
are  the  special cases of \lagger.
The theory \lagger\   is conformally invariant to all orders in $\a'$.
For the purpose of  demonstrating  this  one may ignore the difference between
$\vp'$ and $\vp$ (i.e.  may set $q=0$ or consider $y$ to be non-compact).  Then
 \lagger\  becomes equivalent to \lagge, which is the special case of the
`generalised $F$-model' \twvec,\sonm.   The model
\lagger\  is   the  string theory   corresponding to  the
$D=4$ magnetic backgrounds  \qqqs--\dees.

One may  wonder why  we need to use the dual 2-plane as part of our model, i.e.
why not to start with  a locally flat model  with $\r^{-2} \del \tilde \vp \bd
\tilde \vp$ in \lag  replaced by $\r^{2} \del \tilde \vp \bd \tilde \vp$ and
constant dilaton.
It turns out that in this case the resulting  dual model (i) is not related to
magnetic backgrounds like Melvin one,  and (ii) is not exactly solvable in
contrast to the one  related to  \lag.
To  give an idea why this is so  let us consider the simple case
of the model  with $\a=q=0$. Then  making the duality in $\td \vp$
in \lag\ one finds\foot{As  was mentioned above,  the limit
of non-compact $y$   is the plane-wave model  of \napwi\
a   relation of which to
the flat model by a formal $O(3,3;R)$ duality  was noted in \refs{\kk,\kltspl}.
Our model with $u,v$ defined in \defe\ can be interpreted as a plane wave
moving in the direction of the  compact internal coordinate $y$.}
\eqn\lagnw{ \td I={1\over \pi\alpha '}\int d^2 \s \big(
 \del u \bd v + \beta \r^2  \bd \vp \del u
+ \del \r \bd \r + \r^{2} \del  \vp \bd \vp
+ {\cal R} \p_0 \big) \  . }
Written in terms of $x=x_1 + ix_2= \r e^{i\vp}$ the action  becomes
quadratic  in $x$
\eqn\lagww{ \td I={1\over \pi\alpha '}\int d^2 \s \big[
 \del u \bd v +\ha  i\beta (x\bd x^* - x^* \bd x) \del u
+ \del x \bd x^*  + {\cal R} \p_0 \big] \  ,  }
 and, as a result,  is exactly solvable \ruts.
For example, the torus  partition  is explicitly  computable
since the integral over $v$ constrains $u$ to the zero (winding) mode
value and thus the gaussian  integral over $x$ gives a determinant which can be
expressed  in terms of the Jacobi  $\theta_1$-function.
At the same time, neither the model
 \eqn\lagw{ I'={1\over \pi\alpha '}\int d^2 \s \big(
 \del u \bd v + \beta  \bd \tvp\del u
+ \del \r \bd \r + \r^{2} \del \tilde \vp \bd \tilde \vp
+ {\cal R} \p_0\big) \ , }
nor its $\vp$-dual
\eqn\lagwq{ \td I'={1\over \pi\alpha '}\int d^2 \s \big[
 \del u \bd v + \beta \r^{-2}  \bd \vp\del u
+ \del \r \bd \r + \r^{-2} \del \vp \bd  \vp
+ {\cal R} (\p_0 -  \ln \r) \big] \ , }
become gaussian when expressed in terms of the cartesian coordinates $x_1,x_2$.

\def \X {{\cal X}}

\subsec{Solution of the string model: quantum  Hamiltonian }
Relation to flat space model via formal duality and coordinate shifts
makes possible  to solve the classical  string equations
corresponding to \lagger\   explicitly (the two dual models have related
classical solutions),   expressing the solution in terms of free
fields  satisfying `twisted' boundary conditions \refs{\ruts,\rutsn}.
One can then   proceed with straightforward operator quantisation (fixing,
e.g.,  a light-cone type gauge).
Some of the resulting expressions are  similar to the  ones appearing in the
simpler cases of open string theory in a constant magnetic field  \abo\ or
$R^2/Z_N$ orbifold models \dab.

Introducing the free field  $X=X_1 + iX_2$
such that
\eqn\freee{ L_0=\del_+\rho\del_-\rho+ \rho^2 \del_+\hat \vp \del_-\hat \vp =
\del_+ X \del_- X^* \ ,
\ \ \  \ \ \  X\equiv  \r e^{i\hat \vp} \ ,   }
\eqn\dualsol{ \r^2 = XX^*\ , \ \ \ \hat \vp = {1\over 2i}\ln{X\ov X^*}\ , \ \ \
 X=X_+  (\s_+) + X_- (\s_-) \ , \ \ \s_\pm = \tau \pm \s \ ,  }
we get
\eqn\uuu{\del_\pm  \tilde  \vp= \mp \r^2 \del_\pm  \hat \vp=\pm {i\over
2}(X^*\del_\pm
X-X\del_\pm X ^*)\  , }
\eqn\jjj{
\tilde\vp(\s,\tau  )= 2\pi\a' [J_-(\s_-) - J_+(\s_+)]  +{i\ov 2} \big( X_+
X^*_-
-X_+^*  X_-\big)\ ,}
\eqn\cude{
J_\pm(\s_\pm)\equiv {i\ov 4\pi\a '}\int_0 ^{\s_\pm}d\s_\pm\big( X_\pm \del_\pm
X^*_\pm  -
X_\pm^* \del_\pm  X_\pm\big)\ .  }
One  then finds
\eqn\dualsole{
u=U_+ +U_- -\a\tilde\varphi \ ,\ \  \ \ v=V_+ +V_- -\b \tilde\varphi\  ,
}
where $U_\pm $ and $  V_\pm $ are arbitrary functions of $\s _\pm $,
and
 \eqn\gensol{
x \equiv \r e^{i\vp} = e^{-iq (u + v )} e^{i\a V_- -
i\b U_+    }  X \ . }
The closed string boundary condition
$\ x(\s + \pi, \tau)=x(\s, \tau)$ implies that the free field $X=X_+ +
X_-$ must satisfy  the   ``twisted"    condition
\eqn\bcfz{ X(\s + \pi, \tau)= e^{ i\g \pi } X(\s,\tau)\ , \ \ \  \
 X_\pm  = e^{\pm i\g \s_\pm } \X_\pm  \ ,  \ \ \  \X_\pm (\s_\pm \pm \pi)=
\X_\pm (\s_\pm )\ , }
where $\X_\pm = \X_\pm  (\s_\pm)$ are free  fields with standard periodic
boundary conditions
\eqn\fourie{ \X_+  =  i  \sqrt{\a'/ 2 } \sum_{n=-\infty}^\infty \tilde a_n
\exp (-2in \s_+)
  \  , \ \ \ \
\X_-  =  i  \sqrt{\a'/ 2 } \sum_{n=-\infty}^\infty  a_n
\exp (-2in \s_-)
  \ ,}
and $\g$ is determined by the above relations  and periodicity conditions (see
below).
Then
\eqn\bcfvp{
\tilde\vp (\s +\pi,\tau)=\tilde \vp (\s,\tau) -2\pi\a ' J\ ,\ \ \
J\equiv  J_L + J_R\ ,\   \  J_{L,R}\equiv J_\pm (\pi ) \ .
}
Since  $y=\ha (u+v) $ is compactified on a circle of radius
$R$,
\eqn\bcfuv{
u(\s+\pi, \tau)=u(\s,\tau)+2\pi wR\ ,\   \ \ v(\s+\pi, \tau)=v(\s,\tau)+2\pi w
R\
, \ \ \ w=0, \pm 1, ...,  }
where $w$ is  a winding number.
As a result,
\eqn\uuvv{
U_\pm =\s_\pm p_\pm ^u    + U_\pm '\ ,
\ \ \ \ V_\pm =\s_\pm p_\pm^v    + V_\pm'\ ,
}
\eqn\nombre{
p_\pm^u= \pm ( wR - \a' {\a}J)+p_u  \ ,\ \ \
p_\pm^v= \pm(wR -  \a' {\b  }J)+p_v  \ ,
}
where $U_\pm' $ and $V_\pm' $ are single-valued functions of $\s_\pm $
and $p_u$ and $p_v$   are arbitrary  parameters (related to the   Kaluza-Klein
momentum and the  energy of the string).
Then the   `twist' parameter  $\g$ in \bcfz\ is given by
\eqn\gamsol{
\gamma =(2q + \b-\a )wR+\b p_u+\a p_v
\ . }
\def \t {\tau}
 Evaluating   the classical
stress-energy tensor   on the general solution
\dualsol,\gensol\
one finds that it takes the ``free-theory" form
\eqn\tplus{
T_{\pm\pm}= \del_\pm U_\pm \del_\pm V_\pm   + \del_\pm
X \del_\pm X^*\ . }
This is not surprising since the on-shell values
of the stress-energy tensors in  the two dual \sms\  should be the same.
  It is convenient to fix the light-cone gauge, using the remaining  conformal
 symmetry
to gauge away,    e.g., the non-zero-mode parts   $U'_\pm $.
Then the  classical constraints $T_{--}=T_{++}=0$ can be solved
as
usual and  determine the remaining oscillators of $V'_\pm $
  in terms of the free fields  $X_\pm$.
The classical expressions for the Virasoro operators $L_0, \tilde L_0$
are
\eqn\vira{
L_0\equiv {1\ov 4\pi\a' } \int_0^\pi d\s\  T_{--} =   {p_-^up_- ^v \ov 4\a' }
 +\ha\sum_{n }  \big(n+\ha \gamma \big) ^2 a_n^{*} a_{n} ,
}
 \eqn\soro{
\tilde L_0\equiv { 1\ov 4\pi\a' } \int_0^\pi d\s\  T_{++} =   {p_+^u p_+^v\ov
4\a'} + \ha
\sum_{n}   \big( n-\ha  \gamma  \big) ^2      \td a_n^{*} \td a_{n} \ . }
where $p^{u,v}_\pm$ are given by \nombre\ with
the angular momentum defined in
\bcfvp,\cude\
 having the  following mode expansion:
\eqn\momop{ J=J_R+J_L\ , \ \ \
J_R=-\ha\sum _n (n+\ha\gamma ) a^*_na_n\ ,\ \ J_L=-\ha \sum_n (n-\ha\gamma )
\td a^*_n \td a_n \  . }
One can then quantize the theory by imposing  the canonical commutation
relations $ [P_x(\s,\t), x^*(\s',\t)]= -i\delta(\s -\s')$, etc.
As a result,  $p_u, p_v$  and the Fourier modes  $a_n, \td a_n$
will become  operators  in a  Hilbert  space.
Again,  the duality   between \lagger\ and \lag\
implies that  imposing  the above commutation relations  is equivalent
to demanding the canonical commutation relations
for the  fields $X, X^*$ of the free (but globally non-trivial,
cf.\bcfz) theory,
$[P_X(\s, \tau), X^*(\s', \tau)]= -i\delta(\s
-\s'), $ etc., where  $P_X(\s,\tau)= {1\ov 4\pi \alpha'} \del_\tau X $.
 Using  \bcfz,   we get
\eqn\fff{
[ a_n, a_m^{*}] =    2  (n+ \ha \gamma  )\inv   \delta _{nm}\ , \ \ \
[ \td a_n, \td a_m^{*}] =    2  (n - \ha \gamma )\inv   \delta _{nm } \ .
}
One also finds that   $p_u, p_v $ in \nombre\  and thus $\g$ in \gamsol\
commute with the mode operators.
The  string energy and  the Kaluza-Klein linear momentum operators are given
by
 \eqn\epjr{
E=  \int_0^\pi d\s P_t\ ,\ \ \  \ p_y= \int_0^\pi d\s P_y = { m\ov R}  \ ,\  \
m=0, \pm 1, \pm 2 , ... \ ,
}
\eqn\robo{
E=   { {1\ov 2\a '}} [p_u - p_v  -   \a'(\a + \b ) {\hat J}]   \ ,\ \
\   p_y=  { {1\ov 2\a '}} [ p_u + p_v + \a'   (2q + \b - \a ) \hat J]   \ .  \
}
Here  $\hat J$ is the angular momentum operator  obtained by
`symmetrizing' the classical expression $J=J_R+J_L$ in \momop.
Then
\eqn\gamlim{
\gamma =(2q + \b -\a)wR +\a'[(\a + \b )m R \inv - (\a -\b )E] -  \ha \a 'q(\a +
\b)\hat J\ .
}
The Virasoro operators $\hat L_0$ and $\hat {\td L_0}$
(and thus the quantum Hamiltonian $  \hat H  = \hat L_0 + {\hat {{\td L}}}_0 $)
are  found   by symmetrizing
the mode operator products in \vira,\soro.
In agreement with  the defining relations in \bcfz,   the
expressions for $\hat H$, $\hat J$ and the commutation relations \fff\
are invariant under $\g \to \g +2$ combined with the corresponding renaming
of the mode operators $a_n \to a_{n+1}, \ \td a_n \to \td a_{n-1}$.

The  sectors of states  of the  model  can be  labeled  by  the conserved
quantum numbers:  the   energy $E$, the angular momentum  $\hat J$ in the
$x_1,x_2$ plane,  and the
linear $m/R$ and winding  $wR$  Kaluza-Klein  momenta or ``charges"
(and also  by momenta in additional 22 spatial dimensions which we shall  add
to
saturate the central charge condition).
As in the case of the Landau model or the open string model \abo, the states
with  generic values of $\g$ are ``trapped" by the magnetic field.
 The  states in  the
 ``hyperplanes" in the $(m,w,E,\hat J)$ space with  $|\gamma| = 2n$, $
n=0,1,...,$
are special:
 for them  the  translational invariance on the $(x_1,x_2)$-plane is restored
with the   zero-mode  oscillators $a_0, a_0^*,  \td a_0, \td a_0^*$  being
replaced by the zero mode coordinate and conjugate  linear momentum.

Restricting the consideration to the sector of states with
 $ 0 <  \g  <2 $
 one can   introduce  the  normalized creation and annihilation
operators
 which will be used to define the Fock space of the  model
\eqn\ope{
[ b_{n\pm}, b_{m\pm }^{\dagger}] = \delta _{n m} \ ,\ \
 [\td b_{n\pm}, \td b_{m\pm }^{\dagger}] = \delta _{n m} \ ,\ \
[b_0,b_0^{\dagger}]=1 \ , \ \ [\td b_0,{\td b}_0^{\dagger}]=1 \ , }
where
$
 b_{n+}^{\dagger}= a_{-n} \omega_-  ,\ \    b_{n-}= a_{n} \omega_+\ ,
$ $ b_0=\ha \sqrt{\gamma }a_0
 , \  \ {\td b}_0^{\dagger}=\ha \sqrt{\gamma } \td a_0 ,$  etc.,
$
\omega_\pm \equiv \sqrt {  \ha \big( n \pm \ha {\gamma } \big) }, \
 \ \ \  n=1,2,...$.
Then
the  final expressions for the quantum Virasoro operators and Hamiltonian take
the form \rutsn\
\eqn\hamiltoni{
\hat H= {\hat L}_0 + {\hat {{\td L}}}_0=
 \ha  \a' \big( -E^2 +  p_a^2 + \ha Q_+^2 +
\ha Q_-^2  \big) + N+  {\td N}-2c_0
\  }
$$ -
\a'   [(q +\b)Q_+  + \b  E] J_R - \a'[ (q-\a) Q_-  + \a E] { J}_L $$
$$  +
 \ha \a'q\big [  (q+2\b)  J_R^2 + (q-2\a ) J_L^2
+  2 (q + \b -\a )  J_R  J_L  \big] \ ,   $$
\eqn\lzero { {\hat L}_0 - {\hat {{\td L}}}_0 = N-\td N -mw \ . }
Here $Q_\pm$ are the left and right combinations of the Kaluza-Klein  linear
and winding momenta (which play the role of charges in the present context)
\eqn\defi{  Q_\pm \equiv  {1\ov \sqrt {\a'}}  ({m\ov r} \pm  wr)\  , \ \ \  \ \
\
   \  r\equiv {R\ov \sqrt{\a' }}\ , }
and $p_a, \ a=3, ..., 24\ $ are   momenta  corresponding to  additional  free
spatial dimensions.
$c_0$ is the normal ordering  term\foot{ The  normal ordering constant
is  fixed by  the Virasoro algebra. The free-string
constant in  $\hat L_0$ is shifted  from 1 to
$1-\fourth \gamma (1-\ha\gamma) $ (or to
$1-\fourth \gamma' (1-\ha\gamma') $,  where $\g' = \g - 2k$ and $k$ is  an
integer,
in the case when $2k< \g< 2k +2$).
This corresponds to computing  the infinite sums  using   the generalised
$\zeta$-function
regularisation.
 Similar result is found in the
 open
string theory in  a constant magnetic field  \abo\ and is
characteristic  to the case of   a free scalar field with twisted boundary
conditions.
This shift is also  consistent  with  modular
invariance  of the partition function. }
\eqn\czero{
c_0 \equiv 1- \four \gamma(1-\ha \gamma) ,
}
where $\g$ was given  in \gamlim,
and the operators $N,\td N$  and  the angular momentum  operators  $J_L,J_R$
have the  standard `free-theory' form
($a_{na}, \td a_{na}$ are operators corresponding to  additional free spatial
directions, $a=3,...,24$)
$$
 N= \sum_{n=1}^\infty n ( b^{\dagger}_{n+}b_{n+}+ b^{\dagger}_{n-}b_{n-}
+ a^{\dagger}_{na} a_{na} ) \ , \ \
 {\td N}= \sum_{n=1}^\infty n ( \td b^{\dagger}_{n+}\td b_{n+}+ \td b^{\dagger
}_{n-}\td b_{n-}+\td a^{\dagger}_{na} \td a_{na}   ) , $$
\eqn\angull{ {\hat J}_R= - b^{\dagger}_0 b_0  - \ha  +\sum_{n=1}^\infty \big(
b^{\dagger
}_{n+}b_{n+} - b^{\dagger}_{n-}  b_{n-} \big)\equiv J_R-\ha \to -l_R - \ha +
S_R
 \ ,}
$$
{\hat J}_L= \tilde b^{\dagger}_0 \tilde b_0  + \ha  +\sum_{n=1}^\infty \big(
\tilde b^{\dagger
}_{n+}\tilde b_{n+} - \tilde b^{\dagger}_{n-} \tilde b_{n-} \big)\equiv J_L +
\ha
  \to l_L  +  \ha + S_L
 \ , $$
$$
\hat J= {\hat J}_R+\hat J_L  =J_R + J_L =J  \ . $$
The first line in \hamiltoni\  with $c_0 \to 1$ is the Hamiltonian
of the free string compactified on a circle.
The second line (together with $O(J)$ term from $c_0$)  is the analogue of the
gyromagnetic interaction term
for a particle in a magnetic field.\foot{The $O(EJ_{L,R})$  and $O(\a'E^2)$
terms (explicit in \hamiltoni\ and implicit in $c_0$ through its dependence on
$\g$ \gamlim) reflect
the non-static nature of corresponding backgrounds (related also to the
presence of the antisymmetric tensor).
$O(\a'E^2)$ terms  lead to a rescaling of the coefficient in front of $E^2$.}
 Similar term is present in the   Hamiltonian of conformal model describing
open string  in a constant magnetic field \refs{\abo,\burg,\ferrara}
\eqn\amil{
\hat H^{(open)} =L_0=
 \ha  \a' \big( -E^2 +  p_a^2) + N - c_0   - \g J_R
\ , }
$$ c_0 =1-\four \g  (1- \ha\g ) \ , \ \ \   \g \equiv {2 \ov \pi} |{\rm arctan}
(2\a' \pi Q_1 \b) +
{\rm arctan} (2\a' \pi Q_2  \b )| \ , $$
where $Q_1,Q_2$ are charges at the two ends of the open string, $N$ and $\hat
J_R$ have the same form as in \angull\
and $\b$ is the magnetic field, $F_{ij}=\b \epsilon_{ij}$.
The $O(J^2)$ terms in the  third line of \hamiltoni\  (and in $c_0$)
are special to closed string theory.

The Hamiltonian  \hamiltoni\    is, in general,    of  {\it fourth}
order in creation and annihilation operators.\foot{It is clear from our
construction that \hamiltoni\ is, at the same time, also
the Hamiltonian for the  $\vp$-dual theory \lag\  (the origin of the quartic
terms in $\hat H$ can be  traced, in particular,
 to the presence of the $\a\b \del \td \vp \bd \td \vp $ term in \lag).}
The quartic terms are absent
 when $q=0$, i.e., for example,
in the `constant magnetic field' model \lgg, where one finds \ruts
\eqn\htoni{
\hat H_{\a=q=0} =\ha   \a' \big( -E^2 +  p_a^2 + \ha Q_+^2 +
\ha Q_-^2  \big) + N+  {\td N} -2c_0 } $$  - \
\a' \b   (Q_+  +  E) J_R  \ , \ \ \ \
\ \
\g=\a'\b (Q_+ + E) \ .
$$
In the $a=1$  and $a=\sqrt 3 $ Melvin models \lggm\  and \lggme\ we get
\eqn\hai{
\hat H_{\a=\b=q}=
 \ha  \a' \big( -E^2 +  p_a^2 + \ha Q_+^2 +
\ha Q_-^2  \big) + N+  {\td N}-2c_0
\  }
$$ -
2\a' \b   Q_+   J_R - \a'\b  E { J}  -
 \ha \a'\b^2  J(J-4 J_R )   \ ,   $$
$$ \g = \a'\b Q_+  -\a'\b^2 J \  , \ \ \  J=J_R + J_L \ , $$
\eqn\hamioni{
\hat H_{\a=\b=0} =
 \ha  \a' \big( -E^2 +  p_a^2 + \ha Q_+^2 +
\ha Q_-^2  \big) + N+  {\td N}-2c_0
\  }
$$ -
\a' q (  Q_+  J_R  + Q_-  { J}_L)  +
 \ha \a'q^2  J^2 \ ,  \ \ \  \   \g= 2q wR \  . $$
The  $O(\g^2)$ normal ordering term in $c_0$ in \hamiltoni,\czero\
implies (see \gamlim) that the quantum Hamiltonian contains  $O(\a'^2)$ term
of one  order higher in $\a'$.
The presence of this  higher order term is consistent with
current algebra  approaches in the  two special cases when  our   model
becomes equivalent to  a
WZW or coset model:
 (i)  the non-compact $R=\infty$ limit of the
constant magnetic field model  \lgg\ is equivalent \ruts\ to
the $E^c_2$ WZW model \napwi\  for which the quantum stress tensor
contains  order $1/k \sim \a' $ correction term \refs{\kk,\sfees} (equivalent
to the term  appearing in \htoni\ in this limit);
\ \  (ii) the non-compact limit of the
Melvin model \lggm\  is  related \tsem\  to  a  special limit of the
$SL(2,R)\times R/R$
gauged WZW model, or  to the    $E^c_2/U(1)$ coset  theory,  the quantum
Hamiltonian of which also contains  $O(1/k)$ correction term
\refs{\sfetso,\sfees}.

\lr\nielsen{ N.K. Nielsen and P. Olesen, \jnl \np, B144, 376,  1978;
J. Ambjorn and P. Olesen, \jnl \np,  B315, 606, 1989; \jnl  \np,  B330, 193,
1990.}

The  analogs of the expressions \hamiltoni--\lzero\  in the sectors with
$2k<\gamma <2k +2,\ k=$integer,  can be found in a similar way   by  renaming
the creation and annihilation operators.
The result is the same as  in  \hamiltoni--\lzero\  with the replacement
$ \g \to \g '= \g-2k $ in $c_0$.

\subsec{Spectrum  of states}
Imposing the Virasoro conditions
\eqn\hhrh{ \hat L_0=
{\hat {{\td L}}}_0 =0\ , \ \ \    i.e.,  \ \ \ \  \hat H=0\   ,  \
\ \ \ \  N- {\td N}=mw  \ , }
it is straightforward to compute the spectrum of these models
just as for the free string theory \refs{\ruts,\rutsn}.
Indeed,  the
Hamiltonian \hamiltoni\ is in a {\it diagonal}  form  since  $N, \td N, J_L$
and
$J_R$ are diagonal in Fock space.

The continuous
momenta $p_{1,2}$  corresponding to the zero modes of the
coordinates $x_{1,2}$  of the plane are  effectively replaced  by the integer
eigenvalues
$l_R, l_L =0,1,2,...$ of the zero-mode parts
$b_0^{\dagger}b_0$ and $\td
b_0^{\dagger}\td b_0$   of $\hat J_R$ and
$\hat J_L$. Thus  the `2-plane'  part of  the spectrum is  discrete
in the $0<  \g < 2  $   sector
(but, as mentioned above, becomes   continuous  when  $\g=0$ or $\g=2$).\foot{
 The Hamiltonian for the case of $\g=0$ is obtained by adding
$\ha\a'(p_1^2+p_2^2)$ and replacing
$-b_0^{\dagger}b_0-\ha $ and $\td
b_0^{\dagger}\td b_0 + \ha $   in  $\hat J_R$ and
$\hat J_L$ in \angull\ by one half  of the  center of mass orbital momentum
$(x_1p_2-x_2p_1)$.}
 The  generic string states  are thus  `trapped'  by the flux tube.
This result  is  consistent with   a picture of a charged string moving
 in a   magnetic field
orthogonal to the plane.

The general property  of the spectrum is   the appearance of
new tachyonic instabilities, typically associated with states with
angular momentum aligned along  the magnetic field.
Similar  instabilities are   present in  point-particle field
theories
in  external magnetic fields (and may  lead to a  phase transition with
restoration of  some symmetries, see  \nielsen).
In the context of the
open string theory they were   observed in \abo\
and further investigated in  \ferrara.
The new feature of
 the   closed string theory  is the existence
    of states
 with arbitrarily large charges.
Since the critical
magnetic field at which a given state of a charge $Q$ becomes tachyonic
is of order of $(\a' Q)\inv $ there is  an infinite number  of  tachyonic
instabilities
for any given finite value of the magnetic field.

For example, let us consider a  non-winding
state at zero string excitation
level $S_L=S_R=N=\td N=0$ ($w=0$ tachyon).
The eigen-values of $\hat J_R$ and $\hat J_L$
are $-l_R -\ha $ and $l_L + \ha$  where $l_{L,R}=0,1,2,...$ are the
Landau-level type quantum numbers.
Then  in the $a=1$ Melvin model
$\hat H =0$ reduces to
\eqn\spece{ M^2\equiv E^2- p_a^2=
 - 4{\a'}\inv   + p_y^2  -4\b p_y \hat J_R  + 4\b^2  (\hat J_L + \hat J_R)\hat
J_R
- 2\a' \b^2(p_y - \b \hat J)^2
}
$$ = - 4{\a'}\inv    + p_y^2  + 2\b p_y (2l_R + 1)
- 2\b^2 (l_L-l_R)(2l_R + 1) - 2\a' \b^2[p_y - \b (l_L-l_R)]^2
 \ ,  $$
where  $p_y= m/R$ and it is assumed that
 $ 0< \g=  2\a' \b  [p_y - \b (l_L-l_R)] <2. $
The same expression for the  spectrum (up to the $O(\a')$ correction  coming
from the
$\g^2$ term  in $c_0$ in \hamiltoni) can be found \rutsn\ by
  directly solving (to the leading order in $\a'$) the  tachyon equation
\eqn\kg{\a' [\Delta + O(\a')]  T = 4 T  \ , \ \ \  \Delta = - {1\ov{\sqrt{ -G}
e^{-2\p} }} \del_\m ({\sqrt{ -G}  e^{-2\p} }G^{\m\n} \del_\n) \ .
  }
In the background corresponding to the $D=5$ Melvin model \lggm \kg\ takes the
form
   \eqn\mees{
\big[ - \del_t^2  + \r\inv \del_\r(\r\del_\r)
+ (\r^{-2} + 2\b^2  + \b^4 \r^2) \del_\vp^2 }  $$
\  + \   (1 + \b^2\r^2) \del_y^2
-2\b (1+ \b^2 \r^2) \del_\vp\del_y \big] T=
-4{\a'}\inv  T  \ ,  $$
and \spece\ is reproduced by taking
 $T= \exp (iEt + ip_y y + il\vp ) \ \tilde T(\r)  , \ l = l_L-l_R.$
While in Sections 7,8  and above  we  have assumed that the \sm is defined in
the `leading-order scheme',   once the Hamiltonian of the
corresponding CFT is known explicitly, we may also
 use the `CFT scheme'  in which the form of the
tachyon equation is  not modified  while the \sm background fields  (and thus
the string \sm action)
get $\a'$-corrections (cf. Section 2). For the $a=1$ Melvin model one finds
(cf. \lggm)
\eqn\lggms{L_{\a=\b=q}'    =  - \del t \bd t +  \del \r \bd \r +  {F(\r) \r^2}
(\del \vp  + 2 \b  \del y) ( \bd \vp  + 2 \b  \del y)  } $$
 + F'(\r) \del y \bd y +
 \del x_3 \bd x_3  +   {\cal R} \p(\r)  \ ,  $$
$$  e^{2(\p-\p_0)}=[F(\r) F'(\r)]^{1/2}= \r\inv \sqrt {- G} \  ,
\ \ \ \  \  F'= (1+ \b^2\r^2 -2\a'\b^2)\inv \ . $$
Similar correspondence between the string spectrum and the solution of the
tachyon equation
 is found  also in the constant magnetic field model \lgg\
where the point-particle limit of the Hamiltonian
(obtained by dropping all the oscillator terms and  replacing $J_R$ by its
orbital momentum eigenvalue)  is
\eqn\zerw{  \hat H_0 = {\ha }\a'   \big[- E^2+ { p}_y^2  +  p_a^2
  +2 ( { p}_y +  E)\b (l_R + \ha)
-\ha \a' \b^2(p_y +E)^2 \big] -2       \ .
}
The effect that the magnetic field produces on the energy of a generic
state is a combination of the gyromagnetic Landau-type interaction and the
influence of the
space-time geometry.
To illustrate the presence of the new tachyonic instabilities
 let us  consider the $\a=q=0$ model  and look at   the states which
complete the $SU(2)_{R}$ massless vector
multiplet in the free ($\b =0$)  theory
compactified at the self-dual radius $r=1$. The components with $S_R\neq 0$ are
given by
$
b^{\dag }_{1\pm }|0;m=w=1\rangle  ,\ \ b^{\dag }_{1\pm }|0;m=w=-1\rangle .$
For them $\td N=0$,  $\ J_R=-l_R \pm 1  $, $\ l_R=0,1,2,..., $
and the  energy is
\eqn\sutwor{
  \kappa\big[ E + {\k\inv \b  }({\hat J}_R+\ha \a'\b  Q_+ )\big]^2 =
-4{\a'}\inv   +
{\k\inv } (Q_+ - \b  {\hat J}_R)^2  \ , }
$$ \k\equiv  1 + \ha \a' \b^2 \ , \ \   \ \ \ Q_+ = {1\ov \sqrt{\a'} }
(r+ r\inv ) \ .  $$
At the self-dual radius, $r=1$, an infinitesimal magnetic field
 $\b >0$  makes
the component with $J_R=1$   tachyonic. This instability is the same as
in the
 non-abelian gauge  theory  \nielsen. Away from the self-dual
radius, this state has real  energy for small $\b$ and becomes tachyonic at
 some critical magnetic field.

 Instabilities caused by the linear in $\hat J_{L,R}$ terms
in $\hat H$ are present also in the $\a=\b$ models, in particular, in the
$a=1$ Melvin model.
There  are infinitely many tachyonic charged states at higher
levels.
 Consider, e.g.,  the  level one   state
with $w=0$, $m >0 $, $\  N=\td N=1$,  $l_R=l_L=0, \  S_R=1, S_L=- 1$ (i.e.  $
\hat J_{R,L}=\pm \ha $)
which  corresponds to  a   `massless' scalar
state with  a  Kaluza-Klein charge.
We may assume without  loss of generality that $R >  \sqrt{\a'}$ (if $R<\sqrt
{\a'}$ a similar discussion applies  with  $m$ replaced by  $w$).
Then the  expression for the mass is (cf. \spece)
\eqn\grave{
 M^2=  p_y(p_y -2 \b-2\a'\beta^2 p_y )    \  , \ \ \ \  \ \ p_y =  m/R \ .
}
For $R>>1$, $M^2$ becomes negative when $\b >\b_{\rm cr}\cong \ha p_y$.
For these states   $\g =2\a' \b p_y  $ and thus $\g <2$ if
 $\b >\b_{\rm cr}$ and  $\a' p^2_y <2$ (i.e. $R^2 > \ha  \a' m^2$).
The critical value of the magnetic field goes to zero as $R\to\infty $.
In the noncompact  $R=\infty $ theory  $p_y$ becomes a continuous parameter
representing the momentum of the  `massless' state in the $y$-direction.
Thus the `massless' state
 with  an infinitesimal momentum $p_y$ becomes tachyonic for an infinitesimal
value of $\b $.

It is possible to show  that the  $a=\sqrt {3}$  Melvin  model is stable in the
non-winding ($w=0$, i.e. $\g=0$) sector, i.e. it has  no new  instabilities in
addition to the usual flat space tachyon.  For $w\not=0$ (and $0< \g <2$)
one finds
\eqn\kkme{ M^2 =2{\a'}\inv (-2 + N+\td N)   +   (p_y -  q  \hat J)^2
 +   [{\a'}\inv wR - q (\hat J_R -\hat J_L)]^2 } $$
- \  q^2    (\hat J_R -\hat J_L)^2  -2{\a'}\inv q^2w^2R^2
\ ,  \ \  \   \ \ \g = 2 q w R  \ ,  $$
 and there exists a range of parameters $q,\ R\  $
for which  there is again the same linear instability as in the $a=1$ Melvin
model \grave.

\def \h {\chi}
\def \td {\tilde}
\subsec{Partition function}
Given the explicit expressions for the Virasoro operators \hamiltoni,\lzero\
it is straightforward to compute the partition   function of this conformal
model,
\eqn\zoper{
Z=\int {d^2\t\ov \t_2} \int dE\prod_{a=3}^{24} dp_a
\sum_{m,w=-\infty}^\infty  \Tr \exp \big[ 2\pi i( \t {\hat L_0} - \bar \t {\hat
{\td L}_0} )\big]\ .
}
After the integration  over the energy, momenta, Poisson resummation and
introduction
of two auxiliary variables $\l, \bar \l$
(in order to `split' the $O(J^2)$ terms in ${\hat L_0}$, ${\hat
{\td L}_0} $ to be able to compute the trace over the  oscillators)
one finds \rutsn
\eqn\etr{
 Z(r, \a,\b , q)  =c_1   \int [d^2\t]_1 \ W(r, \a,\b,q|\t, \bar \t) \ ,
}
\eqn\meas{ [d^2\t]_1 \equiv
 {d^2\t \ \tau_2^{-14} }
 e^{4\pi \t_2} |f_0(e^{2\pi i \t})|^{-48}\  , }
where the integrand  $W$ is given  by the sum over windings and two auxiliary
ordinary integrals
\eqn\wew{  W(r, \a,\b,q|\t, \bar \t)\ =  \ r (\a'\a\b\t_2)\inv
\sum_{w,w'=-\infty}^{\infty}
 \int d\l d\bar \l \  }
$$
\times \exp\big( -  \pi (\a' \a\b\t_2 )\inv
 [ \h \td \h   + \sqrt {\a'} r(q + \b )  (w'-  \t w) \td \h
 + \sqrt {\a'} r (  q -\a )      (w'- \bar \t w)\h $$
$$
+\  \a' r^2 q (q + \b -\a)  (w'-  \t w)(w'- \bar \t w)]\ \big) \ $$
$$ \times \
 \exp[{-{\pi  (\h - \td \h)^2 \ov 2 \t_2}}] \ \
 {  \h \td \h |\theta'_1(0|\t )|^2 \ov \theta_1(\h|\tau )
\theta_1 (\td\h |\bar \t ) } \ ,  $$
where
\eqn\ooo{
 \h\equiv - \sqrt{\a'}[ 2  \b \l + q  r (w'-\t w) ] \ ,\ \  \ \
\td \h \equiv - \sqrt{\a'}[ 2 \a\bar  \l   + q  r (w'-\bar\t w)] \ ,
 }
and $\l$ and $\bar \l$ are independent variables having infinite integration
limits.
Like the measure in \etr\  $W$  is $SL(2,Z)$ modular invariant
(to show this one needs to shift $w,w'$ and redefine $\h, \td \h$).
$Z (r, \a,\b,q)$  has several
symmetry properties:
\eqn\syt{ Z(r,\a,\b,q) = Z(r,-\b,-\a ,q)=Z(r,-\a,-\b,-q)=
 Z(r,\b ,\a,-q)\ .  }
It  is also invariant under the  duality in $y$ direction
which transforms the theory with  $y$-period  $2\pi R$ and  parameters
$(\a,\b,q)$ into the theory  with
$y$-period  $2\pi \a'/R$  and  parameters $(q,\b-\a+q, \a)$ or
parameters  $(\a-\b-q,-q,-\b)$,
 \eqn\duy{  Z(r,\a,\b,q) = Z(r\inv, q, \b -\a + q ,\a)   =
Z(r\inv, \a -\b - q , - q  ,-\b)  \ . }
For  $\a=q$   or $\b=-q$
the duality relations \duy\  retain their  standard  `circle' form
 \eqn\duyf{Z(r,\a,\b,\a) = Z(r\inv, \a, \b , \a) \ ,  \ \ \
 Z(r,\a,\b,-\b) = Z(r\inv, \a, \b ,-\b)  \ . }
When $\a=\b=q=0$ the partition function  $Z$ reduces to that of  the  free
string compactified on a circle of radius $R$.
Taking  the limit of the  non-compact $y$-dimension  ($R\to
\infty$) for generic $(\a,\b,q)$  one finds that
$Z$  \etr,\wew\
reduces  to the partition function of the free  bosonic closed string
theory.\foot{This generalizes a similar observation for the $\a=q=0$ model
\ruts. In
the limit $R=\infty$ the
$\a=q=0$ model \lgg\  is equivalent to the model of \napwi\
which has  trivial (free) partition function \kk.}

The   expression for $Z$ \etr,\wew\    simplifies
when at least
one of the   parameters  $\a, \b, q$ or $ q +\b-\a $
 vanishes so that the integrals over $\h, \td \h$ can be computed explicitly.
 For example, in the case when either $\a$ or $\b$ is equal to zero
(which includes the constant magnetic field model and $a=\sqrt 3 $ Melvin
model)
one finds
\eqn\rrtr{
 W(r, \a,\b,q|\t, \bar \t)\vert_{\a\b=0}\   =  \
\ r  \sum_{w,w'=-\infty}^\infty    \exp[{-I_0 (r ) }] \ }
$$ \times  \exp[{-{\pi  (\h _0- \td \h_0)^2 \ov 2 \t_2}}] \
{\   \h_0 \td \h_0 |\theta'_1(0|\t )|^2 \ov
\theta_1(\h_0|\tau )
\theta_1 (\td\h_0 |\bar \t )\ } \ , $$
$$
\h_0  = \sqrt {\a'} (q + \b )  r(w'-\t w) \ , \ \ \ \
\td \h_0 =\sqrt {\a'} (q - \a) r(w'-\bar \t w) \ , \ \  \ \a\b=0 \ ,    $$
\eqn\zeii{ \ I_0 (r) \equiv  \pi r^2  \t_2\inv {
 (w'-\t w)(w'-\bar \t w) } \ . }
The  magnetic instability of these models  (the presence of
tachyons  in the spectrum)
is reflected  in  singularities (or imaginary parts) of $Z$.
This can be seen explicitly  from  the behavior of the integrand
for large $\t_2 $.
The partition function has  new divergences at  critical values of the magnetic
field  parameters  when  the energy develops an imaginary
part.

\newsec{Concluding remarks}
We have discussed several classes of classical
solutions in bosonic string theory which are exact  in $\a'$.
Given that
 string  effective  field equations contain terms of all orders in $\a'$
their general solution can be represented as ($\vp$ stands for a set
of `massless' fields, see \rele): $ \vp=\vp_0 + \a' \vp_1 + \a'^2 \vp_2 + ...
$,
where $\vp_0$ is the leading-order  solution while
$\vp_n$ are, in general,
non-local (involving inverse Laplacians)  functionals of $\vp_0$.
It may happen that for some special $\vp_0$  all  higher order corrections
are {\it local, covariant}   functionals of $\vp_0$. In that case one is able
to change the scheme (i.e.,  to make a local covariant field redefinition) so
that  in the new scheme  $\vp_0$ is actually  an exact solution.
All of the presently explicitly known exact string solutions  are of that type;
 we do not  know an example
of a  solution  with truly non-trivial $\a'$-dependence (the one  which cannot
be absorbed into a local field redefinition).
In general, it seems to be an open question whether  all leading-order  string
solutions (e.g., Schwarzschild)
which can,   of course,   be formally deformed order by order in $\a'$ to make
them satisfy the
full string equations,   do have extensions  to  {\it regular} (satisfying
certain standard axioms)   CFT's which correspond to the
resulting $\a'$-series.\foot{Superstring  solutions  represented by \sms with
$(2,2)$ world sheet supersymmetry are examples where this does seem to  happen.
The corresponding $\beta$-function equations contain non-trivial
$\a'^3$- and higher- order corrections  \gris\ which cannot be redefined away
by a local redefinition of the {\it metric} (otherwise one would  eliminate the
$R^4$-
term in the effective action  \grwi\ and thus change the string S-matrix).  The
fact that there exists a local redefinition of the Kahler potential \nemsen\
seems to
indicate that the $(2,2)$ case is special compared to the general  $(1,1)$ one.
One  expects
that the resulting $\a'$-deformed  solution  corresponds to a regular $(2,2)$
superconformal theory.}

While the   form of  a deformation of a leading-order solution
into an exact one  may not be necessary to know   explicitly
 in the  case of `internal'
string solutions  describing compactified  spatial dimensions
(provided one knows the underlying CFT),   the exact form of the  background
{\it is}  important
in the case of `space-time'  solutions  where one is
interested in large-distance interpretation of a solution.
The determination  of  exact conformal \sm may, in  fact,
be considered as a first step towards finding  the  corresponding CFT.

As we have seen, special chiral null  models
are not modified by $\a'$ corrections.
This suggests that there exist  well-defined CFT's  corresponding to these
models, though
(in contrast to the magnetic flux tube  models)
they
 may be hard to solve in general (i.e.,  for the cases
  which are not, at the same time,  gauged WZW models).
The knowledge of these  CFT's, e.g.,  for
 the fundamental  string and extreme black hole solutions,
would be  important in order  to  be able to address the question  of
singularities.
The latter are determined by the properties of   test
 quantum string propagation in
these backgrounds which are encoded in CFT.
In particular,  CFT determines  the exact form of the
center-of-mass scalar (tachyon) equation;
   conclusions drawn from the analysis of  just the leading order  form  of the
tachyon equation\foot{For the fundamental string background  the leading-order
form of
 this equation has a remarkably simple  form of the Schrodinger equation  for a
charge in the Coulomb potential, suggesting that
the  origin $r=0$   is not  a singular point, see also \horry.}
  may turn out to be misleading.

We have seen that some spherically symmetric solutions  (e.g.,
\sstry, \ffss) having singularity at the origin need
to be supported by sources to satisfy  the conformal invariance conditions at
all points. Does that mean that they are not described by a regular CFT?
A related  issue  is  a breakdown of correspondence between solutions
of $\bar
\b$-function equations and effective string equations at points where
the metric $\k^{ij}$ in \rele\ is degenerate.
For example, the fundamental string solution
needs  a string source
 if viewed
as a solution of the  equations following from the effective action
\refs{\dabha,\gibb} but the corresponding \sm \mof,\fss\ is automatically
conformal at all points  \hrts.

\lr \tow {P.K. Townsend, ``String--membrane duality in seven dimensions",
hep-th/9504095.}

It is sometimes suggested to discriminate between `fundamental' and `solitonic'
 string solutions,  depending on whether they need or  not need to be supported
by sources and  whether the string metric is geodesically complete or not (see,
e.g.,
\tow\ and refs. there).
Both criteria  do not seem to be unambiguous
 as is demonstrated by
   the fundamental string  example.\foot{A peculiarity of the  fundamental
string
 solution is that it admits two possible interpretations. It can be  considered
 as a vacuum solution of the effective equations (an extremal limit of a
general class of charged string solutions \horstr)   which is valid only
outside the core $r=0$.
Alternatively, it can be  derived \gibb\ as a  solution corresponding to the
combined action $\hat S= { g^{-2}_0} S (\vp) + I_{str} (x; \vp),  \ \
I_{str}=I_0(x)  + V_i \vp^i , \ $
containing both the effective action for the background  fields  $\vp^i=
(G,B,\p)$ (condensates of massless string modes)  and the action
$I_{str} $
of a source string interacting with the background.  The static source action
 leads to  the $\d (r)$-term
in the Laplace equation for the bassic function  $F\inv$ (see \mof).
Though such mixture of actions looks strange from the point of view of
perturbative  string theory,
 $\hat S$ can  be interpreted  as  describing a non-perturbative
 `thin handle' (or `wormhole') approximation to quantum string partition
function  \tsmac. Extrema of $\hat S$ can thus be viewed as  some
non-perturbative
solitonic  solutions in string theory.  From this point of view it is not clear
if they can actually be described by a conformal field theory since
what is known only is   that  the solutions of the
tree-level effective equations  $\d S/\d \vp=0$ correspond to  CFT's.}
In particular,   the  question  of singularities should really be addressed at
the CFT level.
These  issues  related to CFT  description of exact  spherically symmetric
solutions, their singularities  and the role
of sources  need  further clarification.

\newsec{Acknowledgements}
The author would like to  acknowledge  useful discussions with J. Russo
and  K. Behrndt and  the support of PPARC
and NATO grant CRG 940870.
\vfill\eject
\listrefs
\end